\documentclass[notitlepage,a4,10.5pt]{article}%,twocolumn]{revtex4}

\usepackage{graphicx}

\usepackage{amsmath}%
\usepackage{amsfonts}%
\usepackage{amssymb}%
\usepackage{mathrsfs}
\usepackage{amsthm}
\usepackage{float}

\usepackage{authblk}

\usepackage[left=2.5cm,right=2.5cm,top=3cm,bottom=3cm]{geometry} 

\usepackage{xcolor}
\usepackage{stmaryrd}

\newcommand{\mc}{\mathcal}
\newcommand{\ms}{\mathscr}
\newcommand{\cp}{\times}

\newcommand{\bol}{\boldsymbol}

\newcommand{\abs}[1]{\left\lvert{#1}\right\rvert}

\newcommand{\lr}[1]{\left({#1}\right)}
\newcommand{\lrs}[1]{\left[{#1}\right]}
\newcommand{\lrc}[1]{\left\{{#1}\right\}}

\newcommand{\mf}{\mathfrak}
\newcommand{\p}{\partial}

\newcommand{\ti}[1]{\textit{#1}}
\newcommand{\tb}[1]{\textbf{#1}}

\newcommand{\eq}[1]{\begin{equation}\begin{split}{#1}\end{split}\end{equation}}
\newcommand{\sys}[2]{\begin{subequations}\begin{align}{#1}\end{align}\label{#2}\end{subequations}}

\newtheorem{mydef}{\textit{Def}}%[section]
\newtheorem{remark}{\textit{Remark}}%[section]
\begin{document}

\title{Scattering Theory in Noncanonical Phase Space:\\ A Drift-Kinetic Collision Operator for Weakly Collisional Plasmas}
%\title{Remarks on the Collision Operator in Noncanonical Phase Space}
\author[1]{Naoki Sato} 
\author[2]{Philip J. Morrison}
\affil[1]{National Institute for Fusion Science, \protect\\ 322-6 Oroshi-cho Toki-city, Gifu 509-5292, Japan \protect\\ Email: sato.naoki@nifs.ac.jp}
%\affil[2]{Graduate School of Frontier Sciences, \protect\\ The University of Tokyo, Kashiwa, Chiba 277-8561, Japan
%\protect \\ Email: saito@ppl.k.u-tokyo.ac.jp}
\affil[2]{Department of Physics and Institute for Fusion Studies, \protect\\ 
University of Texas, Austin, Texas
78712, USA
 \protect \\ Email: morrison@physics.utexas.edu}
\date{\today}
\setcounter{Maxaffil}{0}
\renewcommand\Affilfont{\itshape\small}

%\twocolumn[
 % \begin{@twocolumnfalse}
    \maketitle
    \begin{abstract}
After developing a scattering theory for grazing collisions in general noncanonical phase spaces, we introduce a guiding center collision operator 
in five-dimensional phase space 
designed for plasma regimes characterized by 
long wavelengths (relative to the Larmor radius), low frequencies (relative to the cyclotron frequency), and 
weak collisionality (where repeated Coulomb collisions induce cumulatively small changes in particle magnetic moment). The collision operator is fully determined by the noncanonical Hamiltonian structure of guiding center dynamics and exhibits a metriplectic structure, ensuring the conservation of particle number, momentum, energy, and interior Casimir invariants. It also satisfies an H-theorem, allowing for deviations from Maxwell-Boltzmann statistics due to the nontrivial kernel of the noncanonical guiding center Poisson tensor, spanned by the magnetic moment. We propose that this collision operator and its underlying mathematical structure may offer valuable insights into the study of turbulence, transport, and self-organizing phenomena in both laboratory and astrophysical plasmas.
    \end{abstract}
%\vspace{5mm}
%\end{@twocolumnfalse}
%  ]

%\naoki{To ensure conservation of energy, it may be convenient to use $v_{\parallel}$ instead of $u$}
%\naoki{Do we need a simple recap of the governing equations?}

\section{Introduction}

Plasma physics underlies both astrophysical phenomena and the pursuit of controlled nuclear fusion. A central challenge in the field is understanding plasma turbulence and its impact on confinement—an essential step toward achieving fusion energy \cite{Sche,Kikuchi}. Gyrokinetic theory has enabled significant advances in modeling plasma microturbulence \cite{Hahm88,Sugama,Brizard,Garbet}, and incorporating dissipative (non-ideal) processes within this framework remains an active area of research.

A fundamental question in this context is whether the steady states observed over experimental or observational time scales are accurately described by Maxwell-Boltzmann statistics. In principle, deviations from this distribution—which corresponds to the ultimate maximum entropy state under minimal constraints (typically, conservation of particle number and energy)—may arise when additional conservation laws persist over the system's evolution time scale.

In magnetically confined plasmas, such additional constraints are often associated with adiabatic invariants. Both theoretical studies \cite{YosM,Hel1,Hel2} and experimental observations \cite{Yoshida10,Boxer} suggest that meta-timescales exist over which adiabatic invariants, particularly the magnetic moment $\mu$, remain effectively conserved. This conservation can influence the statistical properties of steady states, as observed in laboratory settings (e.g., levitated dipole experiments \cite{Yoshida10,Boxer}) and astrophysical environments (e.g., formation of radiation belts \cite{SatoPRE}). Analogous deviations from Maxwell-Boltzmann statistics are expected in other magnetized plasmas, such as those confined in tokamaks and stellarators, under suitable plasma regimes.

When such deviations occur, the traditional Landau collision operator \cite{Landau1936,Kampen,Thompson}, which describes relaxation via grazing Coulomb collisions, becomes inadequate. The Landau operator invariably drives the system toward ``thermal death", i.e., a Maxwell-Boltzmann distribution, and therefore cannot capture the self-organized equilibrium states.

The goal of this paper is to develop a scattering theory for grazing collisions in general noncanonical phase spaces, whose structure incorporates adiabatic invariants in the form of Casimir invariants. We aim to construct a collision operator capable of describing relaxation to generalized equilibrium states—statistical steady states constrained by the noncanonical structure \cite{Morrison98,Morrison82} of phase space. We will then apply the theory to grazing Coulomb collisions in guiding center phase space.

Given the inherently out-of-equilibrium nature of many astrophysical and fusion plasmas, the reduced mathematical structure of the guiding center \cite{Cary} and gyrocenter equations of motion, and the complex coupling between the kinetic transport equation and turbulent electromagnetic fields \cite{Morrison13, Pfirsch}, deriving transport operators for drift-kinetic and gyrokinetic theories in a thermodynamically consistent manner remains a significant theoretical challenge. 
Despite these complexities, both Fokker-Planck-type operators \cite{Brizard04,Abel,Li} and Landau-type collision operators \cite{Burby,Hirv,Hir2} have been successfully formulated within the gyrokinetic framework, adhering to both the first law (energy conservation) and the second law (entropy growth) of thermodynamics. 
The core strategy involves accurately describing Coulomb collisions by expressing the Landau collision operator \cite{Landau1936,Kampen,Thompson} in gyrocenter coordinates. 

In this paper, we shift our focus from the microturbulence typically addressed by gyrokinetics to the long-wavelength (relative to the Larmor radius) and low-frequency (relative to the cyclotron frequency) turbulence characteristic of drift-kinetic regimes. This type of turbulence governs large-scale, self-organizing phenomena in both laboratory \cite{Yoshida10} and astrophysical \cite{Ewart22} plasmas. Our aim is to extend the theory of collision operators developed in \cite{SatoMorrison24} for general noncanonical Hamiltonian systems \cite{Morrison98,Morrison82}, deriving a two-species collision operator for weakly collisional guiding center plasmas.

In particular, we develop a general theory of grazing scattering in noncanonical phase space suitable for describing the Coulomb interaction, thereby improving upon the single-species collision operator introduced in~\cite{SatoMorrison24} to accommodate interactions between two species via the Coulomb force, while also incorporating self-consistent electromagnetic fluctuations. 
The resulting collision operator is intrinsically five-dimensional, meaning the kinetic equation, including the collision operator for Coulomb scatterings, does not involve the cyclotron phase and is not obtained through cyclotron averaging of a six-dimensional distribution function separated or expanded into phase-independent and phase-dependent components.

This result is achievable because we restrict our analysis to a long-wavelength, low-frequency, and weakly collisional plasma regime, where repeated Coulomb collisions induce only small cumulative changes in the particles' magnetic moments. Consequently, the magnetic moment remains approximately constant despite collisions, allowing for a complete reduction of the theory to a five-dimensional phase space.

The guiding center collision operator obtained in this study is fully determined by the noncanonical Hamiltonian structure of guiding center dynamics and the Coulomb potential governing interactions among charged particles. Furthermore, the evolution equation for the guiding center distribution function exhibits a \textit{metriplectic structure} \cite{Morrison86,Morrison09}, which is consistent with the conservation of particle number, momentum, energy, and interior Casimir invariants (inherited from microscopic particle dynamics, as discussed in \cite{SatoMorrison24}).

The conservation of interior Casimir invariants, particularly the system's total magnetic moment, leads to intriguing physical consequences: as anticipated, the thermodynamic equilibria arising from the H-theorem associated with the novel collision operator may deviate from Maxwell-Boltzmann statistics. This deviation results from the phase space constraints imposed by the conservation of the magnetic moment, leading to self-organized, inhomogeneous density distributions in configuration space. As such, we expect the derived collision operator to provide valuable insights into the properties of turbulence and transport in laboratory and astrophysical plasmas where self-organization plays a central role. Moreover, it offers an accurate yet computationally efficient kinetic model for numerical simulations of large-scale plasma phenomena.

We also note, as detailed in \cite{SatoMorrison24}, that the present collision operator can be applied in a ``collisionless'' regime, where the typical time between collisions exceeds the system's relaxation time. In this setting, the collision operator mathematically describes binary scatterings between clusters of charged particles.

The organization of this paper is as follows. In Sections~2 and 3, we generalize the theory of collisions in noncanonical phase space developed in~\cite{SatoMorrison24} by  formulating a general notion of grazing scattering \cite{Thirring} for noncanonical Hamiltonian systems. In addition, we relax certain conditions on the interaction potential energy, which will later prove essential for accurately describing Coulomb collisions.
The general form of the collision operator and the corresponding transport equation for the distribution function are presented, and the associated conservation laws, $H$-theorem, and equilibrium states are also discussed. 
In Section 4 we explain how the theory applies to grazing Coulomb collisions in canonical phase space, and how the resulting collision operator is related to the Landau operator.  
In Section 5 we review the noncanonical Hamiltonian structure governing the dynamics of a charged particle in the guiding center framework. Section 6 outlines the treatment of electromagnetic potentials within this theory. In Section 7, we quantitatively define the plasma regime under consideration, characterized by long wavelengths, low frequencies, and weak collisionality. Section 8 focuses on the reduction of the noncanonical Hamiltonian structure for a two-species plasma. In Section 9, we derive a guiding center collision operator by applying the framework developed in Sections 2 and 3 for constructing collision operators in noncanonical Hamiltonian systems. Section 10 demonstrates that the derived collision operator is consistent with the conservation of particle number, momentum, energy, and interior Casimir invariants, while also proving an H-theorem, ensuring compatibility with the laws of thermodynamics. In Section 11, leveraging the H-theorem, we infer the structure of thermodynamic equilibria and find that conservation of the magnetic moment, resulting from the nontrivial kernel of the guiding center Poisson tensor, leads to deviations from Maxwell-Boltzmann statistics. As a result, the particle density distribution exhibits a self-organized profile, with higher concentrations in regions of stronger magnetic field. In Section 12, we reveal that the derived collision operator exhibits a metriplectic structure, combining a noncanonical Poisson bracket and a dissipative bracket. 
In section 13, we draw comparisons between the Landau collision operator \cite{Landau1936,Kampen,Thompson,Lenard} for grazing Coulomb collisions in canonical phase space, its gyrokinetic formulation \cite{Burby,Hirv,Hir2,Brizard04}, and the collision operator derived in this study (equation \eqref{f1t2}) for Coulomb collisions in guiding center phase space. Additionally, we derive the linearized version of equation \eqref{f1t2}, equation \eqref{df11}, highlighting its connection to linearized model collision operators commonly used in gyrokinetic theory \cite{Sugama09,Sugama19}. 
Finally, concluding remarks are provided in Section 14.

\section{A collision operator for grazing scatterings in noncanonical phase space}

The purpose of this section is to derive a collision operator for grazing collisions in a general noncanonical phase space that improves upon the construction presented in~\cite{SatoMorrison24}, making it more suitable for accurately describing Coulomb collisions in reduced phase spaces, such as guiding-center phase space. Specifically, we generalize the derivation of the collision operator by removing the assumption that the interaction potential \( V(\bol{z}_a, \bol{z}_b) \) is
a function of the difference $\bol{z}_a-\bol{z}_b$ of the phase space coordinates \( \bol{z}_a \) and \( \bol{z}_b \) of the two colliding particles. We also relax the elastic scattering condition (Eq.~(18) in~\cite{SatoMorrison24}), which does not hold exactly when \( V \) corresponds to the Coulomb potential energy, by introducing the weaker notion of \ti{grazing scattering} in noncanonical phase space. These refinements will be essential in Section~9 for deriving the collision operator applicable to a two-species guiding-center plasma.

%A scattering process is an event in which an initial state involving free particles (or waves) evolves into a final state involving outgoing free particles (or waves), after an interaction governed by the underlying physical laws. 

We begin by considering a single particle with coordinates \( \bol{z} = (z^1, \dots, z^n) \in M \), where \( M\subset\mathbb{R}^n \) is the phase space domain with smooth boundary $\p M$. The equations of motion for the particle take the noncanonical Hamiltonian form
\begin{equation}
\dot{\bol{z}} = \mc{J} \cdot \partial_{\bol{z}} H,
\end{equation}
where \( \mc{J}(\bol{z}) \) is the Poisson \( 2 \)-tensor and \( H(\bol{z}) \) is the single-particle Hamiltonian.
Here, $\cdot$ denotes a contraction, while $\p_{\bol{z}}=\p/\p\bol{z}$. 

Let \( f(\bol{z}, t) \) denote the distribution function for an ensemble of \( N \) particles. Since we aim to formulate a kinetic equation for \( f \) that is consistent with both the first and second laws of thermodynamics, it is advantageous to adopt coordinates \( \bol{z} \) that define an invariant measure, allowing for a well-defined entropy functional. 
In what follows, we therefore assume that the coordinates \( \bol{z} \) define an invariant measure for any choice of Hamiltonian \( H \), that is,
\begin{equation}
\mf{L}_{\dot{\bol{z}}} \, d\bol{z} = 0 \quad \forall H 
\quad \iff \quad \frac{\partial \mc{J}^{ij}}{\partial z^i} = 0,~~~~j=1,...,n,\label{IM0}
\end{equation}
where \( \mf{L} \) denotes the Lie derivative and the summation convention on repeated indexes is used. 
We recall that, by the Lie–Darboux theorem~\cite{Arnold,deLeon,Littlejohn}, such invariant measures are always locally available.

Let
\eq{
\delta\bol{z}_a=\bol{z}_a'-\bol{z}_a,~~~~\delta\bol{z}_b=\bol{z}_b'-\bol{z}_b,
}
denote the phase space displacement caused by a binary collision in which two particles  located to  $\lr{\bol{z}_a,\bol{z}_b}$ are scattered at $\lr{\bol{z}_a',\bol{z}_b'}$. 
%In the following, we assume that the displacements $\delta\bol{z}_a$ and $\delta\bol{z}_b$ are small. 
Let $\mc{V}\lr{\bol{z}_a,\bol{z}_b;\bol{z}_a',\bol{z}_b'}=\mc{V}\lr{\bol{z}_b,\bol{z}_a;\bol{z}_b',\bol{z}_a'}$ denote the scattering volume density per unit time. The term $\mc{V}\,d\bol{z}_b\,d\bol{z}_a'\,d\bol{z}_b'$ represents the phase space  volume 
swept by a particle at $\bol{z}_a$ within a 
unit time interval, which is accessible for collisions. The collision operator of the system can be expressed as
\eq{
\mc{C}\lr{f,f}=\int\mc{V}\lr{\bol{z}_a,\bol{z}_b;\bol{z}_a',\bol{z}_b'}\lrs{f\lr{\bol{z}_a',t}f\lr{\bol{z}_b',t}-f\lr{\bol{z}_a,t}f\lr{\bol{z}_b,t}}d\bol{z}_bd\bol{z}_a'd\bol{z}_b',
}
where integrals are performed on the whole phase space.  %and $f_a$ and $f_b$ denote the colliding  pair guiding center distribution functions. 
In the following, we set $f_a'=f\lr{\bol{z}_a',t}$,  $f_b'=f\lr{\bol{z}_b',t}$, $f_a=f\lr{\bol{z}_a,t}$, and $f_b=f\lr{\bol{z}_b,t}$. A similar notation will be used for derivatives, e.g., $\p f_a/\p\bol{z}_a=\p f/\p\bol{z}\rvert_{\bol{z}_a}$. 
Note that the Boltzmann collision operator corresponds to the case $\bol{z}=\lr{\bol{v},\bol{q}}$, with $\bol{v}$ and $\bol{q}$ the particle velocity and position respectively, and 
\eq{
\mc{V} =m^{-6}\sigma\abs{\bol{v}_a-\bol{v}_b}\delta\lr{\bol{q}_a-\bol{q}_b}\delta\lr{\bol{q}_a'-\bol{q}_b'}\delta\lr{\bol{q}_a-\bol{q}_a'},\label{mcV}
}
with $m$ the particle mass and $\sigma\lr{\bol{v}_a,\bol{v}_b;\bol{v}_a',\bol{v}_b'}$ the scattering cross section.  

In the weakly collisional regime under consideration, the dispacements $\delta\bol{z}_a$ and $\delta\bol{z}_b$ are small (they result in minor changes of the particles trajectories). Hence, the integrand may be expanded in powers of $\delta\bol{z}_a$ and $\delta\bol{z}_b$. At second order, one can verify that
\eq{
\mc{C}\lr{f,f}=&\int\mc{V}\lrs{f_a\frac{\p f_b}{\p\bol{z}_b}\cdot\delta\bol{z}_b+f_b\frac{\p f_a}{\p\bol{z}_a}\cdot\delta\bol{z}_a}d\bol{z}_bd\bol{z}_a'd\bol{z}_b'\\&+\int\mc{V}\lrs{\frac{1}{2}f_a\delta\bol{z}_b\cdot\frac{\p^2 f_b}{\p\bol{z}_b^2}\cdot\delta\bol{z}_b+\frac{\p f_a}{\p\bol{z}_a}\cdot\delta\bol{z}_a
\frac{\p f_b}{\p\bol{z}_b}\cdot\delta\bol{z}_b+\frac{1}{2}f_b\delta\bol{z}_a\cdot\frac{\p^2 f_a}{\p\bol{z}_a^2}\cdot\delta\bol{z}_a
}d\bol{z}_bd\bol{z}_a'd\bol{z}_b',
}
which can be rearranged as follows:
\eq{
\mc{C}\lr{f,f}=&
\int{\mc{V}}\lrs{f_a\frac{\p f_b}{\p\bol{z}_b}\cdot\delta\bol{z}_b+f_b\frac{\p f_a}{\p\bol{z}_a}\cdot\delta\bol{z}_a}\,d\bol{z}_bd\bol{z}_a'd\bol{z}_b'\\&-\frac{1}{2}\int f_a\frac{\p f_b}{\p\bol{z}_b}\cdot\lrs{\frac{\p}{\p\bol{z}_b}\cdot\lr{{\mc{V}}\delta\bol{z}_b\delta\bol{z}_b}+\frac{\p}{\p\bol{z}_a}\cdot\lr{\mc{V}\delta\bol{z}_a\delta\bol{z}_b}}\,d\bol{z}_bd\bol{z}_a'd\bol{z}_b'\\&-\frac{1}{2}\int f_b\frac{\p f_a}{\p\bol{z}_a}\cdot\lrs{\frac{\p}{\p\bol{z}_b}\cdot\lr{\mc{V}\delta\bol{z}_b\delta\bol{z}_a}+\frac{\p}{\p\bol{z}_a}\cdot\lr{\mc{V}\delta\bol{z}_a\delta\bol{z}_a}}
\,d\bol{z}_bd\bol{z}_a'd\bol{z}_b'\\
&+\frac{1}{2}\int\frac{\p}{\p\bol{z}_b}\cdot\lrs{f_af_b\mc{V}\lr{\delta\bol{z}_b\frac{\p \log f_b}{\p\bol{z}_b}\cdot\delta\bol{z}_b+
\delta\bol{z}_b\frac{\p \log f_a}{\p\bol{z}_a}\cdot\delta\bol{z}_a
}}\,d\bol{z}_bd\bol{z}_a'd\bol{z}_b'\\
&\frac{1}{2}\frac{\p}{\p\bol{z}_a}\cdot\int f_af_b\mc{V}\lr{\delta\bol{z}_a\delta\bol{z}_b\cdot\frac{\p\log f_b}{\p\bol{z}_b}+\delta\bol{z}_a\delta\bol{z}_a\cdot\frac{\p\log f_a}{\p\bol{z}_a}}\,d\bol{z}_bd\bol{z}_a'd\bol{z}_b'.\label{C1}
}
The fourth term on the right-hand side of \eqref{C1} vanishes by appropriate choice of boundary conditions. For example, %we may demand that the scattering frequency $\mc{V}$ to vanish on the boundary. 
%Alternatively, 
we may assume that the distribution function $f_b$ vanishes on the boundary under the condition that  $\p\log f_b/\p\bol{z}_b$ does not diverge there. 
Similarly, if %$\mc{V}$ or 
$f_a$  vanishes on the boundary, integrating equation \eqref{C1} with respect to $d\bol{z}_a$ gives
\eq{
0=&\int f_a\frac{\p f_b}{\p\bol{z}_b}\cdot\lrs{\mc{V}\delta\bol{z}_b-\frac{1}2{}\frac{\p}{\p\bol{z}_b}\cdot\lr{{\mc{V}}\delta\bol{z}_b\delta\bol{z}_b}-\frac{1}{2}\frac{\p}{\p\bol{z}_a}\cdot\lr{\mc{V}\delta\bol{z}_a\delta\bol{z}_b}}\,d\bol{z}_ad\bol{z}_bd\bol{z}_a'd\bol{z}_b'\\&+
\int f_b\frac{\p f_a}{\p\bol{z}_a}\cdot\lrs{\mc{V}\delta\bol{z}_a-\frac{1}{2}\frac{\p}{\p\bol{z}_b}\cdot\lr{{\mc{V}}\delta\bol{z}_b\delta\bol{z}_a}-\frac{1}{2}\frac{\p}{\p\bol{z}_a}\cdot\lr{\mc{V}\delta\bol{z}_a\delta\bol{z}_a}}\,d\bol{z}_ad\bol{z}_bd\bol{z}_a'd\bol{z}_b',\label{dNdtcoll}
}
where we used the fact that the particle number is preserved by collisions ($\int\mc{C}\lr{f_a,f_b}d\bol{z}_a=0$) and eliminated  boundary integrals.
Now observe that, 
at a given instant, 
%although $f_a$ and $f_b$ are statistically correlated (the shape of one is affected by the shape of the other, as well as the scattering frequency $\mc{V}$, and the ) 
there is no restriction on the shape of the distributions $f_a$ and $f_b$, and that the terms inside the square brackets are independent of $f_a$ and $f_b$. 
For \eqref{dNdtcoll} to vanish for arbitrary $f_a$ and $f_b$, we should therefore enforce the following conditions on $\mc{V}$, 
\sys{
&\int\lrs{\mc{V}\delta\bol{z}_b-\frac{1}2{}\frac{\p}{\p\bol{z}_b}\cdot\lr{{\mc{V}}\delta\bol{z}_b\delta\bol{z}_b}-\frac{1}{2}\frac{\p}{\p\bol{z}_a}\cdot\lr{\mc{V}\delta\bol{z}_a\delta\bol{z}_b}}\,d\bol{z}_a'd\bol{z}_b'=0,\\
&\int\lrs{\mc{V}\delta\bol{z}_a-\frac{1}{2}\frac{\p}{\p\bol{z}_b}\cdot\lr{{\mc{V}}\delta\bol{z}_b\delta\bol{z}_a}-\frac{1}{2}\frac{\p}{\p\bol{z}_a}\cdot\lr{\mc{V}\delta\bol{z}_a\delta\bol{z}_a}}\,d\bol{z}_a'd\bol{z}_b'=0,
}{V}
to ensure conservation of particle number. 
The collision operator \eqref{C1} thus reduces to
\eq{
\mc{C}\lr{f,f}=\frac{1}{2}\frac{\p}{\p\bol{z}_a}\cdot\int f_af_b\mc{V}\lr{\delta\bol{z}_a\delta\bol{z}_b\cdot\frac{\p\log f_b}{\p\bol{z}_b}+\delta\bol{z}_a\delta\bol{z}_a\cdot\frac{\p\log f_a}{\p\bol{z}_a}}\,d\bol{z}_bd\bol{z}_a'd\bol{z}_b'.\label{C2}
}
%\section{Grazing collisions in noncanonical phase space}
Next, we define the average energy $E$ of the two interacting particles according to 
\eq{
E\lr{\bol{z}_a,\bol{z}_b,t}=H_a\lr{\bol{z}_a}+\Phi_a\lr{\bol{z}_a,t}+H_b\lr{\bol{z}_b}+{\Phi}_b\lr{\bol{z}_b,t}, 
}
where  
\eq{
\Phi_{a}\lr{\bol{z}_a,t}=\int f\lr{\bol{z}_b,t}V\lr{\bol{z}_a,\bol{z}_b}\,d\bol{z}_b,~~~~\Phi_{b}\lr{\bol{z}_{b},t}=\int f\lr{\bol{z}_a,t}V\lr{\bol{z}_b,\bol{z}_a}\,d\bol{z}_a,
}
are the ensemble averages of the binary  interaction  potential energy $V\lr{\bol{z}_a,\bol{z}_b}$. 
A grazing scattering between two particles governed by the interaction potential \( V \) is one that results in a small variation in the average energy \( E \)
along the particle orbits $\bol{z}_a\lr{t}$ and $\bol{z}_b\lr{t}$:

\begin{mydef}
A grazing scattering in noncanonical phase space is defined by the property
\eq{
\frac{1}{E_-}\int_{-\infty}^{+\infty}\frac{dE}{dt}\,dt
%=1-\frac{E_{+}}{E_{-}}
=\frac{1}{E_-}\int_{-\infty}^{+\infty}\frac{\p E}{\p t}\,dt+O\lr{\varepsilon}
%=1-\frac{E_{\infty}}{E_0}=O\lr{\epsilon}
,\label{grazing}
}
where $E_-=\lim_{t\rightarrow -\infty}E\lr{\bol{z}_a\lr{t},\bol{z}_b\lr{t},t}$,
%$E_{+}=\lim_{t\rightarrow+\infty}E\lr{t}$, 
and $\varepsilon>0$ is a small positive constant (ordering parameter). 
\end{mydef}

We now observe that equation~\eqref{C2} applies to general interaction potential energies, as the form of \( V \) has not been specified. This generality will be exploited in Section~9 to derive the collision operator for a two-species guiding-center plasma, where multiple particles interact via the Coulomb force.

For the time being, however, we specialize to the case in which the interaction potential energy takes the form 
\eq{V = V(\abs{\bol{z}_a - \bol{z}_b}),} a setting that is appropriate for understanding the properties of the collision operator in a single-species context. 
We also assume that the interaction potential vanishes at infinity, i.e.,
\begin{equation}
\lim_{\abs{\bol{z}_a - \bol{z}_b} \rightarrow +\infty} V = 0,
\end{equation}
and, for physical reasons that will become clear later, we also require that the effective range \( \ell_i \) of the interaction in the \( z^i \)-direction is much shorter than the characteristic scale \( L_i \) associated with the ideal part of the dynamics in that direction, and that the collision time \( \tau_c \) is much shorter than the characteristic time scale $T$ of the system. 
In formulae:
\eq{
\frac{H_\sigma+\Phi_\sigma}{\p_{z^i_\sigma}\lr{H_\sigma+\Phi_\sigma}}\sim\frac{\mc{J}_{\sigma}^{jk}}{\p_{z^i_\sigma}\mc{J}^{jk}_\sigma}\sim L_i\gg \ell_i,~~~~\frac{L_i}{\mc{J}_\sigma^{ij}\p_{z^j_\sigma}\lr{H_\sigma+\Phi_\sigma}}\sim T\gg\tau_c,~~~~\sigma=a,b,~~~~i,j,k=1,...,n,
}
and
\eq{
\frac{\ell_i}{L}\sim\frac{\tau_c}{T}\sim\epsilon, 
}
where $\epsilon$ is a small positive constant. 
Under these assumptions the grazing scattering condition \eqref{grazing} 
for a collision starting at time $t_0$ 
becomes
\eq{
\frac{1}{E_{t_0}}\int_{t_0}^{t_0+\tau_c}\frac{dE}{dt}\,dt=
%\frac{E_{t_0+\tau_c}}{E_{t_0}}-1=
\frac{1}{E_{t_0}}\int_{t_0}^{t_0+\tau_c}\frac{\p E}{\p t}\,dt+O\lr{\epsilon^2}, \label{grazing2}
}
where we set $\varepsilon=\epsilon^2$ to take into account the fact that the time interval $\tau_c$ is small, and we used the notation $E_t=E\lr{t}$. 
Now notice that
\eq{
\frac{dE}{dt}-\frac{\p E}{\p t}=\p_{\bol{z}_a}\lr{H_a+\Phi_a}\cdot\mc{J}_a\cdot\p_{\bol{z}_a}V-\p_{\bol{z}_b}\lr{H_b+\Phi_b}\cdot\mc{J}_b\cdot\p_{\bol{z}_b}V,
}
%Now notice that, due to the brevity of the interaction, we have
%\eq{
%\int_{t_0}^{t_0+\tau_c}T\frac{\p\Phi_a}{\p t}\frac{dt}{T}=O\lr{\epsilon},~~~~a=1,2. 
%}
where $\mc{J}_a=\mc{J}\lr{\bol{z}_a}$ and so on. 
Using the fact that $\p_{\bol{z}_a}V=-\p_{\bol{z}_b}V$, the grazing scattering condition \eqref{grazing2} thus takes the form 
\eq{
\frac{1}{E_{t_0}}\int_{t_0}^{t_0+\tau_c}\lrs{\mc{J}_a\cdot\p_{\bol{z}_a}\lr{H_a+\Phi_a}-\mc{J}_b\cdot\p_{\bol{z}_b}\lr{H_b+\Phi_b}}\cdot\p_{\bol{z}_a}V\,dt=O\lr{\epsilon^2}.
}
Let us define the vector field 
\eq{
\bol{\xi}=\mc{J}_a\cdot\p_{\bol{z}_a}\lr{H_a+\Phi_a}-\mc{J}_b\cdot\p_{\bol{z}_b}\lr{H_b+\Phi_b}.\label{xi}
}
Then, we arrive at 
\eq{
\frac{1}{E_{t_0}}\int_{t_0}^{t_0+\tau_c}{\p_{\bol{z}_a}}V\cdot\bol{\xi}\,dt=O\lr{\epsilon^2}.\label{grazing3} 
}
Equation \eqref{grazing3} expresses the fact that the effect of the component of the force $\p_{\bol{z}_a}V$ along the vector field $\bol{\xi}$ is small. 
Specifically, the phase space displacement $\delta\bol{z}_\sigma$, $\sigma=a,b$, caused by a collision can be expressed as
\eq{
\delta\bol{z}_\sigma=\int_{t_0}^{t_0+\tau_c}\mc{J}_\sigma\cdot\p_{\bol{z}_\sigma}V\,dt=\mc{J}_\sigma\lr{\bol{z}_\sigma\lr{t_0}}\cdot\int_{t_0}^{t_0+\tau_c}\p_{\bol{z}_\sigma 
}V\,dt+O\lr{\epsilon^2}.
}
Now define
\eq{
\p_{\bol{z}_\sigma}^{\perp}V=\p_{\bol{z}_\sigma}V-\frac{\p_{\bol{z}_\sigma}V\cdot\bol{\xi}}{\bol{\xi}^2}\bol{\xi},
}
and note that equation \eqref{grazing3} implies
\eq{
\int_{t_0}^{t_0+\tau_c}\p_{\bol{z}_\sigma}V\,dt=\int_{t_0}^{t_0+\tau_c}\p_{\bol{z}_\sigma}^{\perp}V\,dt+O\lr{\epsilon^2}=\lr{I-\frac{\bol{\xi}\bol{\xi}}{\bol{\xi}^2}}_{t_0}\cdot\int_{t_0}^{t_0+\tau_c}\p_{\bol{z}_{\sigma}}V\,dt+O\lr{\epsilon^2}, \label{ord}
}
where $I$ is the identity matrix. 
It follows that, 
%at leading order, 
\eq{
\delta\bol{z}_\sigma=\mc{J}_{\sigma}\cdot\int_{\tau_c}\p_{\bol{z}_{\sigma}}V\,dt+O\lr{\epsilon^2}=
\mc{J}_{\sigma}\cdot\mathbb{P}^{\perp}\cdot\int_{\tau_c}\p_{\bol{z}_{\sigma}}V\,dt+O\lr{\epsilon^2}, 
%\mc{J}_\sigma\int_{\tau_c}\p^{\perp}_{\bol{z}_\sigma}V\,dt,
}
where $\mc{J}_\sigma$ is evaluated at $\bol{z}_\sigma\lr{t_0}$, we defined the projector  
\eq{
\mathbb{P}^{\perp}=\lr{I-\frac{\bol{\xi}\bol{\xi}}{\bol{\xi}^2}}_{t_0}
}
and we introduced the simplified notation $\int_{\tau_c}=\int_{t_0}^{t_0+\tau_c}$. 
Note that the property $\p_{\bol{z}_a}V=-\p_{\bol{z}_b}V$ implies
\eq{
\delta\bol{z}=\delta\bol{z}_a=-\delta\bol{z}_b+O\lr{\epsilon^2}.
}
Hence, introducing the 
\textit{interaction tensor}
\eq{
\Pi=\int\frac{1}{2}\mc{V}\lr{
\mathbb{P}^{\perp}\cdot
\int_{\tau_c}\frac{\p V}{\p \bol{z}_a}\,dt}\lr{ \mathbb{P}^{\perp}\cdot\int_{\tau_c}\frac{\p V}{\p\bol{z}_a}\,dt}d\bol{z}_a'd\bol{z}_b',\label{IT1}
}
the collision operator \eqref{C2}
becomes
\eq{
C\lr{f,f}=\frac{\p}{\p\bol{z}_a}\cdot\lrs{ f_a\mc{J}_a\cdot\int f_b\Pi\cdot\lr{\mc{J}_b\cdot\frac{\p\log f_b}{\p\bol{z}_b}-\mc{J}_a\cdot\frac{\p\log f_a}{\p\bol{z}_a}}\,d\bol{z}_b}.\label{COp0}
}
The full kinetic equation for the distribution function $f_a\lr{\bol{z}_a,t}$ therefore reads 
\eq{
\frac{\p f_a}{\p t}=\frac{\p}{\p \bol{z}_a}\cdot\lrc{f_a\mc{J}_a\cdot\lrs{-\frac{\p \lr{H_a+\Phi_a}}{\p\bol{z}_a}+\int f_b\Pi\cdot\lr{\mc{J}_b\cdot\frac{\p\log f_b}{\p\bol{z}_b}-\mc{J}_a\cdot\frac{\p\log f_a}{\p\bol{z}_a}}\,d\bol{z}_a}},\label{dfdt0}
}
If the collision time $\tau_c$ is short enough to allow a meaningful expansion of $\mathbb{P}^{\perp}\cdot\int_{\tau_c}\p_{\bol{z}}V\,dt$ in powers of $\tau_c$,    
we may define the \textit{scattering frequency}
\eq{
\Gamma=\int\mc{V}\,d\bol{z}_a'd\bol{z}_b',
}
and introduce a simplified  
interaction tensor 
\eq{
\Pi_{\epsilon^2}=\int\Pi\,d\bol{z}_a'd\bol{z}_b'+O\lr{\epsilon^3}=\frac{1}{2}\tau_c^2\Gamma\lr{
\mathbb{P}^{\perp}\cdot
\frac{\p V}{\p\bol{z}_a}}\lr{\mathbb{P}^{\perp}\cdot\frac{\p V}{\p\bol{z}_a}}. \label{IT2}
}
In cases of practical interest, 
one would expect that, due to integration, the diagonal terms of the tensor $\int_{\tau_c}\p_{\bol{z}_a}V\,dt\int_{\tau_c}\p_{\bol{z}_a}V\,dt$ 
in equation \eqref{IT1} to be dominant, leading to an alternative simplified form for the interaction tensor,
\eq{
\Pi^{\perp}=D\,\mathbb{P}^{\perp},~~~~D=\Gamma \Delta^2,\label{D}
}
where the function $D$ plays a role analogous to a diffusion coefficient,   
the function $\Delta\lr{\bol{z}_a,\bol{z}_b}$ represents the characteristic 
size of the impulse $\int_{\tau_c}\p_{\bol{z}_a}V\,dt$ caused by a scattering event, and, for this expression, the coordinates $\bol{z}_a$ are assumed to be normalized (i.e., without physical dimensions).  

The appearance of equations \eqref{COp0} and \eqref{dfdt0} can be slightly simplified by replacing $f_a$ with $f$ and $f_b$ with $f'$ and so on, where the prime symbol $'$ indicates evaluation at $\bol{z}'$. Then,  
the collision operator has the form,  
\eq{
C\lr{f,f}=\frac{\p}{\p\bol{z}}\cdot\lrs{ f\mc{J}\cdot\int f'\Pi\cdot\lr{\mc{J}'\cdot\frac{\p\log f'}{\p\bol{z}'}-\mc{J}\cdot\frac{\p\log f}{\p\bol{z}}}\,d\bol{z}'},\label{Cop}
}
while the 
full kinetic equation for the distribution function $f\lr{\bol{z},t}$ takes the form 
\eq{
\frac{\p f}{\p t}=\frac{\p}{\p \bol{z}}\cdot\lrc{f\mc{J}\cdot\lrs{-\frac{\p \lr{H+\Phi}}{\p\bol{z}}+\int f'\Pi\cdot\lr{\mc{J}'\cdot\frac{\p\log f'}{\p\bol{z}'}-\mc{J}\cdot\frac{\p\log f}{\p\bol{z}}}\,d\bol{z}'}}.\label{dfdt}
}
Finally, we remark that equations \eqref{Cop} and \eqref{dfdt} apply to the case in which $f$ and $f'$ describe collisions between different statistical ensembles, i.e., $f' \neq f\lr{\bol{z}', t}$.  
In such a case, the equation for $f'$ can be obtained by replacing $f$ with $f'$ and so on in equation \eqref{dfdt}.

%We conclude this section by observing that if the collision time $\tau_c$ is short enough to allow a meaningful expansion of $\mathbb{P}^{\perp}\int_{\tau_c}\p_{\bol{z}}V\,dt$ in powers of $\tau_c$, the interaction tensor may be expressed in the simplified leading order form
%\eq{
%\Pi=\frac{1}{2}\tau_c^2\Gamma\lr{
%\mathbb{P}^{\perp}\cdot
%\frac{\p V}{\p\bol{z}}}\lr{\mathbb{P}^{\perp}\cdot\frac{\p V}{\p\bol{z}}}. \label{IT2}
%}

\section{Conservation laws, entropy production, and  equilibria}
Equation~\eqref{dfdt} preserves the particle number, total energy, and the interior Casimir invariants (the Casimir invariants induced on the field theory by the Casimir invariants of single particle  dynamics).  
Moreover, any additional quantity conserved during both isolated particle motion and binary collisions gives rise to a corresponding macroscopic conservation law for equation~\eqref{dfdt}. For instance, if angular momentum is conserved during both the single-particle dynamics and each collision event, then the total angular momentum is a conserved quantity of the system~\eqref{dfdt}.

Equation~\eqref{dfdt} also satisfies an $H$-theorem, from which thermodynamic equilibria can be explicitly determined. These equilibria generally deviate from the standard Maxwell--Boltzmann distribution due to the conservation of interior Casimir invariants.

In this section, we verify these conservation laws and derive the associated entropy production and equilibrium states. 
%\naoki{We emphasize that the same properties 
%apply to the parent equation \eqref{dfdt0}.} 

\subsection{Conservation of particle number}
Equation \eqref{dfdt} is in divergence form. 
Hence, conservation of total probability (particle number)
\eq{
\mc{N}=\int f\,d\bol{z},
}
follows by application of Gauss's theorem under appropriate boundary conditions. 
Let $M\ni\bol{z}$ denote the phase space domain occupied by the system. Define 
the effective phase space velocity 
\eq{
\bol{Z}=\mc{J}\cdot\lrs{\frac{\p \lr{H+\Phi}}{\p\bol{z}}-\int f'\Pi\cdot\lr{\mc{J}'\cdot\frac{\p\log f'}{\p\bol{z}'}-\mc{J}\cdot\frac{\p\log f}{\p\bol{z}}}\,d\bol{z}'}.\label{Zeff}
}
Then, the boundary conditions required for conservation of $\mc{N}$ are
\eq{
f\bol{Z}\cdot\bol{n}=
%f\mc{J}\cdot\lrs{\frac{\p \lr{H+\Phi}}{\p\bol{z}}+\int f'\Pi\cdot\lr{\mc{J}'\cdot\frac{\p\log f'}{\p\bol{z}'}-\mc{J}\cdot\frac{\p\log f}{\p\bol{z}}}\,d\bol{z}'}\cdot\bol{n}=
0~~~~{\rm on}~~\p M,\label{bcN}
}
where $\bol{n}$ denotes the unit outward normal of the bounding surface $\p M$.
Physically, equation \eqref{bcN} expresses the fact that the normal  component of the  effective phase space flux $f\bol{Z}$ must vanish on $\p M$ (particles do not escape from $M$). 

\subsection{Conservation of energy}
The total energy of the system is given by sum of the energies of the interacting pairs, 
\eq{
\mathscr{H}=\int ff'\lr{H+V+H'}\,d\bol{z}d\bol{z}'=\int f\lr{H+\frac{1}{2}\Phi}\,d\bol{z}+\int f'\lr{H'+\frac{1}{2}\Phi'}\,d\bol{z}'.\label{Energy}
%=\int \lr{H_a+\Phi_a}f_a\,d\bol{z}_a+\int \lr{H_a+\Phi_b}f_b\,d\bol{z}_b.
}
We note that this expression yields twice the Hamiltonian of the Vlasov equation,
$
\int f\left(H + {\Phi}/{2}\right)\, d\boldsymbol{z},
$
when \( H = m \boldsymbol{v}^2/2 \) is the kinetic energy and \( V = {q^2}/{4\pi\epsilon_0 \lvert \boldsymbol{x} - \boldsymbol{x}' \rvert} \) is the Coulomb potential energy.

\begin{remark}
Note that in equation \eqref{Energy} $\int f\lr{H+\Phi/2}\,d\bol{z}=\int f'\lr{H'+\Phi'/2}\,d\bol{z}'$ when $f$ and $f'$ pertain to the same statistical ensemble, and that 
including the term involving primed quantities in the expression of $\ms{H}$ simplifies the algebra 
associated with the evaluation of the rate of change of $\ms{H}$. 
In the following, functionals $\ms{F}[f]$ of $f$ will often be expressed as the sum of identical terms, one in terms of $f$ and the other in terms of $f'$, $\ms{F}[f]=\ms{F}[f]/2+\ms{F}[f']/2$, in order to 
simplify the evaluation of $d\ms{F}/dt$. 
When $f$ and $f'$ represent distinct statistical ensembles, 
the form \eqref{Energy} also provides the relevant invariant. 
\end{remark}

An explicit evaluation of the rate of change in $\mathscr{H}$ shows that
\eq{
\frac{d\mathscr{H}}{dt}=&\int \frac{\p f}{\p t}\lr{H+\Phi}\,d\bol{z}+
%\int\frac{\p f}{\p t}\,d\bol{z}
\frac{d\mc{N}}{dt}\int f'H'\,d\bol{z}'+\int \frac{ \p f'}{\p t}\lr{H'+\Phi'}\,d\bol{z}'+\frac{d\mc{N}}{dt}\int fH\,d\bol{z}
%\int \frac{\p f'}{\p t}\,d\bol{z}'
\\
=&\int\frac{\p}{\p\bol{z}}\cdot\lrc{\lr{H+\Phi}f\mc{J}\cdot\lrs{-\frac{\p\lr{H+\Phi}}{\p\bol{z}}
+\int f'\Pi\cdot\lr{\mc{J}'\cdot\frac{\p\log f'}{\p\bol{z}'}-\mc{J}\cdot\frac{\p\log f}{\p\bol{z}}}\,d\bol{z}'
}}\,d\bol{z}\\
&-\int ff'\frac{\p\lr{H+\Phi}}{\p\bol{z}}\cdot \mc{J}\cdot \Pi\cdot\lr{\mc{J}'\cdot\frac{\p\log f'}{\p\bol{z}'}-\mc{J}\cdot\frac{\p\log f}{\p\bol{z}}}\,d\bol{z}d\bol{z}'\\
&\int\frac{\p}{\p\bol{z}'}\cdot\lrc{\lr{H'+\Phi'}f'\mc{J}'\cdot\lrs{-\frac{\p\lr{H'+\Phi'}}{\p\bol{z}'}
+\int f\Pi\cdot\lr{\mc{J}\cdot\frac{\p\log f}{\p\bol{z}}-\mc{J}'\cdot\frac{\p\log f'}{\p\bol{z}'}}\,d\bol{z}
}}\,d\bol{z}'\\
&-\int ff'\frac{\p\lr{H'+\Phi'}}{\p\bol{z}'}\cdot \mc{J}'\cdot \Pi\cdot\lr{\mc{J}\cdot\frac{\p\log f}{\p\bol{z}}-\mc{J}'\cdot\frac{\p\log f'}{\p\bol{z}'}}\,d\bol{z}d\bol{z}'\\
=&\int ff'\bol{\xi}\cdot\Pi\cdot\lr{\mc{J}'\cdot\frac{\p\log f'}{\p\bol{z}'}-\mc{J}\cdot\frac{\p\log f}{\p\bol{z}}}\,d\bol{z}d\bol{z}'=0,
}
where we used the facts that 
$\mc{N}=1$, 
$\p_{\bol{z}}V=-\p_{\bol{z}'}V$, $\bol{\xi}=-\bol{\xi}'$,  
{$\Pi\lr{\bol{z},\bol{z}'}=\Pi\lr{\bol{z}',\bol{z}}$}, 
and $\bol{\xi}\cdot\Pi=0$, 
and eliminated boundary integrals through the boundary condition \eqref{bcN}. 

\subsection{Conservation of interior Casimir invariants}
Let $C^k$, $k=1,...,m$, denote the $m=n-{\rm rank}\lr{\mc{J}}$ Casimir invariants spanning the kernel of the Poisson tensor $\mc{J}$. The interior Casimir invariants are the functionals
\eq{
\mathscr{C}^k=
\int fC^k\,d\bol{z}.
%+
%\int f'C^{k}{}'\,d\bol{z}'.
}
It follows that
\eq{
\frac{d\mathscr{C}^k}{dt}=&
\int\frac{\p}{\p\bol{z}}\cdot\lr{C^kf\mc{J}\cdot\lrs{-\frac{\p\lr{H+\Phi}}{\p\bol{z}}+\int f'\Pi\cdot\lr{\mc{J}'\cdot\frac{\p\log f'}{\p\bol{z}'}-\mc{J}\cdot\frac{\p\log f}{\p\bol{z}}}\,d\bol{z}'
}}\,d\bol{z}\\
&-\int f\frac{\p C^k}{\p\bol{z}}\cdot\mc{J}\cdot\lrs{-
\frac{\p\lr{H+\Phi}}{\p\bol{z}}+\int f'\Pi\cdot\lr{\mc{J}'\cdot\frac{\p\log f'}{\p\bol{z}'}-\mc{J}\cdot\frac{\p\log f}{\p\bol{z}}}\,d\bol{z}'
}\,d\bol{z}=0,
}
where we used the boundary condition \eqref{bcN} to eliminate boundary integrals, and the fact that $\mc{J}\cdot\p_{\bol{z}}C^k=\bol{0}$. 

\subsection{Other invariants}
Let $\mf{p}\lr{\bol{z}}$ denote some physical quantity, and consider the observable 
\eq{
\mathscr{P}=\int f\mf{p}\,d\bol{z}+\int f'\mf{p}'\,d\bol{z}'.\label{msP}
}
Applying boundary conditions \eqref{bcN}, we have
\eq{
\frac{d\mathscr{P}}{dt}=&
\int \frac{\p}{\p\bol{z}}\cdot\lrc{\mf{p}f\mc{J}\cdot\lrs{-
\frac{\p\lr{H+\Phi}}{\p\bol{z}}+\int f'\Pi\cdot\lr{\mc{J}'\cdot\frac{\p\log f'}{\p\bol{z}'}-\mc{J}\cdot\frac{\p\log f}{\p\bol{z}}}\,d\bol{z}'
}}\,d\bol{z}\\
&-\int\frac{\p\mathfrak{p}}{\p\bol{z}}\cdot f\mc{J}\cdot\lrs{-\frac{\p\lr{H+\Phi}}{\p\bol{z}}+\int f'\Pi\cdot\lr{\mc{J}'\cdot\frac{\p\log f'}{\p\bol{z}'}-\mc{J}\cdot\frac{\p\log f}{\p\bol{z}}}\,d\bol{z}'}\,d\bol{z}\\
&+\int \frac{\p }{\p\bol{z}'}\cdot\lrc{\mf{p}'f'\mc{J}'\cdot\lrs{-
\frac{\p\lr{H'+\Phi'}}{\p\bol{z}'}+\int f\Pi\cdot\lr{\mc{J}\cdot\frac{\p\log f}{\p\bol{z}}-\mc{J}'\cdot\frac{\p\log f'}{\p\bol{z}'}}\,d\bol{z}
}}\,d\bol{z}'\\
&-\int\frac{\p\mathfrak{p}'}{\p\bol{z}'}\cdot f'\mc{J}'\cdot\lrs{-\frac{\p\lr{H'+\Phi'}}{\p\bol{z}'}+\int f\Pi\cdot\lr{\mc{J}\cdot\frac{\p\log f}{\p\bol{z}}-\mc{J}'\cdot\frac{\p\log f'}{\p\bol{z}'}}\,d\bol{z}}\,d\bol{z}'\\
=&\int f\frac{\p\mf{p}}{\p\bol{z}}\cdot\mc{J}\cdot\frac{\p\lr{H+\Phi}}{\p\bol{z}}\,d\bol{z}+
\int f'\frac{\p\mf{p}'}{\p\bol{z}'}\cdot\mc{J}'\cdot\frac{\p\lr{H'+\Phi'}}{\p\bol{z}'}\,d\bol{z}'\\
&+\int ff'\lr{\mc{J}\cdot\frac{\p\mf{p}}{\p\bol{z}}-\mc{J}'\cdot\frac{\p\mf{p}'}{\p\bol{z}'}}\cdot\Pi\cdot\lr{\mc{J}'\cdot\frac{\p\log f'}{\p\bol{z}'}-\mc{J}\cdot\frac{\p\log f}{\p\bol{z}}}\,d\bol{z}d\bol{z}'.
}
From this equation, we see that a sufficient condition for $\mathscr{P}$ to qualify as an invariant is that $\mf{p}$ is a constant of motion of single particle dynamics,
\eq{
\frac{\p\mf{p}}{\p\bol{z}}\cdot\mc{J}\cdot\frac{\p\lr{H+\Phi}}{\p\bol{z}}=0,
}
and that $\mf{p}+\mf{p}'$ is preserved during a collision,
\eq{
\lr{\mc{J}\cdot\frac{\p\mf{p}}{\p\bol{z}}-\mc{J}'\cdot\frac{\p\mf{p}'}{\p\bol{z}'}}\cdot\mathbb{P}^{\perp}\cdot\int_{\tau_c}\p_{\bol{z}}V\,dt=0.\label{xip}
}
As an example, consider 
conservation of the total $x$-momentum $\mathscr{P}$ corresponding to 
$\mf{p}=p_x$ in a phase space $M$ with canonical  coordinates $\bol{z}=\lr{\bol{p},\bol{q}=\bol{x}}$ and symplectic Poisson tensor $\mc{J}=\mc{J}_c$. 
We have
\eq{
\frac{d\mathscr{P}}{dt}=
-\int f\frac{\p \lr{H+\Phi}}{\p x}\,d\bol{z}-\int f'\frac{\p\lr{H'+\Phi'}}{\p {x}'}\,d\bol{z}',
}
which vanishes if, for example, $H=H\lr{\bol{p}}$ is the kinetic energy, 
and $\Phi\lr{y,z,t}$ is
an $x$-symmetric electrostatic potential.

\subsection{H-theorem and thermodynamic equilibria}
The kinetic equation \eqref{dfdt} satisfies an $H$-theorem. 
The entropy measure of the system is given by the functional  
\eq{
S=-\int f\log f\,d\bol{z}-\int f'\log f'\,d\bol{z}'.
}
%Recall that, as explained in Remark 1 above, the two integrals are identical, 
%the primed part being used to simplify the algebra in the evaluation of $dS/dt$.
Using the boundary conditions \eqref{bcN}, the rate of change in $S$ can be evaluated as
\eq{
\frac{dS}{dt}=&-\int \frac{\p f}{\p t}\lr{1+\log f}\,d\bol{z}
-\int \frac{\p f'}{\p t}\lr{1+\log f'}\,d\bol{z}'\\
=&-\int \bol{Z}\cdot\frac{\p f}{\p\bol{z}}\,d\bol{z}-\int \bol{Z}'\cdot\frac{\p f'}{\p\bol{z}'}\,d\bol{z}'\\
=&-\int \frac{\p f}{\p\bol{z}}\cdot{\mc{J}\cdot\frac{\p\lr{H+\Phi}}{\p\bol{z}}}\,d\bol{z}
-\int 
\frac{\p f'}{\p\bol{z}'}\cdot
{\mc{J}'\cdot\frac{\p\lr{H'+\Phi'}}{\p\bol{z}'}}\,d\bol{z}'\\
&+\int ff'\lr{\mc{J}'\cdot\frac{\p\log f'}{\p\bol{z}'}-\mc{J}\cdot\frac{\p\log f}{\p\bol{z}}}\cdot\Pi\cdot\lr{\mc{J}'\cdot\frac{\p\log f'}{\p\bol{z}'}-\mc{J}\cdot\frac{\p\log f}{\p\bol{z}}}\,d\bol{z}d\bol{z}'.
}
Now recall that $\bol{z}$ defines an invariant measure, in the sense of equation \eqref{IM0}. 
Furthermore, 
since we want 
the system to be thermodynamically isolated 
regardless of the magnitude of collisions, we require the ideal part $\mc{J}\cdot\p_{\bol{z}}\lr{H+\Phi}$ of the effective phase space velocity $\bol{Z}$ to be tangent to the boundary, i.e., 
\eq{
\bol{n}\cdot\mc{J}\cdot\frac{\p\lr{H+\Phi}}{\p\bol{z}}=0~~~~{\rm on}~~\p M.\label{bcN2}
}
Then, 
\eq{
\int_M \frac{\p f}{\p\bol{z}}\cdot\mc{J}\cdot\frac{\p\lr{H+\Phi}}{\p\bol{z}}\,d\bol{z}=\int_{\p M}f\bol{n}\cdot\mc{J}\cdot\frac{\p\lr{H+\Phi}}{\p\bol{z}}\,dS=0,
}
where $dS$ denotes the surface element on $\p M$. 
It follows that 
\eq{
\frac{dS}{dt}=
\frac{1}{2}\int ff'\mc{V}\lrs{\lr{\mc{J}'\cdot\frac{\p\log f'}{\p\bol{z}'}-\mc{J}\cdot\frac{\p\log f}{\p\bol{z}}}\cdot\mathbb{P}^{\perp}\cdot\int_{\tau_c}\p_{\bol{z}}V\,dt}^2\,d\bol{z}d\bol{z}'d\bol{z}''d\bol{z}'''\geq 0,
}
where we assumed that $f$, $f'$, and $\mc{V}\lr{\bol{z},\bol{z}';\bol{z}'',\bol{z}'''}\geq 0$ and substituted the expression \eqref{IT1} for the interaction tensor. 

If the system achieves a state of thermodynamic equilibrium, we must have
\eq{
\lim_{t\rightarrow+\infty}\frac{dS}{dt}=0\iff \lim_{t\rightarrow +\infty} \lr{\mc{J}'\cdot\frac{\p\log f'}{\p\bol{z}'}-\mc{J}\cdot\frac{\p\log f}{\p\bol{z}}}\cdot\mathbb{P}^{\perp}\cdot\int_{\tau_c}\p_{\bol{z}}V\,dt=0.\label{theq}
}
Let
\eq{
f_{\infty}\lr{\bol{z}}=\lim_{t\rightarrow+\infty}f,  
}
denote the equilibrium distribution function and define 
\eq{
\bol{\Delta}_{\infty}=\mc{J}'\cdot\frac{\p\log f'_{\infty}}{\p\bol{z}'}-\mc{J}\cdot\frac{\p\log f_{\infty}}{\p\bol{z}}.
}
Equation \eqref{theq} is equivalent to
\eq{
%\mc{J}'\cdot\frac{\p\log f'_{\infty}}{\p\bol{z}'}-\mc{J}\cdot\frac{\p\log f_{\infty}}{\p\bol{z}}
\bol{\Delta}_{\infty}\in {\rm ker}\lr{\Pi},
}
and can be satisfied whenever
\eq{
\bol{\Delta}_{\infty}=\beta\bol{\xi}+\gamma_i\bol{\xi}_{\mf{p}^i},~~~~\beta,\gamma_i\in\mathbb{R},\label{kerpi}
}
where $\bol{\xi}\in {\rm ker}\lr{\mathbb{P}^{\perp}}$ 
is the vector field \eqref{xi} and $\bol{\xi}_{\mf{p}^i}=\mc{J}\cdot\p_{\bol{z}}\mf{p}^{i}-\mc{J}'\cdot\p_{\bol{z}'}\mf{p}^i{}'\in {\rm ker}\lr{\Pi}$ are other elements of the kernel of the interaction tensor  $\Pi$ associated with additional scattering invariants $\mf{p}$ as described by equation \eqref{xip}. 
From equation \eqref{kerpi}, we thus obtain the family of thermodynamic equilibria
\eq{
\log f_{\infty}=-\beta\lr{H+\Phi}-\gamma_i\mf{p}^i+g\lr{\bol{C}},\label{theq2}
}
where $\bol{C}=\lr{C^1,...,C^k}$ are the Casimir invariants of the Poisson tensor $\mc{J}$, and the function  $g\lr{\bol{C}}$ is arbitrary (in the sense that any choice of $g$ corresponds to a steady solution). In practice, the function $g$ can be determined from the initial conditions for $f$; more details and examples on how to determine $g$ will be given at the end of Section~11. 
We remark that the equilibria given in equation~\eqref{theq2} deviate from Maxwell--Boltzmann statistics due to the presence of the function \( g(\bol{C}) \), which reflects the conservation of Casimir invariants, and the nontrivial structure of the invariant measure \( d\bol{z} \), which generally differs from the standard configuration space measure, such as \( d\bol{v}\, d\bol{x} \).

\section{Grazing Coulomb collisions in canonical phase space}
It is useful to consider the form of the collision operator \eqref{COp0} and the kinetic equation \eqref{dfdt0} when $V=\kappa/\abs{\bol{q}_a-\bol{q}_b}$ is the Coulomb potential, with $\kappa\in\mathbb{R}_{\geq 0}$, $H=m\bol{v}^2/2$ the kinetic energy of a particle with mass $m$, $\Phi\lr{\bol{q},t}$ the electrostatic potential, and the phase space is canonical, with phase space measure  $d\bol{z}=d\bol{p}\,d\bol{q}=m^3d\bol{v}\,d\bol{q}$ and symplectic Poisson tensor $\mc{J}=\mc{J}_c$. 
%Note that, since the potential \( V \) is singular at \( \bol{q}_a = \bol{q}_b \), the calculation is more straightforward using equation~\eqref{IT1} (rather than~\eqref{IT2}), where the interaction tensor \( \Pi \) is expressed in terms of the finite impulse \( \int_{\tau_c} \partial_{\bol{z}_a} V\, dt \). 
Recalling the expression of the scattering volume density in this context, equation \eqref{mcV}, we have
\eq{
\mc{C}\lr{f,f}=&\frac{1}{2}\frac{\p}{\p\bol{z}_a}\cdot\left[ f_a\mc{J}_c\cdot\int f_b \frac{\sigma\abs{\bol{v}_a-\bol{v}_b}}{m^6}\delta\lr{\bol{q}_a-\bol{q}_b}\delta\lr{\bol{q}_a'-\bol{q}_b'}\delta\lr{\bol{q}_a-\bol{q}_a'}  \lr{\mathbb{P}^{\perp}\cdot\int_{\tau_c}\frac{\p V}{\p\bol{z}_a}\,dt}\lr{\mathbb{P}^{\perp}\cdot\int_{\tau_c}\frac{\p V}{\p\bol{z}_a}\,dt}\right.\\&\left.\cdot\mc{J}_c\cdot\lr{\frac{\p\log f_b}{\p\bol{z}_b}-\frac{\p\log f_a}{\p\bol{z}_a}}\,d\bol{z}_bd\bol{z}_a'd\bol{z}_b'\right].}
As shown in Section~6 of~\cite{SatoMorrison24}, if the projector \( \mathbb{P}^{\perp} \) is removed from this equation—which is justified under the present ordering assumption given in equation~\eqref{ord}—then the collision operator reduces to the Landau collision operator via the standard procedure involving the Coulomb logarithm.

It is, however, instructive to examine the implications of retaining the projector \( \mathbb{P}^{\perp} \): doing so yields a collision operator that is mathematically analogous to the Landau operator, but obtained without invoking a minimum deflection angle, i.e., a cutoff at the Coulomb logarithm. 

To see this, we begin by observing that the impulse caused by a Coulomb scattering is given by 
\eq{\delta\bol{z}=\lr{m\delta\bol{v},\bol{0}}=m\lr{\bol{v}'-\bol{v},\bol{0}}=-\lr{\int_{\tau_c}\frac{\p V}{\p\bol{q}}\,dt,\bol{0}}.} 
Let
$\mf{f}\lr{\bol{v},\bol{q},t}=m^3f\lr{\bol{p},\bol{q},t}$ denote the distribution function on the measure $d\bol{v}d\bol{q}$. 
Hence, we may write 
\eq{C\lr{\mf{f},\mf{f}}=&\frac{1}{2}\frac{\p}{\p\bol{z}_a}\cdot\left[\mf{f}_a\mc{J}_c\cdot\int \mf{f}_b\sigma\abs{\bol{v}_a-\bol{v}_b}
\delta\lr{\bol{q}_a-\bol{q}_b}{\mathbb{P}^{\perp}\cdot\lr{\mc{J}_c\cdot\delta\bol{z}_a}}{\lr{\mc{J}_c\cdot\delta\bol{z}_a}}\cdot\mathbb{P}^{\perp}\right.\\&\left.\cdot\mc{J}_c\cdot\lr{\frac{\p\log \mf{f}_b}{\p\bol{z}_b}-\frac{\p\log \mf{f}_a}{\p\bol{z}_a}}\,d\bol{v}_bd\bol{q}_bd\bol{v}_a'd\bol{v}_b'\right].
}
It is now convenient to introduce the Coulomb interaction tensor 
\eq{
\Pi_C=\frac{1}{2}\abs{\bol{v}_a-\bol{v}_b}\int\sigma\delta\bol{v}_a\delta\bol{v}_a\,d\bol{v}_a'd\bol{v}_b', \label{PiC}
}
set $\mf{f}_{b}=\mf{f}\lr{\bol{v}_b,\bol{q}_a,t}$, and define the velocity space projector
\eq{
\mathbb{P}^{\perp}_v=I_v-\frac{\lr{\bol{v}_a-\bol{v}_b}\lr{\bol{v}_a-\bol{v}_b}}{\abs{\bol{v}_a-\bol{v}_b}^2},
}
with $I_v$ the $3\times 3$ identity matrix. 
The collision operator then becomes 
\eq{
C\lr{\mf{f},\mf{f}}=&\frac{\p}{\p\bol{v}_a}\cdot\lrs{\mf{f}_a\int \mf{f}_{b}{\mathbb{P}^{\perp}_v\cdot\Pi_C\cdot\mathbb{P}^{\perp}_v\cdot\lr{\frac{\p\log \mf{f}_{a}}{\p\bol{v}_a}-\frac{\p\log \mf{f}_b}{\p\bol{v}_b}}\,d\bol{v}_b}
}.\label{CffL}}
Next, observe that the dominant contributions to the integral in equation~\eqref{PiC} arise from the diagonal components of the tensor \( \delta\bol{v}_a \delta\bol{v}_a \), since the off-diagonal terms \( \delta v_a^i \delta v_a^j \) for \( i \neq j \) are expected to largely cancel out upon integration. 
Let \( r_{\mathfrak{n}} = \mf{n}^{-1/3} \) denote the characteristic spatial separation between particles in a system with spatial density \( \mf{n}(\bol{q}, t) \), and assume that particles have a characteristic energy \( \mc{E} = 3k_B T/2 \), where \( k_B \) is the Boltzmann constant and \( T \) is the temperature. 
Under these assumptions, we may approximate the tensor \( \Pi_C \) as
\begin{equation}
\Pi_C \approx \frac{1}{2} |\bol{v}_a - \bol{v}_b|\, r_{\mathfrak{n}}^2 \left( \frac{\kappa}{r_{\mathfrak{n}}^2} \cdot \frac{r_{\mathfrak{n}}}{m \sqrt{3k_B T / m}} \right)^2 I_v
= \frac{m |\bol{v}_a - \bol{v}_b|}{6 k_B T} \left( \frac{\kappa}{m} \right)^2 I_v.
\end{equation}
The corresponding collision operator takes the simplified form
\eq{
C\lr{\mf{f},\mf{f}}=\lr{\frac{\kappa}{m}}^2\frac{\p}{\p\bol{v}_a}\cdot\lrs{\mf{f}_a\int \mf{f}_{b}\frac{m\abs{\bol{v}_a-\bol{v}_b}}{6k_BT}
\mathbb{P}_v^{\perp}\cdot\lr{\frac{\p\log\mf{f}_{a}}{\p\bol{v}_a}-\frac{\p\log\mf{f}_b}{\p\bol{v}_b}}\,d\bol{v}_b
}.\label{COpC}
}
Remarkably, the Landau collision operator can be obtained through the substitution 
\eq{
\frac{m\abs{\bol{v}_a-\bol{v}_b}}{6k_BT}\mapsto 
\frac{2\pi\log\Lambda}{\abs{\bol{v}_a-\bol{v}_b}},
}
where $\log\Lambda$ is the Coulomb logarithm. 

\begin{remark}
We observe that the reason why there was no need to introduce a cutoff for the deflection angle caused by a Coulomb collision lies in the fact that, in the collision operator  \eqref{CffL},   
the projector $\mathbb{P}^{\perp}_v$ appears in front of the interaction tensor $\Pi_C$ of \eqref{PiC} due to the grazing scattering ansatz (recall Def.~1). 
In fact, the usual derivation of the Landau operators involves the approximation of the interaction tensor $\Pi_C$, 
and, in particular, of the integral 
$\mc{I} = \int \delta\bol{v}\, \delta\bol{v} \, {d\Omega}/{\sin^4\lr{\chi/2}}$ within $\Pi_C$, 
where $\chi$ is the deflection angle, $\delta\bol{v}$ the velocity displacement, and $d\Omega$ the infinitesimal solid angle (see, e.g., \cite{Fitz}). 
This approximation assumes that $\chi \in [\chi_{\rm min}, \chi_{\rm max}]$ is small and bounded above and below. 
It is from this approximation of $\Pi_C$ that one recovers the projector $\mathbb{P}^{\perp}_v$ in the standard derivation, 
effectively introducing the grazing scattering ansatz at this stage. 
However, the evaluation of $\mc{I}$ is not needed in our construction, effectively bypassing the mathematical singularity of the cross-section $\sigma$ near $\chi = 0$. 
More precisely, the possibility of expressing the scattering volume density per unit time $\mc{V}$ in the spatially localized form \eqref{mcV} hinges upon the assumption 
that the inter-particle distance is small compared to the system size, 
a condition that is inconsistent with $\chi \rightarrow 0$, which corresponds to a diverging impact parameter. 
The cross-section $\sigma$ appearing in the interaction tensor \eqref{PiC} does not include small-angle deflections, 
and therefore $\Pi_C$ does not diverge as $\chi \rightarrow 0$. 
\end{remark}

We conclude this section with a few remarks on the properties of the collision operator
\eqref{COpC} and the associated kinetic equation
\eq{
\frac{\p\mf{f}}{\p t}=\frac{1}{m}\frac{\p}{\p\bol{\bol{v}}}\cdot\lr{\mf{f}\frac{\p\Phi}{\p{\bol{q}}}}-\frac{\p}{\p\bol{q}}\cdot\lr{\mf{f}\bol{v}}+\lr{\frac{\kappa}{m}}^2\frac{\p}{\p\bol{v}}\cdot\lrs{\mf{f}\int \mf{f}'\frac{m\abs{\bol{v}-\bol{v}'}}{6k_BT}
\mathbb{P}_v^{\perp}\cdot\lr{\frac{\p\log\mf{f}}{\p\bol{v}}-\frac{\p\log\mf{f}'}{\p\bol{v}'}}\,d\bol{v}'
},\label{dfdtC}}
where $\mf{f}'=\mf{f}\lr{\bol{v}',\bol{q},t}$. 
Equation~\eqref{dfdtC} preserves the total particle number under suitable boundary conditions, due to its divergence form. Discarding boundary integrals as usual, conservation of energy can be shown as follows:
\eq{
\frac{d\mathscr{H}}{dt}=&\int\frac{\p \mf{f}}{\p t}\lr{\frac{1}{2}m\bol{v}^2+\Phi}\,d\bol{v}d\bol{q}+
\int\frac{\p \mf{f}'}{\p t}\lr{\frac{1}{2}m\bol{v}^2{}'+\Phi}\,d\bol{v}'d\bol{q}\\
=&-{\kappa}^2\int  \mf{f}\mf{f}'\frac{\abs{\bol{v}-\bol{v}'}}{6k_BT}\lr{\bol{v}-\bol{v}'}\cdot\mathbb{P}^{\perp}_v\cdot\lr{\frac{\p\log \mf{f}}{\p\bol{v}}-\frac{\p\log\mf{f}'}{\p\bol{v}'}}\,d\bol{v}d\bol{v}'d\bol{q}=0, 
}
where we used the fact that $\lr{\bol{v}-\bol{v}'}\cdot\mathbb{P}^{\perp}_v=\bol{0}$. 
Equation \eqref{dfdtC} does not conserve interior Casimir invariants because the underlying canonical particle dynamics does not possess Casimir invariants. 
Finally, the entropy law is given by
\eq{
\frac{dS}{dt}=&\lr{\frac{\kappa}{m}}^2\int \mf{f}\mf{f}'\frac{m\abs{\bol{v}-\bol{v}'}}{6k_BT}\lrs{\mathbb{P}^{\perp}_v\cdot\lr{\frac{\p\log\mf{f}'}{\p\bol{v}'}-\frac{\p\log\mf{f}}{\p\bol{v}}}}^2\,d\bol{v}d\bol{v}'d\bol{q}\geq 0. 
}
Hence, the Maxwell-Boltzmann distribution
\eq{
\log\mf{f}_{\infty}=-\beta\lr{\frac{1}{2}m\bol{v}^2+\Phi}+\log A,  
}
where $A$ is a normalization constant,  
is a thermodynamic equilibrium.

\section{Noncanonical Hamiltonian structure of guiding center dynamics}

Let $\lr{x,y,z,u,\mu,\theta}$ denote guiding center coordinates \cite{Cary} in a domain $M=\Omega\times\mathbb{R}\times[0,+\infty)\times[0,2\pi]$ with $\Omega\subset\mathbb{R}^3$ a smooth bounded domain with boundary $\p\Omega$. 
Let $\bol{b}=\bol{B}_0/B_0$ denote the unit vector along the equilibrium (time-independent) magnetic field $\bol{B}_0\neq\bol{0}$, and $\omega_c=qB_0/m$ the cyclotron frequency with $q$ and $m$ the particle charge and mass respectively.   
In these coordinates, the Poisson tensor has expression
\begin{equation}
\mc{J}_{\rm GC}=
\begin{bmatrix}
0&-\frac{b_z}{qB_{\parallel}^*}&\frac{b_y}{qB_{\parallel}^*}&\frac{B_x^{*}}{mB_{\parallel}^*}&0&0\\
\frac{b_z}{qB_{\parallel}^*}&0&-\frac{b_x}{qB_{\parallel}^{\ast}}&\frac{B_y^{\ast}}{mB_{\parallel}^{\ast}}&0&0\\
-\frac{b_y}{qB_{\parallel}^{\ast}}&\frac{b_x}{qB_{\parallel}^{\ast}}&0&\frac{B_z^{\ast}}{mB_{\parallel}^{\ast}}&0&0\\
-\frac{B_{x}^{\ast}}{mB_{\parallel}^{\ast}}&-\frac{B_y^{\ast}}{mB_{\parallel}^{\ast}}&-\frac{B_z^{\ast}}{mB_{\parallel}^{\ast}}&0&0&0\\
0&0&0&0&0&-\frac{\omega_c}{B_0}\\
0&0&0&0&\frac{\omega_c}{B_0}&0
\end{bmatrix},\label{J} 
\end{equation}
where ${B}_{\parallel}^{\ast}=\bol{B}^{\ast}\cdot\bol{b}$ and $\bol{B}^{\ast}%\lr{\bol{X}}
=\bol{B}_0+\frac{B_0u}{\omega_c}\nabla
%_{\bol{X}}
\cp\bol{b}$. 
%, with $\bol{X}$ guiding-center coordinates and $\nabla_{\bol{X}}$ the associated gradient operator. 
%\naoki{
%For the scope of the present paper, we shall identify the guiding center gradient operator $\nabla_{\bol{X}}$ with the guiding center gradient operator $\nabla$, under the assumption that their difference is $O\lr{\epsilon}$, with $\epsilon>0$ the small ordering parameter of the gyrokinetic ordering.
%} 
Given $F,G\in C^{\infty}\lr{M}$, the Poisson bracket associated with the Poisson tensor \eqref{J} can be written as
\eq{
\lrc{F,G}_{\rm GC}=
\frac{\omega_c}{B_0}\lr{F_{\theta}G_{\mu}-F_{\mu}G_{\theta}}+\frac{\bol{B}^*}{mB_{\parallel}^*}\cdot\lr{G_u\nabla F-F_u\nabla G}-\frac{1}{qB_{\parallel}^*}\bol{b}\cdot\nabla F\cp\nabla G,\label{PB}
}
where lower indexes are used to denote  partial derivatives, e.g. $G_u=\p G/\p u$. 
In this setting the guiding center Hamiltonian is independent of the cyclotron phase $\theta$ and has expression
\eq{
H_{\rm GC}=\frac{1}{2}mu^2+\mu B_0+q\Psi,
}
where $q\Psi=q\Phi-quA_{\parallel}$
%\frac{q}{2\pi}\int_0^{2\pi}\lr{\Phi-uA_{\parallel}}d\theta$ 
denotes the cyclotron phase averaged generalized potential energy,  $q\Phi$ the electric potential energy, and $\bol{B}=\bol{B}_0+\bol{B}_1$ the total magnetic field with perturbation $\bol{B}_1=\nabla\cp\lr{A_{\parallel}\bol{b}}$ \cite{Garbet}. 
Recall also that the parallel momentum $mu=mv_{\parallel}+qA_{\parallel}$ includes a kinetic part associated with the parallel velocity $v_{\parallel}$ and a magnetic part associated with $A_{\parallel}$.  
The Noether invariant associated with the $\theta$-symmetry is the magnetic moment $\mu$. 

\section{Electromagnetic potentials}
In the following analysis, we neglect displacement currents, as they are considered negligible for the low-frequency phenomena under investigation, which occur at frequencies much lower than the cyclotron frequency. Additionally, we omit finite Larmor radius (FLR) effects, as our focus will be on turbulence with wavelengths much larger than the Larmor radius. 

%streamline the exposition. However, these effects can be reinstated if necessary (see, e.g., \cite{Garbet} for the expression of Poisson-Amp\`ere equations including FLR effects).

%For simplicity, in the following we neglect %finite Larmor radius (FLR)  effects and 
%displacement currents, under the assumption that they are negligible for the type of low frequency (compared to the cyclotron frequency) phenomena examined here. 
%In particular, we will approximate the guiding center position with the guiding center position,    $\bol{X}=\bol{x}=\lr{x,y,z}$. 
Let $q_s$, $m_s$, $f_s\lr{x,y,z,u,\mu,t}$, and
\eq{n_s\lr{x,y,z,t}=%m^2_s
\int_{\mathbb{R}}\int_{0}^{+\infty}f_sB_{\parallel}^*du d\mu,} 
denote the charge, mass, distribution function, and number density of species $s$. 
Note that the invariant measure with respect to which $f_s$ is defined is given by the volume element (with units of $\mathrm{kg} \cdot \mathrm{m}^6 \cdot \mathrm{s}^{-3}$)
\eq{
%m^2_s
B_{\parallel}^*dxdydzdud\mu.
}
For the purpose of the present study, it will be sufficient to consider 2 species, $s=1,2$. 
Then, the electric field $\bol{E}=-\nabla\Phi$ is determined by Poisson's equation 
\eq{
\Delta\Phi=-\frac{1}{\epsilon_0}\sum_sq_sn_s~~~~{\rm in }~~\Omega, \label{Phi}
}
under suitable boundary conditions on $\p\Omega$. Here, $\epsilon_0$ is the vacuum permittivity. 
Let's assume that $\bol{B}_0$ is a vacuum field in the confinement region $\Omega$, that is $\nabla\cp\bol{B}_0=\bol{0}$ in $\Omega$. This is true if the coils generating $\bol{B}_0$ are located outside $\Omega$. The magnetic field in $\Omega$ can then be obtained from Amp\`ere's law,
\eq{
\nabla\cp\nabla\cp\lr{A_{\parallel}\bol{b}}=\mu_0 \sum_s\bol{j}_s,
}
where $\bol{j}_s$ is the current density of species $s$. 
Dotting the left-hand side of this equation with the unit vector $\bol{b}$ gives
\eq{
\bol{b}\cdot\nabla\cp\nabla\cp\lr{A_{\parallel}\bol{b}}=-\Delta A_{\parallel}+\lr{\bol{b}\cdot\nabla}^2A_{\parallel}+\lr{\nabla\cdot\bol{b}}\lr{\bol{b}\cdot\nabla A_{\parallel}}+A_{\parallel}\bol{b}\cdot\nabla\cp\nabla\cp\bol{b}.
}
Now assume that the parallel and perpendicular wavenumber scales as $k_{\parallel}/k_{\perp}\sim O\lr{\epsilon}$  
, where $\epsilon>0$ is a small ordering parameter associated with the drift-kinetic ordering. 
Since $\bol{b}\cdot\nabla\sim k_{\parallel}$, it follows that, at leading order,  
\eq{
\Delta A_{\parallel}=A_{\parallel}\bol{b}\cdot\nabla\cp\nabla\cp\bol{b}-\mu_0\sum_{s}q_sn_s\lr{\bar{u}_s-\frac{q_s}{m_s}A_{\parallel}}~~~~{\rm in}~~\Omega,\label{Apar}
}
with
\eq{
\bar{u}_s=\frac{1}{n_s}\int u f_sB_{\parallel}^*dud\mu,
}
the local average of $u$ for species $s$. Under suitable boundary conditions, equation \eqref{Apar} determines $A_{\parallel}$.  

Finally, observe that since $\nabla\cp\bol{B}_0=\bol{0}$ in $\Omega$, 
locally there exists a scalar function $\zeta$ such that $\bol{b}=\nabla\zeta/\abs{\nabla\zeta}$ (Poincar\'e lemma). Hence $\bol{b}\cdot\nabla \cp\bol{b}=0$ implying $B_{\parallel}^*=B_0$ in $\Omega$. 

\section{A weakly collisional plasma regime}

In order to obtain a collision operator for the present guiding center system according to the theory developed above, some physical assumptions are necessary with regard to the type of collision process driving dissipation.
We will be concerned with a weakly collisional plasma regime such that the change in  magnetic moment after a collision event is small, and the cumulative magnetic moment change resulting from repeated  collisions is negligible. 
The first requirement can be satisfied if the charged particles have a high enough kinetic energy, their characteristic spatial distance $r_{s}\sim n_s^{-1/3}$ is large enough, and collision events are localized, i.e. they occur over spacetime  scales that are small compared to the characteristic spacetime scales of ideal (collisionless)  dynamics. 
To see this explicitly, denote with $T_s$ the temperature of species $s$. Further assume that $q_s$ has the same order of magnitude for all $s$. Then, the distance $r_C$ such that a Coulomb scattering results in a significant change in particle energy can be estimated as
\eq{
r_C\sim \frac{q^2_s}{4\pi\epsilon_0 k_BT_s}.
}
%where $k_B$ is the Boltzmann constant. 
A condition required for the particle energy change to be negligible is therefore
$r_s\gg r_C$. 
For example, consider a plasma consisting of electrons $(s=1)$ and protons $(s=2)$ and set $T_1=T_2=10\,{\rm keV}$ and $n_1=n_2=10^{21}\,{\rm m^{-3}}$. We have $r_s\sim 10^{-7}\,{\rm m}$ as well as $r_C\sim 10^{-13}\,{\rm m}$. Let $L_B\sim 1/\abs{\nabla B_0}$ the characteristic spatial scale of the background magnetic field. 
If 
\eq{L_B\gg r_s\gg r_C,\label{ord1}}
Coulomb collisions are spatially localized, with the background magnetic field remaining essentially constant during a scattering event. This implies that the magnetic moment is also preserved, since
\eq{
0\approx H_{GC}-H_{GC}'\approx \lr{\mu-\mu'}B_0,
}
where the $'$ denotes the value after the collision. 
We also remark that in such regime collisions are localized in time because $L_B/\sqrt{k_BT_s/m_s}\gg r_C/\sqrt{k_BT_s/m_s}$. 

The second requirement of negligible cumulative change in magnetic moment can be satisfied provided that
\begin{equation}
\sum_{i=1}^N (\mu_i - \mu_i') 
%\ll \sum_{i=1}^N |\mu_i - \mu_i'|
\approx 0,\label{cum}
\end{equation}
where $i = 1, \ldots, N$ denotes the $i$th collision event, with $N$ being the total number of collisions occurring, on average, over the time scale $\tau$ during which the system is observed.

Equation \eqref{cum} indicates that, in addition to being small, changes in the magnetic moment should not be biased but should instead exhibit a compensating effect, resulting in the conservation of the magnetic moment even after a large number $N \gg 1$ of collisions. This hypothesis seems physically reasonable, especially close to thermodynamic equilibrium, since localized Coulomb scatterings can be expected to behave predominantly as a random process, sometimes increasing the perpendicular component of the particle energy and sometimes decreasing it.

In the following, we shall therefore assume that the conditions \eqref{ord1} and \eqref{cum} hold. 

\section{Two-species reduced noncanonical Hamiltonian structure}
Because the Poisson tensor \eqref{J} is a function of $q_s$ and $m_s$, 
calculations are greatly simplified if the charged particles of the $2$ species $s=1,2$ are treated in pairs. 
It is also convenient to use phase space coordinates whose  corresponding volume element defines an invariant measure (the  Jacobian factor is constant). 
Neglecting the cyclotron phase, we therefore choose the following  reduced phase space variables
\eq{
\bol{z}=\lr{x_1,y_1,z_1,u_1,\eta_1,x_2,y_2,z_2,u_2,\eta_2}, 
}
where $\eta_s=\mu_sB_{\parallel s}^\ast=\mu_sB_{0s}$, $s=1,2$, and the lower index applied to a function indicates evaluation 
at the corresponding particle phase space position, e.g. $B_1={B}\lr{\bol{x}_1}$, with $\bol{x}_1=\lr{x_1,y_1,z_1}$. 
The pair guiding center Hamiltonian takes the form,
\eq{
\mathscr{H}=\mc{H}+\mc{Q}=\frac{1}{2}m_1u_1^2+\frac{1}{2}m_2u_2^2+\eta_1+\eta_2+q_1\Phi_1+q_2\Phi_2-q_1u_1A_{\parallel 1}-q_2u_2A_{\parallel 2}+v, \label{pairH}
}
where 
%we introduced the notation $\bol{x}_1=\lr{x_1,y_1,z_1}$, 
\eq{
v=\frac{q_1^2}{2m_1}A_{\parallel 1}^2+
\frac{q_2^2}{2m_2}A_{\parallel 2}^2+\lambda.
}
Here, the first two terms are higher-order contributions that must however be kept to later ensure conservation of total energy, while $\lambda=\lambda_1\lr{\bol{x}_1}+\lambda_2\lr{\bol{x}_2}$ allows for some additional potential energy. Note that we defined the average electric potential energy  \eq{\mc{Q}=q_1\Phi_1+q_2\Phi_2} 
and the difference 
\eq{\mc{H}=\mathscr{H}-\mc{Q}=\frac{1}{2}m_1u_1^2+\frac{1}{2}m_2u_2^2+\eta_1+\eta_2-q_1u_1A_{\parallel 1}-q_2u_2A_{\parallel 2}+v,}
for later convenience.
%of the interaction between the $2$ particles. 
%For the Coulomb force, we have $v=q_1q_2/4\pi\epsilon_0\abs{\bol{x}_1-\bol{x}_2}$. 
In the new variables, the pair guiding center reduced Poisson tensor can be obtained by replacing $\nabla$ with $\nabla+\eta\nabla\log B_0\frac{\p}{\p\eta}$ in eq. \eqref{PB}. In particular, the single particle reduced guiding center Poisson bracket now reads as  
\eq{
\lrc{F,G}_{\rm GC}=&\frac{\bol{B}^*}{mB_{0}}\cdot\lrs{G_u\nabla F-F_u\nabla G+\eta\lr{G_uF_{\eta}-F_uG_{\eta}}\nabla\log B_0}\\&-\frac{1}{qB_0}\bol{b}\cdot\lrs{\nabla F\cp\nabla G+\eta\nabla\log B_0\cp\lr{F_{\eta}\nabla G-G_\eta\nabla F}},
}
leading to the guiding center reduced Poisson tensor 
\begin{equation}
\begin{split}
\mc{J}_{\rm GC}=\begin{bmatrix}
0&-\frac{b_{z}}{qB_{0}}&\frac{b_{y}}{qB_{0}}&\frac{B_{x}^{*}}{mB_{0}}&\frac{\eta}{qB_0^2}\bol{b}\cp\nabla B_0\cdot\nabla x\\
\frac{b_{z}}{qB_{0}}&0&-\frac{b_{x}}{qB_{0}}&\frac{B_{y}^{\ast}}{mB_{0}}&\frac{\eta}{qB_0^2}\bol{b}\cp\nabla B_0\cdot\nabla y\\
-\frac{b_{y}}{qB_{0}}&\frac{b_{x}}{qB_{0}}&0&\frac{B_{z}^{\ast}}{mB_{0}}&\frac{\eta}{qB_0^2}\bol{b}\cp\nabla B_0\cdot\nabla z\\
-\frac{B_{x}^{\ast}}{mB_{0}}&-\frac{B_{y}^{\ast}}{mB_{0}}&-\frac{B_{z}^{\ast}}{mB_{0}}&0&-\eta\frac{\bol{B}^{\ast}\cdot\nabla B_0}{mB_0^2}\\
\frac{\eta}{qB_0^2}\bol{b}\cp\nabla x\cdot\nabla B_0&\frac{\eta}{qB_0^2}\bol{b}\cp\nabla y\cdot\nabla B_0&\frac{\eta}{qB_0^2}\bol{b}\cp\nabla z\cdot\nabla B_0&\eta\frac{\bol{B}^{\ast}\cdot\nabla B_0}{mB_0^2}&0
\end{bmatrix}.\label{Jgc}
\end{split}
\end{equation}
The pair guiding center reduced Poisson tensor thus takes the form
\eq{
\mf{J}_{\rm GC}=\mc{J}_{\rm GC 1}+\mc{J}_{\rm GC 2},
}
or, in matrix form, 
\begin{equation}
\mf{J}_{\rm GC}=
\begin{bmatrix}
\mc{J}_{\rm GC 1}&\bol{0}_5\\
\bol{0}_5&\mc{J}_{\rm GC 2}
\end{bmatrix},\label{J2}
\end{equation}
where $\bol{0}_5$ is the $5\times 5$ null matrix. 
Note that the equations of motion can be expressed as  
\eq{
\dot{\bol{z}}= \mf{J}_{\rm GC}\cdot\p_{\bol{z}}\lr{\mc{H+Q}}. 
}
Noting that
\eq{
\p_{\bol{z}}\lr{\mc{H+Q}}=\lr{\nabla_1\lr{q_1\Phi_1-q_1u_1A_{\parallel 1}+v},m_1u_1-q_1A_{\parallel 1},1,\nabla_2\lr{q_2\Phi_2-q_2u_2A_{\parallel 2}+v},m_2u_2-q_2A_{\parallel 2},1 
},
}
where $\nabla_s$ denotes the spatial gradient with respect to $\bol{x}_s$, $s=1,2$, we have
\sys{
&\dot{\bol{x}}_s=\frac{1}{q_sB_{0s}}\bol{b}_s\cp\nabla_s\lr{q_s\Phi_s-q_su_sA_{\parallel s}+v}+\frac{m_su_s-q_sA_{\parallel s}}{m_sB_{0s}}\bol{B}^*_s+\frac{\eta_s}{q_sB_{0s}^2}\bol{b}_s\cp\nabla_s B_{0s},\\
&\dot{u}_s=-\frac{\bol{B}^{\ast}_s\cdot\nabla_s\lr{q_s\Phi_s-q_su_sA_{\parallel s}+v}}{m_sB_{0s}}-\eta_s\frac{\bol{B}_s^*\cdot\nabla_sB_{0s}}{m_sB_{0s}^2},\\
&\dot{\eta}_s=\frac{\eta_s}{q_sB_{0s}^2}\bol{b}_s\cp\nabla_s\lr{q_s\Phi_s-q_su_sA_{\parallel s}+v}\cdot\nabla B_{0s}+\eta_s\frac{\bol{B}_s^*\cdot\nabla_sB_{0s}}{m_sB_{0s}^2}\lr{m_su_s-q_sA_{\parallel s}}.
}{EoM}
Define $\bol{\zeta}_s=\lr{\bol{x}_s,u_s,\eta_s}$. 
From system \eqref{EoM} it follows that
\eq{
\p_{\bol{\zeta}_s}\cdot\dot{\bol{\zeta}}_s=&\nabla_s\cdot\dot{\bol{x}}_s+{\p_{u_s} \dot{u}_s}+{\p_{\eta_s}\dot{\eta}_s}\\=&
\frac{1}{q_sB_{0s}}\nabla_s\cp\bol{b}_s\cdot\nabla_s\sigma_s+\frac{1}{q_s}\nabla_s\lr{\frac{1}{B_{0s}}}\cp\bol{b}_s\cdot\nabla_s\sigma_s+\nabla_s\lr{\frac{m_su_s-q_sA_{\parallel s}}{m_sB_{0s}}}\cdot\bol{B}^*_s\\&-\frac{\eta_s}{q_s}\nabla_s\cp\bol{b}_s\cdot\nabla_s\lr{\frac{1}{B_{0s}}}-\frac{1}{q_sB_{0s}}\nabla_s\cp\bol{b}_s\cdot\nabla_s\sigma_s+\frac{q_s}{m_sB_{0s}}\bol{B}^*_s\cdot\nabla_sA_{\parallel s}\\&-\frac{\eta_s}{q_sB_{0s}^2}\nabla_s\cp\bol{b}_s\cdot\nabla_sB_{0s}+\frac{1}{q_sB_{0s}^2}\bol{b}_s\cp\nabla_s\sigma_s\cdot\nabla_sB_{0s}+\frac{m_su_s-q_sA_{\parallel s}}{m_sB_{0s}^2}\bol{B}^*_s\cdot\nabla_sB_{0s}\\=&0,
}
where $\sigma_s=q_s\Phi_s-q_su_sA_{\parallel s}+v$. 
Hence, each volume element
\eq{
d\bol{\zeta}_s=d\bol{x}_sdu_sd\eta_s, \label{IMs}
}
defines an invariant measure. In this notation, $d\bol{x}_s=dx_sdy_sdz_s$.

%\textcolor{red}{Poisson bracket sign}

\section{Two-species guiding center collision operator}
Let $f\lr{\bol{z},t}$ denote the pair guiding center distribution function with respect to the phase space (invariant) measure
\eq{
d\bol{z}=d\bol{\zeta}_1d\bol{\zeta}_2=d\bol{x}_1du_1d\eta_1d\bol{x}_2du_2d\eta_2.\label{IM}
} 
Note that $f$, $f_1$, and $f_2$ are related by
\eq{
f_1=\int f\,dx_2dy_2dz_2du_2d\eta_2,~~~~f_2=\int f\,dx_1dy_1dz_1du_1d\eta_1.
}
Suppose that $2$ pairs of charged particles with phase space coordinates $\bol{z}_a$ and $\bol{z}_b$ interact through a potential energy $V\lr{\bol{z}_a,\bol{z}_b}$. For the Coulomb interaction,
\eq{
V=\frac{1}{4\pi\epsilon_0}\lr{\frac{q_1^2}{\abs{\bol{x}_{1a}-\bol{x}_{1b}}}
%+\frac{q_1q_2}{\abs{\bol{x}_{1a}-\bol{x}_{2a}}}
+\frac{q_1q_2}{\abs{\bol{x}_{1a}-\bol{x}_{2b}}}
+\frac{q_1q_2}{\abs{\bol{x}_{2a}-\bol{x}_{1b}}}
%+\frac{q_1q_2}{\abs{\bol{x}_{1b}-\bol{x}_{2b}}}
+\frac{q_2^2}{\abs{\bol{x}_{2a}-\bol{x}_{2b}}}},\label{VC}
}
where $\bol{x}_{1b}$ denotes the configuration space position of particle $1$ of the pair at $\bol{z}_b$, and so on. 
Then, the evolution equation for the distribution function $f$ can be obtained with aid of the theory developed in Section 2. 
Indeed, the collision operator of the system is given by equation \eqref{C2}. 
However, the potential energy in equation~\eqref{VC} cannot be expressed solely as a function of \( \bol{z}_a - \bol{z}_b \), implying that the simplifications leading to equation~\eqref{dfdt0} do not apply in this case. Consequently, some refinements are required in order to derive the correct kinetic equation for \( f \).

%On the other hand, due to the scale separation between ideal dynamics and collisions, we have 
%\eq{\mf{J}_{\rm GC}\lr{\bol{z}_a}
%=\mf{J}_{{\rm GC}a}
%\approx\mf{J}_{\rm GC}\lr{\bol{z}_b}%=\mf{J}_{{\rm GC}b}
%,} as well as
%\eq{
%\delta\bol{z}_a\approx\tau_c\mf{J}_{{\rm GC}a}\cdot\frac{\p V}{\p\bol{z}_a},~~~~\delta\bol{z}_b\approx\tau_c\mf{J}_{{\rm GC}b}\cdot\frac{\p V}{\p\bol{z}_b},\label{IM} 
%}
%where $\tau_c$ is a physical constant representing the characteristic time interval between collisions and we introduced the notation $\mf{J}_{{\rm GC}a}=\mf{J}_{{\rm GC}}\lr{\bol{z}_a}$. 
%It is also worth observing that, using the expressions \eqref{IM}, we have  
%\eq{
%\frac{\p}{\p\bol{z}_a}\cdot\delta\bol{z}_a=0,~~~~\frac{\p}{\p\bol{z}_b}\cdot\delta\bol{z}_b=0.
%}

We begin by examining the implications of the grazing scattering condition \eqref{grazing} for the phase space  displacement $\delta\bol{z}$. 
Our working assumption is that 
the dominant type of collisions involves only 2 particles, while collisions involving more than 2 particles are `rare.' 
Hence, if we define
\eq{
V_{11}=\frac{q_1^2}{4\pi\epsilon_0\abs{\bol{x}_{1a}-\bol{x}_{1b}}},~~~~V_{12}=\frac{q_1q_2}{4\pi\epsilon_0\abs{\bol{x}_{1a}-\bol{x}_{2b}}},~~~~V_{21}=\frac{q_1q_2}{4\pi\epsilon_0\abs{\bol{x}_{2a}-\bol{x}_{1b}}},~~~~V_{22}=\frac{q_2^2}{4\pi\epsilon_0\abs{\bol{x}_{2a}-\bol{x}_{2b}}},
}
only one of the four potential energies above will be at play during a collision event, and each of these potentials are of the type $V_{ij}=V_{ij}\lr{\abs{\bol{x}_{ia}-\bol{x}_{jb}}}$, $i,j=1,2$. The grazing scattering condition therefore amounts to
\eq{
\frac{1}{E_{t_0}}\int_{\tau_c}\bol{\xi}_{ij}\cdot\frac{\p V_{ij}}{\p\bol{\zeta}_{ai}}\,dt=O\lr{\epsilon^2},~~~~\bol{\xi}_{ij}=\mc{J}_{{\rm GC}a}\cdot\frac{\p\lr{\mc{H}+\mc{Q}}}{\p\bol{\zeta}_{ai}}-\mc{J}_{{\rm GC}bj}\cdot\frac{\p\lr{\mc{H}+\mc{Q}}}{\p\bol{\zeta}_{bj}},~~~~i,j=1,2.
}
At leading order, the phase space displacements  
can thus be expressed as
\begin{equation}
\begin{split}
&\delta\bol{z}_{a}=\mf{J}_{{\rm GC}a}
%\cdot
%\mathbb{P}^{\perp}
\cdot
\begin{bmatrix}
\mathbb{P}^{\perp}_{11}\cdot\int_{\tau_c}\p_{\bol{\zeta}_{a1}}V_{11}\,dt+\mathbb{P}^{\perp}_{12}\cdot\int_{\tau_c}\p_{\bol{\zeta}_{a1}}V_{12}\,dt\\
\mathbb{P}^{\perp}_{21}\cdot\int_{\tau_c}\p_{\bol{\zeta}_{a2}}V_{21}\,dt+\mathbb{P}^{\perp}_{22}\cdot\int_{\tau_c}\p_{\bol{\zeta}_{a2}}V_{22}\,dt
\end{bmatrix}
%+O\lr{\epsilon^2}
,\\
&\delta\bol{z}_{b}=\mf{J}_{{\rm GC}b}
%\cdot
%\mathbb{P}^{\perp}
\cdot
\begin{bmatrix}
\mathbb{P}^{\perp}_{11}\cdot\int_{\tau_c}\p_{\bol{\zeta}_{b1}}V_{11}\,dt+\mathbb{P}^{\perp}_{21}\cdot\int_{\tau_c}\p_{\bol{\zeta}_{b1}}V_{21}\,dt\\
\mathbb{P}^{\perp}_{12}\cdot\int_{\tau_c}\p_{\bol{\zeta}_{b2}}V_{12}\,dt+\mathbb{P}^{\perp}_{22}\cdot\int_{\tau_c}\p_{\bol{\zeta}_{b2}}V_{22}\,dt
\end{bmatrix}
%+O\lr{\epsilon^2}
, \label{dzdz}
\end{split}
\end{equation}
where now the projectors $\mathbb{P}^{\perp}_{ij}$ are given by %\begin{equation}
%\mathbb{P}^{\perp}=\begin{bmatrix}
%\mathbb{P}_{11}^{\perp}&\mathbb{P}_{12}^{\perp}\\
%\mathbb{P}_{12}^{\perp}&\mathbb{P}_{22}^{\perp}
%\end{bmatrix},
%\end{equation} 
%where
\eq{
\mathbb{P}_{ij}^{\perp}=\lr{1-\frac{\bol{\xi}_{ij}\bol{\xi}_{ij}}{\bol{\xi}_{ij}^2}},
%~~~~\bol{\xi}_{ij}=\mc{J}_{{\rm GC}ai}\cdot\p_{\bol{\zeta}_{ai}}\lr{\mc{H}+\mc{Q}}-\mc{J}_{{\rm GC}bj}\cdot\p_{\bol{\zeta}_{bj}}\lr{\mc{H}+\mc{Q}},
~~~~i,j=1,2.
}
To simplify the notation, let us write $\int\p_{\bol{\zeta}_{a1}}V_{11}=\int_{\tau_c}\p_{\bol{\zeta}_{a1}}V_{11}\,dt$ and so on. 
Introducing the interaction tensors
\sys{
&\Pi_{ab}=-\frac{1}{2}\int\mc{V}
{
\begin{bmatrix}
\mathbb{P}^{\perp}_{11}\cdot\int\p_{\bol{\zeta}_{a1}}V_{11}+\mathbb{P}^{\perp}_{12}\cdot\int\p_{\bol{\zeta}_{a1}}V_{12}\\
\mathbb{P}^{\perp}_{21}\cdot\int\p_{\bol{\zeta}_{a2}}V_{21}+\mathbb{P}^{\perp}_{22}\cdot\int\p_{\bol{\zeta}_{a2}}V_{22}
\end{bmatrix}
}
{
\begin{bmatrix}
\mathbb{P}^{\perp}_{11}\cdot\int\p_{\bol{\zeta}_{b1}}V_{11}+\mathbb{P}^{\perp}_{21}\cdot\int\p_{\bol{\zeta}_{b1}}V_{21}\\
\mathbb{P}^{\perp}_{12}\cdot\int\p_{\bol{\zeta}_{b2}}V_{12}+\mathbb{P}^{\perp}_{22}\cdot\int\p_{\bol{\zeta}_{b2}}V_{22}
\end{bmatrix}}d\bol{z}_a'd\bol{z}_b',
\\&\Pi_{aa}=\frac{1}{2}\int\mc{V}{
\begin{bmatrix}
\mathbb{P}^{\perp}_{11}\cdot\int\p_{\bol{\zeta}_{a1}}V_{11}+\mathbb{P}^{\perp}_{12}\cdot\int\p_{\bol{\zeta}_{a1}}V_{12}\\
\mathbb{P}^{\perp}_{21}\cdot\int\p_{\bol{\zeta}_{a2}}V_{21}+\mathbb{P}^{\perp}_{22}\cdot\int\p_{\bol{\zeta}_{a2}}V_{22}
\end{bmatrix}
}{
\begin{bmatrix}
\mathbb{P}^{\perp}_{11}\cdot\int\p_{\bol{\zeta}_{a1}}V_{11}+\mathbb{P}^{\perp}_{12}\cdot\int\p_{\bol{\zeta}_{a1}}V_{12}\\
\mathbb{P}^{\perp}_{21}\cdot\int\p_{\bol{\zeta}_{a2}}V_{21}+\mathbb{P}^{\perp}_{22}\cdot\int\p_{\bol{\zeta}_{a2}}V_{22}
\end{bmatrix}
}d\bol{z}_a'd\bol{z}_b',
}{Piab}
%where
%\eq{
%\Gamma=\int\mc{V}d\bol{z}_a'd\bol{z}_b',
%}
%is the scattering frequency, we arrive at the following expression for 
at leading order, the collision operator becomes  
\eq{
\mc{C}\lr{f,f}=\frac{\p}{\p\bol{z}_a}\cdot \lrs{f_a\mf{J}_{{\rm GC}a}\cdot\int f_b\lr{\Pi_{ab}\cdot\mf{J}_{{\rm GC}b}\cdot\frac{\p\log f_b}{\p\bol{z}_b}-\Pi_{aa}\cdot\mf{J}_{{\rm GC}a}\cdot\frac{\p\log f_a}{\p\bol{z}_a}}\,d\bol{z}_b}\label{C3}.
}
The appearance of this equation can be slightly simplified by setting $\bol{z}=\bol{z}_a$, $\mf{J}_{{\rm GC}}=\mf{J}_{{\rm GC}a}$, $f=f_a$,  $\bol{z}'=\bol{z}_b$, $\mf{J}_{{\rm GC}}'=\mf{J}_{{\rm GC}b}$, $f'=f_b$, $\Pi'=\Pi_{ab}$, and $\Pi=\Pi_{aa}$. We have
\eq{
\mc{C}\lr{f,f}=\frac{\p}{\p\bol{z}}\cdot \lrs{f\mf{J}_{{\rm GC}}\cdot\int f'\lr{\Pi'\cdot\mf{J}'_{{\rm GC}}\cdot\frac{\p\log f'}{\p\bol{z}'}-\Pi\cdot\mf{J}_{{\rm GC}}\cdot\frac{\p\log f}{\p\bol{z}}}\,d\bol{z}'}.\label{C4}
}
The full evolution equation for the distribution function $f$ therefore takes the form
%vanishes for collisions between a pair inside $M$ and a pair on the boundary $\p M$, 
%\eq{
%\Gamma\lr{\bol{z}_a,\bol{z}_b}=0~~~~\bol{z}_a\in M,~~~~\bol{z}_b\in\p M.
%}
%Let $\p M$ denote the boundary 
%of the phase space domain $M$, with unit outward normal $\bol{n}$. Assume the boundary condition
%\eq{
%{\Gamma\lr{f_a\delta\bol{z}_a\frac{\p f_b}{\p\bol{z}_b}\cdot\delta\bol{z}_b+
%f_b\delta\bol{z}_b\frac{\p f_a}{\p\bol{z}_a}\cdot\delta\bol{z}_a
%}}\cdot\bol{n}=0,~~~~\bol{z}_b\in\p M.
%}
\eq{
\frac{\p f}{\p t}=\frac{\p}{\p\bol{z}}\cdot\lrc{f\mf{J}_{{\rm GC}}\cdot\lrs{-\frac{\p\lr{\mc{H}+\mc{Q}}}{\p\bol{z}}+\int f'\lr{\Pi'\cdot\mf{J}'_{{\rm GC}}\cdot\frac{\p\log f'}{\p\bol{z}'}-\Pi\cdot\mf{J}_{{\rm GC}}\cdot\frac{\p\log f}{\p\bol{z}}}d\bol{z}'}}. \label{ft}
}
In this notation,
\eq{
\mc{Q}=\int f'V\lr{\bol{z},\bol{z}'}d\bol{z}',
}
is the average potential energy of the interaction between particle pairs. Recall that, for the Coulomb interaction, the potential energy $\mc{Q}$ is related to electric potential $\Phi$ by
\eq{
\mc{Q}=q_1\Phi_{1}+q_2\Phi_{2},
}
with 
\eq{
\Phi_{1}=\frac{q_1}{4\pi\epsilon_0}\int \frac{n_{1}'}{\abs{\bol{x}_{1}-\bol{x}_{1}'}}d\bol{x}_{1}'+\frac{q_2}{4\pi\epsilon_0}\int \frac{n_{2}'}{\abs{\bol{x}_{1}-\bol{x}_{2}'}}d\bol{x}_{2}'=\Phi\lr{\bol{x}_{1},t},
}
and a similar expression for $\Phi_{2}$. Here, $n_{s}'=n_{s}\lr{\bol{x}_{s}',t}$ denotes the number density of species $s$ at $\bol{x}_{s}'$. 
%Here, we introduced the notation $d\bol{x}_1'=dx_1'dy_1'dz_1'$. 
Notice that equation \eqref{ft} defines a closed system once coupled with the Maxwell's equations \eqref{Phi} and \eqref{Apar}. 
%Observe also that equation \eqref{ft} has a general nature, i.e.  one would arrive at the same expression for a general system endowed with arbitrary Hamiltonian $\ms{H}=\mc{H}+\mc{Q}$, interaction potential $V$, and Poisson tensors $\mf{J}$. 

The evolution equations for the single species distribution functions $f_1$ and $f_2$ can be obtained by integrating the evolution equation for $f$. For example, integrating \eqref{ft} with respect to $\bol{\zeta}_2=\lr{\bol{x}_2,u_2,\mu_2}$ and assuming boundary integrals to vanish, we have
\eq{
\frac{\p f_1}{\p t}=-\frac{\p}{\p\bol{\zeta}_1}\cdot\lrs{f_1\mc{J}_{\rm GC1}\cdot\frac{\p\lr{{h}_{1}+q_1\Phi_1+{v}_1}}{\p\bol{\zeta}_1}}+\int \mc{C}\lr{f,f}\,d\bol{\zeta}_2,\label{ft1}
}
where the function $v_1$ is defined by 
\eq{
f_1\frac{\p v_1}{\p\bol{\zeta}_1}=\int f\frac{\p v}{\p\bol{\zeta}_1}\,d\bol{\zeta}_2,
}
and we defined \eq{h_s=\frac{1}{2}m_su_s^2+\eta_s-q_su_sA_{\parallel s}, ~~~~s=1,2.}
We remark however that, in general, the evaluation of the collision operator in eq. \eqref{ft1} requires knowledge of $f$. This is a consequence of the fact that $f_1$ and $f_2$ are not statistically independent. Nevertheless, if one approximates the pair distribution function $f$  as $f\lr{\bol{\zeta}_1,\bol{\zeta}_2,t}=f_1\lr{\bol{\zeta}_1,t}f_2\lr{\bol{\zeta}_2,t}$, one recovers two coupled equations for $f_1$ and $f_2$ that define a closed system (together with Maxwell's equations \eqref{Phi} and \eqref{Apar}). 
To see this, define  
\sys{
\Pi_{11}=&\frac{1}{2}\int f_2f_2'\mc{V}
\lr{\mathbb{P}^{\perp}_{11}\cdot\int_{\tau_c}\p_{\bol{\zeta}_1}V_{11}\,dt}
\lr{\mathbb{P}^{\perp}_{11}\cdot\int_{\tau_c}\p_{\bol{\zeta}_1}V_{11}\,dt}
\,d\bol{z}''d\bol{z}'''d\bol{\zeta}_2d\bol{\zeta}_2',\\
=&\frac{1}{2}\int \mc{V}_{11}\lr{\mathbb{P}^{\perp}_{11}\cdot\int_{\tau_c}\p_{\bol{\zeta}_1}V_{11}\,dt}\lr{\mathbb{P}^{\perp}_{11}\cdot\int_{\tau_c}\p_{\bol{\zeta}_1}V_{11}\,dt}d\bol{\zeta}_1''d\bol{\zeta}_1''',~~~~\mc{V}_{11}=
\int f_2f_2'\mc{V}d\bol{\zeta}_2d\bol{\zeta}_2'd\bol{\zeta}_2''d\bol{\zeta}_2''',\\
%&\Pi_{11}'=
%-\frac{1}{2}\int\mc{V}
%\lr{\int_{\tau_c}\mathbb{P}^{\perp}_{11}\cdot\p_{\bol{\zeta}_1}V_{11}\,dt}
%\lr{\int_{\tau_c}\mathbb{P}^{\perp}_{11}\cdot\p_{\bol{\zeta}_1'}V_{11}\,dt}
%\,d\bol{z}_a'd\bol{z}_b'
%,\\
%&\Pi_{12}=\frac{1}{2}
%\int\mc{V}
%,\\
\Pi_{12}'=&-\frac{1}{2}
\int f_2f_1'\mc{V}\lr{\mathbb{P}^{\perp}_{12}\cdot\int_{\tau_c}\p_{\bol{\zeta}_1}V_{12}\,dt}
\lr{\mathbb{P}^{\perp}_{12}\cdot\int_{\tau_c}\p_{\bol{\zeta}_2'}V_{12}\,dt}\,d\bol{z}''d\bol{z}'''d\bol{\zeta}_2d\bol{\zeta}_1'
,\\
=&-\frac{1}{2}\int \mc{V}_{12}'\lr{\mathbb{P}^{\perp}_{12}\cdot\int_{\tau_c}\p_{\bol{\zeta}_1}V_{12}\,dt}\lr{\mathbb{P}^{\perp}_{12}\cdot\int_{\tau_c}\p_{\bol{\zeta}_2'}V_{12}\,dt}d\bol{\zeta}_1''d\bol{\zeta}_2''',~~~~\mc{V}_{12}'=
\int f_2f_1'\mc{V}d\bol{\zeta}_2d\bol{\zeta}_1'd\bol{\zeta}_2''d\bol{\zeta}_1''',
}{allPi}
with $\mc{V}=\mc{V}\lr{\bol{z},\bol{z}';\bol{z}'',\bol{z}'''}$. 
Neglecting all terms corresponding to collisions among more than 2 particles, from equation \eqref{ft} 
we have 
\eq{
&\frac{\p f_1}{\p t}=
-\frac{\p}{\p\bol{\zeta}_1}\cdot\lrs{f_1\mc{J}_{{\rm GC1}}\cdot\frac{\p\lr{{h}_1+q_1\Phi_1+{v}_1}}
{\p\bol{\zeta}_1}}\\&
+\frac{\p}{\p\bol{\zeta}_1}\cdot\lrs{
f_1\mc{J}_{{\rm GC}1}\cdot\int f_1'\Pi_{11}\cdot\lr{\mc{J}_{{\rm  GC}1}'\cdot\frac{\p\log f_1'}{\p\bol{\zeta}_1'}-\mc{J}_{{\rm GC}1}\cdot\frac{\p\log f_1}{\p\bol{\zeta}_1}}\,d\bol{\zeta}_1'
}\\
&+\frac{\p}{\p\bol{\zeta}_1}\cdot\lrs{
f_1\mc{J}_{{\rm GC}1}\cdot\int f_2'\Pi_{12}'\cdot\lr{\mc{J}_{{\rm  GC}2}'\cdot\frac{\p\log f_2'}{\p\bol{\zeta}_2'}-\mc{J}_{{\rm GC}1}\cdot\frac{\p\log f_1}{\p\bol{\zeta}_1}}\,d\bol{\zeta}_2'
}.
\label{f1t2}} 
A similar equation holds for $f_2$ (just replace $1$ with $2$ and $2$ with $1$).
Observe that equation \eqref{f1t2} contains two collision operators, the first expressing binary collisions between particles of the same species, the other representing binary collisions between particles belonging to different species. 
Finally, when $f_2=0$, we recover the single-species collision operator \eqref{Cop}.

%We also remark that $\Pi_{12}'$

%Note that each equation is independent of the other (collisions occur only among particles belonging to the same species), as one should expect from the hypothesis $f=f_1f_2$. 

\section{Conservation laws and entropy production for a two-species guiding center plasma}
In this section we show that equations \eqref{ft} and \eqref{f1t2} preserve total particle number, energy, and interior Casimir invariants, and that they satisfy H-theorems. 
We remark that additional conservation laws, such as total momentum, are not discussed here as proofs are similar to conservation of energy and have already been discussed in Sec. 3  in the single species setting. 

\subsection{Conservation of particle number}
Because \eqref{ft} and \eqref{f1t2} are in divergence form, it is clear that both  equations preserve the total probabilities (particle numbers) 
\eq{ \mc{N}=\int f\,d\bol{z}=1,~~~~\mc{N}_1=\int f_1\,d\bol{\zeta}_1=1,
}
under suitable boundary conditions. 
Of course, a corresponding invariant $\mc{N}_2$ exists for the evolution equation satisfied by $f_2$. 
Note that the domain in which equation \eqref{ft} is solved is given by $\mc{M}=\mc{M}_1\times\mc{M}_2$ with $\mc{M}_1=\mc{M}_2=\Omega\times\mathbb{R}\times [0,+\infty)$. In the following we shall denote with $\p\mc{M}$, $\p\mc{M}_1$, and $\p\mc{M}_2$ the boundaries (including the limit at infinity when appropriate) of $\mc{M}$, $\mc{M}_1$,  and $\mc{M}_2$ respectively. 
%\naoki{A possible choice of boundary conditions is then given by $f=0$ and $f_1=0$ on $\p\mc{M}$ and $\p\mc{M}_1$ respectively.}

\subsection{Conservation of energy}
Let us first focus on the energy conservation law for equation \eqref{ft}. 
In this case, the energy functional is given by the ensemble averaged energy of two colliding pairs of particles, plus the electromagnetic energy. The latter is given, up to time-independent terms, by the spatial integrals of the leading order terms of the magnetic perturbation energy density $\abs{\nabla\times\lr{A_{\parallel}\bol{b}}}^2/2\mu_0$. We have   
\eq{
\mf{H}=&\int ff'\lr{\mc{H}+V+\mc{H}'}\,d\bol{z}d\bol{z}'+\frac{1}{\mu_0}\int\lr{
\abs{\nabla A_{\parallel}}^2+A_{\parallel}^2\bol{b}\cdot\nabla\cp\nabla\cp\bol{b}
}\,d\bol{x}\\=&2\int f\mc{H}\,d\bol{z}+\int f\mc{Q}\,d\bol{z}
+\frac{1}{\mu_0}\int\lr{\abs{\nabla A_{\parallel}}^2+A_{\parallel}^2\bol{b}\cdot\nabla\cp\nabla\cp\bol{b}}\,d\bol{x},\label{Hft}
}
where $\mc{H}=\mc{H}\lr{\bol{z}}$ and $\mc{H}'=\mc{H}\lr{\bol{z}'}$ with $\mc{H}=\mathscr{H}-\mc{Q}$ the pair energy encountered in equation \eqref{pairH}. 
To see that \eqref{Hft} is an invariant of \eqref{ft}, %it is useful to perform the change of variables $\lr{x_s,y_s,z_s,u_s,\eta_s}\rightarrow\lr{x_s,y_s,z_s,v_{\parallel s},\eta_s}$. Note that the invariant measure remains unchanged, that is
%\eq{
%d\bol{z}=d\bol{\zeta}_1d\bol{\zeta}_2=d\bol{x}_1dv_{\parallel 1}d\eta_1d\bol{x}_2dv_{\parallel 2}d\eta_2.
%}
%Next,
we first observe that
\eq{
\mc{H}=\frac{1}{2m_1}\lr{m_1u_1-q_1A_{\parallel 1}}^2+\frac{1}{2m_2}\lr{m_2u_2-q_2A_{\parallel 2}}^2+\eta_1+\eta_2+\lambda.  
%=\frac{1}{2}m_1v_{\parallel 1}^2+\frac{1}{2}m_2v_{\parallel 2}^2+\eta_1+\eta_2+\lambda.
}
Hence, 
%$\p\mc{H}/\p t=0$ so that 
\eq{
\frac{\p\mc{H}}{\p t}=-q_1v_{\parallel 1}\frac{\p A_{\parallel 1}}{\p t}-q_2v_{\parallel 2}\frac{\p A_{\parallel 2}}{\p t}. 
}
It follows that
\eq{
&\int f\frac{\p\mc{H}}{\p t}\,d\bol{z}+\frac{1}{2\mu_0}\frac{\p}{\p t}\int\lr{
\abs{\nabla A_{\parallel}}^2+A_{\parallel}^2\bol{b}\cdot\nabla\times\nabla\times\bol{b}
}\,d\bol{x}\\&=-\int n_1q_1\bar{v}_{\parallel 1}\frac{\p A_{\parallel 1}}{\p t}\,d\bol{x}_1-\int n_2q_2\bar{v}_{\parallel 2}\frac{\p A_{\parallel 2}}{\p t}\,d\bol{x}_2-\frac{1}{\mu_0}\int \lr{\Delta A_{\parallel}-A_{\parallel}\bol{b}\cdot\nabla\cp\nabla\cp\bol{b}}\frac{\p A_{\parallel}}{\p t}\,d\bol{x}=0,\label{htapar} 
}
where we used equation \eqref{Apar} and
eliminated boundary integrals. 
We therefore have 
\eq{
\frac{d\mf{H}}{dt}=&\int\frac{\p f}{\p t}\mc{H}\,d\bol{z}+\int\frac{\p f}{\p t}\mc{Q}\,d\bol{z}+\int\frac{\p f'}{\p t}\mc{Q}'\,d\bol{z}'+\int\frac{\p f'}{\p t}\mc{H}'\,d\bol{z}'\\
=&\int \lrs{f\mf{J}_{\rm GC}\cdot\frac{\p\lr{\mc{H}+\mc{Q}}}{\p\bol{z}}\cdot\int f'\lr{\Pi'\cdot\mf{J}'_{\rm GC}\cdot\frac{\p\log f'}{\p\bol{z}'}-\Pi\cdot\mf{J}_{\rm GC}\cdot\frac{\p\log f}{\p\bol{z}}}\,d\bol{z}'}\,d\bol{z}\\
&+\int \lrs{f'\mf{J}'_{\rm GC}\cdot\frac{\p\lr{\mc{H}'+\mc{Q}'}}{\p\bol{z}'}\cdot\int f\lr{\tilde{\Pi}\cdot\mf{J}_{\rm GC}\cdot\frac{\p\log f}{\p\bol{z}}-\tilde{\Pi}'\cdot\mf{J}'_{\rm GC}\cdot\frac{\p\log f'}{\p\bol{z}'}}\,d\bol{z}}\,d\bol{z}'.
%\\
%=&\int ff'\lrs{\mf{J}_{\rm GC}\cdot\frac{\p\lr{\mc{H}+\mc{Q}}}{\p\bol{z}}-\mf{J}'_{\rm GC}\cdot\frac{\p\lr{\mc{H}'+\mc{Q}'}}{\p\bol{z}'}}\cdot\lr{\Pi'\cdot\mf{J}'_{\rm GC}\cdot\frac{\p\log f'}{\p\bol{z}'}-\Pi\cdot\mf{J}_{\rm GC}\cdot\frac{\p\log f}{\p\bol{z}}}\,d\bol{z}d\bol{z}'
\\
=&
-\frac{1}{2}\int ff'\mc{V}
\lr{\delta\bol{z}\cdot\frac{\p\log f}{\p\bol{z}}+\delta\bol{z}'\cdot\frac{\p\log f'}{\p\bol{z}'}}\lrs{\delta\bol{z}'\cdot\frac{\p\lr{\mc{H}'+\mc{Q}'}}{\p\bol{z}'}+\delta\bol{z}\cdot\frac{\p\lr{\mc{H}+\mc{Q}}}{\p\bol{z}}}\,d\bol{z}d\bol{z}'d\bol{z}''d\bol{z}'''\\
=&\frac{1}{2}\int ff'\mc{V}
\lr{\delta\bol{z}\cdot\frac{\p\log f}{\p\bol{z}}+\delta\bol{z}'\cdot\frac{\p\log f'}{\p\bol{z}'}}
\lrs{
%\lr{\mathbb{P}_{11}^{\perp}\cdot\int\p_{\bol{\zeta}_1}V_{11}}\cdot\bol{\xi}_{11}
%+\lr{\mathbb{P}_{12}^{\perp}\cdot\int\p_{\bol{\zeta}_1}V_{12}}\cdot\bol{\xi}_{12}
%+\lr{\mathbb{P}_{21}^{\perp}\cdot\int\p_{\bol{\zeta}_2}V_{21}}\cdot\bol{\xi}_{21}
%+\lr{\mathbb{P}_{22}^{\perp}\cdot\int\p_{\bol{\zeta}_2}V_{22}}\cdot\bol{\xi}_{22}
\sum_{i,j=1}^2\lr{\mathbb{P}_{ij}^{\perp}\cdot\int\p_{\bol{\zeta}_i}V_{ij}}\cdot\bol{\xi}_{ij}
}
\,d\bol{z}d\bol{z}'d\bol{z}''d\bol{z}'''=0
,
%-\frac{1}{2}\tau_c^2\int ff'\Gamma\lrs{\mf{J}_{\rm GC}\cdot\frac{\p\lr{\mc{H}+\mc{Q}}}{\p\bol{z}}\cdot\frac{\p V}{\p\bol{z}}+\mf{J}'_{\rm GC}\cdot\frac{\p\lr{\mc{H}'+\mc{Q}'}}{\p\bol{z}'}\cdot\frac{\p V}{\p\bol{z}'}}\\&\lr{\frac{\p V}{\p\bol{z}}\cdot\mf{J}_{\rm GC}\cdot\frac{\p\log f}{\p\bol{z}}+\frac{\p V}{\p\bol{z}'}\cdot\mf{J}'_{\rm GC}\cdot\frac{\p\log f'}{\p\bol{z}'}}\,d\bol{z}d\bol{z}',
}
where we performed integration by parts, eliminated boundary integrals, %(a posssible choice of boundary conditions is again $f=0$ on $\p\mc{M}$)
and defined
\eq{
\tilde{\Pi}=
%-\frac{1}{2}\tau_c^2\Gamma\frac{\p V}{\p\bol{z}'}\frac{\p V}{\p\bol{z}}
\Pi_{ba}
,~~~~\tilde{\Pi}'=
%\frac{1}{2}\tau_c^2\Gamma\frac{\p V}{\p\bol{z}'}\frac{\p V}{\p\bol{z}'}
\Pi_{bb}. 
}
Here, we also used the fact that the scattering 
%frequency is symmetric, 
volume density per unit time $\mc{V}$ is symmetric in the 
order of the particle pairs, 
i.e., $\mc{V}\lr{\bol{z}_a,\bol{z}_b;\bol{z}_a',\bol{z}_b'}=\mc{V}\lr{\bol{z}_b,\bol{z}_a;\bol{z}_b',\bol{z}_a'}$, and the fact that $\mathbb{P}^{\perp}_{ij}\cdot\bol{\xi}_{ij}=0$, $i,j=1,2$, by construction. 
%$\Gamma\lr{\bol{z},\bol{z}'}=\Gamma\lr{\bol{z}',\bol{z}}$. 
%with $\Gamma'=\Gamma\lr{\bol{z}',\bol{z}}$. 

%However, by hypothesis, the total energy of two colliding pairs is preserved. Furthermore, due to the scale separation between ideal dynamics and collisions, the phase space gradients of the Hamiltonians $\mc{H}$ and $\mc{H}'$ and the rate of change in the electrostatic energy $\p\mc{Q}/\p t$ are negligible during binary interactions, leading to the elastic scattering condition 
%\eq{
%\frac{d}{dt}\lr{\mc{H}+\mc{Q}+\mc{H}'+\mc{Q}'}=
%\frac{\p\lr{\mc{H}+\mc{Q}}}{\p\bol{z}}\cdot\mf{J}_{\rm GC}\cdot\frac{\p V}{\p\bol{z}}+
%\frac{\p\lr{\mc{H}'+\mc{Q}'}}{\p\bol{z}'}\cdot\mf{J}'_{\rm GC}\cdot\frac{\p V}{\p\bol{z}'}=0.\label{elsc}
%}
%It follows that the energy functional $\mf{H}$ defined in \eqref{Hft} is an  invariant of \eqref{ft}. 
%It can be shown that the elastic scattering condition \eqref{elsc} reduces to the  conservation of the modulus of the relative velocity in the case of grazing Coulomb collisions in canonical phase space (on this point, see \cite{SatoMorrison24}).

We now move to the energy invariant associated with the evolution equation \eqref{f1t2} for $f_1$ and the corresponding equation for $f_2$. 
Now recall that \eqref{f1t2} was derived under the assumption that $f=f_1f_2$. We therefore expect that the relevant energy functional $\mf{H}_{12}$ can be obtained by substituting $f=f_1f_2$ in eq.  \eqref{Hft},
\eq{
\mf{H}_{12}=&\int f_1{h_1}\,d\bol{\zeta}_1+\int f_1f_2v\,d\bol{\zeta}_1d\bol{\zeta}_2+\int f_2{h_2}\,d\bol{\zeta}_2
+\int f_1'{h_1}'\,d\bol{\zeta}_1'+\int f_1'f_2'v'\,d\bol{\zeta}_1'd\bol{\zeta}_2'+\int f_2'{h_2}'\,d\bol{\zeta}_2'
\\&+\int f_1f_2'V_{12}\,d\bol{\zeta}_1d\bol{\zeta}_2'+\int f_1f_1'V_{11}\,d\bol{\zeta}_1d\bol{\zeta}_1'+\int f_2f_1'V_{21}\,d\bol{\zeta}_1'd\bol{\zeta}_2+\int f_2f_2'V_{22}\,d\bol{\zeta}_2d\bol{\zeta}_2'\\&+\frac{1}{\mu_0}\int\lr{
\abs{\nabla A_{\parallel}}^2+A_{\parallel}^2\bol{b}\cdot\nabla\cp\nabla\cp\bol{b}
}\,d\bol{x}\\
=&\int f_1\lr{2h_1+q_1\Phi_1}\,d\bol{\zeta}_1+2\int f_1f_2v\,d\bol{\zeta}_1d\bol{\zeta}_2+\int f_2\lr{2h_2+q_2\Phi_2}\,d\bol{\zeta}_2
\\&+\frac{1}{\mu_0}\int\lr{
\abs{\nabla A_{\parallel}}^2+A_{\parallel}^2\bol{b}\cdot\nabla\cp\nabla\cp\bol{b}
}\,d\bol{x}
.\label{H12}
}
As in the case of Eq.~\eqref{ft}, the change in magnetic energy is canceled by the terms involving $\partial(h_s + v_s)/\partial t$, $s = 1, 2$.  
In particular, the same calculation as in Eq.~\eqref{htapar} applies. 
%where we used the notation
%\eq{
%V_{11}=\frac{q_1^2}{4\pi\epsilon_0\abs{\bol{x}_1-\bol{x}_1'}},~~~~V_{12}=\frac{q_1q_2}{4\pi\epsilon_0\abs{\bol{x}_1-\bol{x}_2'}},~~~~V_{21}=\frac{q_1q_2}{4\pi\epsilon_0\abs{\bol{x}_2-\bol{x}_1'}},~~~~V_{22}=\frac{q_2^2}{4\pi\epsilon_0\abs{\bol{x}_2-\bol{x}_2'}}.
%}
It follows that
\eq{
\frac{d\mf{H}_{12}}{dt}=&2\int \frac{\p f_1}{\p t}\lr{h_1+q_1\Phi_1+v_1}\,d\bol{\zeta}_1+2\int \frac{\p f_2}{\p t}\lr{h_2+q_2\Phi_2+v_2}\,d\bol{\zeta}_2\\=&\int \lrs{f_1\mc{J}_{\rm GC1}\cdot\frac{\p\lr{h_1+q_1\Phi_1+v_1}}{\p\bol{\zeta}_1}\cdot\int f_1'{\Pi}_{11}\cdot\lr{\mc{J}_{\rm GC1}'\cdot\frac{\p\log f_1'}{\p\bol{\zeta}_1'}-\mc{J}_{\rm GC1}\cdot\frac{\p\log f_1}{\p\bol{\zeta}_1}}\,d\bol{\zeta}_1'}\,d\bol{\zeta}_1
\\&+
\int \lrs{f_1\mc{J}_{\rm GC1}\cdot\frac{\p\lr{h_1+q_1\Phi_1+v_1}}{\p\bol{\zeta}_1}\cdot\int f_2'{\Pi}_{12}'\cdot\lr{\mc{J}_{\rm GC2}'\cdot\frac{\p\log f_2'}{\p\bol{\zeta}_2'}-\mc{J}_{\rm GC1}\cdot\frac{\p\log f_1}{\p\bol{\zeta}_1}}\,d\bol{\zeta}'_2}\,d\bol{\zeta}_1\\
&+\int \lrs{f_2\mc{J}_{\rm GC2}\cdot\frac{\p\lr{h_2+q_2\Phi_2+v_2}}{\p\bol{\zeta}_2}\cdot\int f_2'{\Pi}_{22}\cdot\lr{\mc{J}_{\rm GC2}'\cdot\frac{\p\log f_2'}{\p\bol{\zeta}_2'}-\mc{J}_{\rm GC2}\cdot\frac{\p\log f_2}{\p\bol{\zeta}_2}}\,d\bol{\zeta}'_2}\,d\bol{\zeta}_2
\\&+
\int \lrs{f_2\mc{J}_{\rm GC2}\cdot\frac{\p\lr{h_2+q_2\Phi_2+v_2}}{\p\bol{\zeta}_2}\cdot\int f_1'{\Pi}_{21}'\cdot\lr{\mc{J}_{\rm GC1}'\cdot\frac{\p\log f_1'}{\p\bol{\zeta}_1'}-\mc{J}_{\rm GC2}\cdot\frac{\p\log f_2}{\p\bol{\zeta}_2}}\,d\bol{\zeta}'_1}\,d\bol{\zeta}_2
\\&+\int \lrs{f_1'\mc{J}'_{\rm GC1}\cdot\frac{\p\lr{h'_1+q_1\Phi'_1+v'_1}}{\p\bol{\zeta}'_1}\cdot\int f_1{\Pi}_{11}'\cdot\lr{\mc{J}_{\rm GC1}\cdot\frac{\p\log f_1}{\p\bol{\zeta}_1}-\mc{J}'_{\rm GC1}\cdot\frac{\p\log f_1'}{\p\bol{\zeta}_1'}}\,d\bol{\zeta}_1}\,d\bol{\zeta}_1'
\\&+
\int \lrs{f_1'\mc{J}'_{\rm GC1}\cdot\frac{\p\lr{h_1'+q_1\Phi_1'+v_1'}}{\p\bol{\zeta}_1'}\cdot\int f_2{\Pi}_{12}\cdot\lr{\mc{J}_{\rm GC2}\cdot\frac{\p\log f_2}{\p\bol{\zeta}_2}-\mc{J}'_{\rm GC1}\cdot\frac{\p\log f_1'}{\p\bol{\zeta}_1'}}\,d\bol{\zeta}_2}\,d\bol{\zeta}_1'\\
&+\int \lrs{f_2'\mc{J}'_{\rm GC2}\cdot\frac{\p\lr{h_2'+q_2\Phi_2'+v_2'}}{\p\bol{\zeta}_2'}\cdot\int f_2{\Pi}_{22}'\cdot\lr{\mc{J}_{\rm GC2}\cdot\frac{\p\log f_2}{\p\bol{\zeta}_2}-\mc{J}'_{\rm GC2}\cdot\frac{\p\log f_2'}{\p\bol{\zeta}_2'}}\,d\bol{\zeta}_2}\,d\bol{\zeta}_2'
\\&+
\int \lrs{f_2'\mc{J}'_{\rm GC2}\cdot\frac{\p\lr{h_2'+q_2\Phi'_2+v'_2}}{\p\bol{\zeta}'_2}\cdot\int f_1{\Pi}_{21}\cdot\lr{\mc{J}_{\rm GC1}\cdot\frac{\p\log f_1}{\p\bol{\zeta}_1}-\mc{J}'_{\rm GC2}\cdot\frac{\p\log f'_2}{\p\bol{\zeta}_2'}}\,d\bol{\zeta}_1}\,d\bol{\zeta}_2'\\
=&
\sum_{i,j=1}^2\frac{1}{2}\int f_1f_1'f_2f_2'\mc{V}\,{\bol{\xi}_{ij}\cdot\lr{\mathbb{P}^{\perp}_{ij}\cdot\int\p_{\bol{\zeta}_i}V_{ij}}}\lr{\mathbb{P}^{\perp}_{ij}\cdot\int\p_{\bol{\zeta}_i}V_{ij}}\\&\cdot\lr{\mc{J}_{{\rm GC}j}'\cdot\frac{\p\log f_j'}{\p\bol{\zeta}_j'}-\mc{J}_{{\rm GC}i}\cdot\frac{\p \log f_i}{\p\bol{\zeta}_i}}\,d\bol{z}d\bol{z}'d\bol{z}''d\bol{z}'''=0,\label{dh12dt}
}
where the tensors $\Pi_{ij}$ and $\Pi_{ij}'$ are those appearing in the evolution equations for $f_1$, $f_1'$, $f_2$, and $f_2'$ and we used the fact that $\mathbb{P}^{\perp}_{ij}\cdot\bol{\xi}_{ij}=\bol{0}$, $i,j=1,2$. We have thus shown that the energy $\mf{H}_{12}$ is constant. 

\subsection{Conservation of interior Casimir invariants}
Equations \eqref{ft} and \eqref{f1t2} inherit a Casimir invariant from the Casimir invariant (the magnetic moment)  of the Poisson tensor of guiding center dynamics.  
This type of Casimir invariant induced by microscopic equations of motion on the field theory are called interior Casimir invariants \cite{SatoMorrison24}. 
First, consider equation \eqref{ft}. The relevant Casimir invariants are the total magnetic moments of the $2$ species, 
\eq{
\ms{M}=\int fg\lr{\frac{\eta_1}{B_0},\frac{\eta_2}{B_0}}\,d\bol{z},
%\mf{M}_1=\int f\frac{\eta_1}{B_0}\,d\bol{z}d\bol{z}',~~~~
%\mf{M}_2=\int f\frac{\eta_2}{B_0}\,d\bol{z}d\bol{z}'.
\label{mu12}
}
where $g$ is an arbitrary function of the magnetic moments $\mu_s=\eta_s/B_0$, $s=1,2$. 
Let us verify that \eqref{mu12} is  preserved by equation \eqref{ft}. 
Performing integration by parts and eliminating boundary integrals, we have
\eq{
\frac{d\ms{M}}{dt}=\int f\lrs{-\frac{\p\lr{\mc{H}+\mc{Q}}}{\p\bol{z}}+\int f'\lr{\Pi'\cdot\mf{J}'_{\rm GC}\cdot\frac{\p\log f'}{\p\bol{z}'}-\Pi\cdot\mf{J}_{\rm GC}\cdot\frac{\p\log f}{\p\bol{z}}}\,d\bol{z}'}\cdot\mf{J}_{\rm GC}\cdot\frac{\p g}{\p\bol{z}}\,d\bol{z}=0, 
}
where we used the fact that $\mf{J}_{\rm GC}\cdot\p g/\p\bol{z}=\bol{0}$. 

We now return to equation \eqref{f1t2}. The Casimir invariant in this case can be obtained by substituting $f=f_1f_2$ into equation \eqref{mu12}, 
\eq{
\ms{M}_{12}=\int f_1f_2 g\lr{\frac{\eta_1}{B_0},\frac{\eta_2}{B_0}}\,d\bol{\zeta}_1d\bol{\zeta}_2.
}
Again, performing integration by parts and dropping boundary integrals, we have 
\eq{
\frac{d\ms{M}_{12}}{dt}=&-\int f_1f_2\frac{\p\lr{h_1+q_1\Phi_1+v_1}}{\p\bol{\zeta}_1}\cdot\mc{J}_{\rm GC1}\cdot\frac{\p g}{\p\bol{\zeta}_1}\,d\bol{\zeta}_1d\bol{\zeta}_2\\&+
\int f_1f_2f_1'{\Pi}_{11}\cdot\lr{\mc{J}_{\rm GC1}'\cdot\frac{\p\log f_1'}{\p\bol{\zeta}_1'}-\mc{J}_{\rm GC1}\cdot\frac{\p\log f_1}{\p\bol{\zeta}_1}}\cdot\mc{J}_{\rm GC1}\cdot\frac{\p g}{\p\bol{\zeta}_1}\,d\bol{\zeta}_1'd\bol{\zeta}_1d\bol{\zeta}_2\\
&+\int f_1f_2f_2'{\Pi}'_{12}\cdot\lr{\mc{J}_{\rm GC2}'\cdot\frac{\p\log f_2'}{\p\bol{\zeta}_2'}-\mc{J}_{\rm GC1}\cdot\frac{\p\log f_1}{\p\bol{\zeta}_1}}\cdot\mc{J}_{\rm GC1}\cdot\frac{\p g}{\p\bol{\zeta}_1}\,d\bol{\zeta}_2'd\bol{\zeta}_1d\bol{\zeta}_2\\
&-\int f_1f_2\frac{\p\lr{h_2+q_2\Phi_2+v_2}}{\p\bol{\zeta}_2}\cdot\mc{J}_{\rm GC2}\cdot\frac{\p g}{\p\bol{\zeta}_2}\,d\bol{\zeta}_1d\bol{\zeta}_2\\&+
\int f_1f_2f_2'{\Pi}_{22}\cdot\lr{\mc{J}_{\rm GC2}'\cdot\frac{\p\log f_2'}{\p\bol{\zeta}_2'}-\mc{J}_{\rm GC2}\cdot\frac{\p\log f_2}{\p\bol{\zeta}_2}}\cdot\mc{J}_{\rm GC2}\cdot\frac{\p g}{\p\bol{\zeta}_2}\,d\bol{\zeta}_2'd\bol{\zeta}_1d\bol{\zeta}_2\\
&+\int f_1f_2f_1'{\Pi}'_{21}\cdot\lr{\mc{J}_{\rm GC1}'\cdot\frac{\p\log f_1'}{\p\bol{\zeta}_1'}-\mc{J}_{\rm GC2}\cdot\frac{\p\log f_2}{\p\bol{\zeta}_2}}\cdot\mc{J}_{\rm GC2}\cdot\frac{\p g}{\p\bol{\zeta}_2}\,d\bol{\zeta}_1'd\bol{\zeta}_1d\bol{\zeta}_2=0,
}
where we used the fact that $\mc{J}_{\rm GCs}\cdot{\p g}/{\p\bol{\zeta}_s}=0$, $s=1,2$.

\subsection{Entropy production}
The Shannon entropy measure associated with the distribution function $f$ of equation \eqref{ft} is given by
\eq{
S=-\int f\log f\,d\bol{z}
-\int f'\log f'\,d\bol{z}'.\label{S}
}
The rate of change in $S$ can be evaluated as follows:
\eq{
\frac{dS}{dt}=&\int f\mf{J}_{\rm GC}\cdot\lrs{-\frac{\p\lr{\mc{H}+\mc{Q}}}{\p\bol{z}}+\int f'\lr{\Pi'\cdot\mf{J}_{\rm GC}'\cdot\frac{\p\log f'}{\p\bol{z}'}-\Pi\cdot\mf{J}_{\rm GC}\cdot\frac{\p\log f}{\p\bol{z}}}\,d\bol{z}'}\cdot\frac{\p\log f}{\p\bol{z}}\,d\bol{z}\\
&+\int f'\mf{J}_{\rm GC}'\cdot\lrs{-\frac{\p\lr{\mc{H}'+\mc{Q}'}}{\p\bol{z}'}+\int f\lr{\tilde{\Pi}\cdot\mf{J}_{\rm GC}\cdot\frac{\p\log f}{\p\bol{z}}-\tilde{\Pi}'\cdot\mf{J}'_{\rm GC}\cdot\frac{\p\log f'}{\p\bol{z}'}}\,d\bol{z}}\cdot\frac{\p\log f'}{\p\bol{z}'}\,d\bol{z}'
\\
=&%\int \Pi'\cdot\mf{J}'_{\rm GC}\cdot\frac{\p f'}{\p\bol{z}'}
%\frac{1}{2}\tau_c^2\int ff'\Gamma\lr{\frac{\p V}{\p\bol{z}'}\cdot\mf{J}'_{\rm GC}\cdot\frac{\p\log f'}{\p\bol{z}'}+\frac{\p V}{\p\bol{z}}\cdot\mf{J}_{\rm GC}\cdot\frac{\p\log f}{\p\bol{z}}}^2\,d\bol{z}d\bol{z}'\geq 0
\frac{1}{2}\int ff'\mc{V}\lr{\delta\bol{z}'\cdot\frac{\p\log f'}{\p\bol{z}'}-\delta\bol{z}\cdot\frac{\p\log f}{\p\bol{z}}}^2\,d\bol{z}d\bol{z}'d\bol{z}''d\bol{z}'''\\
=&\frac{1}{2}\int ff'\mc{V}\lrs{\sum_{ij=1}^2\lr{\int_{\tau_c}\p_{\bol{\zeta}_i}V_{ij}\,dt}\cdot\mathbb{P}^{\perp}_{ij}\cdot\lr{\mc{J}_{{\rm GC}i}\cdot\frac{\p\log f}{\p\bol{\zeta}_i}-\mc{J}_{{\rm GC}j}'\cdot\frac{\p\log f'}{\p\bol{\zeta}_j'}}}^2\,d\bol{z}d\bol{z}'d\bol{z}''d\bol{z}'''
\geq 0
,\label{dSdt}
}
where we performed integration by parts, eliminated surface integrals, used the fact that 
\eq{\frac{\p}{\p\bol{z}}\cdot\lrs{\mf{J}_{\rm GC}\cdot\frac{\p\lr{\mc{H}+\mc{Q}}}{\p\bol{z}}}=\frac{\p}{\p\bol{z}'}\cdot\lrs{\mf{J}_{\rm GC}'\cdot\frac{\p\lr{\mc{H}'+\mc{Q}'}}{\p\bol{z}'}}=0,}
and assumed that $ff'\geq 0$.

Now consider equation \eqref{f1t2}. Again, the entropy measure for this equation can be obtained by substituting $f=f_1f_2$ into \eqref{S}. The result is
\eq{
S_{12}=%-\int f_1f_2\log\lr{f_1f_2}\,d\bol{\zeta}_1d\bol{\zeta}_2-\int f_1'f_2'\log \lr{f_1'f_2'}\,d\bol{\zeta}_1'd\bol{\zeta}_2'=
-\int f_1\log f_1\,d\bol{\zeta}_1
-\int f_2\log f_2\,d\bol{\zeta}_2
-\int f_1'\log f_1'\,d\bol{\zeta}_1'
-\int f_2'\log f_2'\,d\bol{\zeta}_2'.\label{S12}
}
%where we dropped terms of the type $\int \log f_1\,d\bol{\zeta}_1$ as their rate of change is a vanishing boundary integral. 
Performing integration by parts and eliminating boundary integrals, it can be shown that
\eq{
\frac{dS_{12}}{dt}=&\int f_1f_1'\mc{J}_{\rm GC1}\cdot{\Pi}_{11}\cdot\lr{\mc{J}_{\rm GC1}'\cdot\frac{\p\log f_1'}{\p\bol{\zeta}_1'}-\mc{J}_{\rm GC1}\cdot\frac{\p\log f_1}{\p\bol{\zeta}_1}}\cdot\frac{\p\log f_1}{\p\bol{\zeta}_1}\,d\bol{\zeta}_1'd\bol{\zeta}_1\\
&+\int f_1f_2'\mc{J}_{\rm GC1}\cdot{\Pi}_{12}'\cdot\lr{\mc{J}_{\rm GC2}'\cdot\frac{\p\log f_2'}{\p\bol{\zeta}_2'}-\mc{J}_{\rm GC1}\cdot\frac{\p\log f_1}{\p\bol{\zeta}_1}}\cdot\frac{\p\log f_1}{\p\bol{\zeta}_1}\,d\bol{\zeta}_2'd\bol{\zeta}_1\\
&+\int f_2f_2'\mc{J}_{\rm GC2}\cdot{\Pi}_{22}\cdot\lr{\mc{J}_{\rm GC2}'\cdot\frac{\p\log f_2'}{\p\bol{\zeta}_2'}-\mc{J}_{\rm GC2}\cdot\frac{\p\log f_2}{\p\bol{\zeta}_2}}\cdot\frac{\p\log f_2}{\p\bol{\zeta}_2}\,d\bol{\zeta}_2'd\bol{\zeta}_2\\
&+\int f_2f_1'\mc{J}_{\rm GC2}\cdot{\Pi}_{21}'\cdot\lr{\mc{J}_{\rm GC1}'\cdot\frac{\p\log f_1'}{\p\bol{\zeta}_1'}-\mc{J}_{\rm GC2}\cdot\frac{\p\log f_2}{\p\bol{\zeta}_2}}\cdot\frac{\p\log f_2}{\p\bol{\zeta}_2}\,d\bol{\zeta}_1'd\bol{\zeta}_2\\
&+\int f_1f_1'\mc{J}_{\rm GC1}'\cdot{\Pi}_{11}'\cdot\lr{\mc{J}_{\rm GC1}\cdot\frac{\p\log f_1}{\p\bol{\zeta}_1}-\mc{J}'_{\rm GC1}\cdot\frac{\p\log f_1'}{\p\bol{\zeta}_1'}}\cdot\frac{\p\log f_1'}{\p\bol{\zeta}_1'}\,d\bol{\zeta}_1'd\bol{\zeta}_1\\
&+\int f_1'f_2\mc{J}_{\rm GC1}'\cdot{\Pi}_{12}\cdot\lr{\mc{J}_{\rm GC2}\cdot\frac{\p\log f_2}{\p\bol{\zeta}_2}-\mc{J}_{\rm GC1}'\cdot\frac{\p\log f_1'}{\p\bol{\zeta}_1'}}\cdot\frac{\p\log f_1'}{\p\bol{\zeta}_1'}\,d\bol{\zeta}_2d\bol{\zeta}_1'\\
&+\int f_2f_2'\mc{J}_{\rm GC2}'\cdot{\Pi}_{22}'\cdot\lr{\mc{J}_{\rm GC2}\cdot\frac{\p\log f_2}{\p\bol{\zeta}_2}-\mc{J}_{\rm GC2}'\cdot\frac{\p\log f_2'}{\p\bol{\zeta}_2'}}\cdot\frac{\p\log f_2'}{\p\bol{\zeta}_2'}\,d\bol{\zeta}_2'd\bol{\zeta}_2\\
&+\int f_2'f_1\mc{J}_{\rm GC2}'\cdot{\Pi}_{21}\cdot\lr{\mc{J}_{\rm GC1}\cdot\frac{\p\log f_1}{\p\bol{\zeta}_1}-\mc{J}_{\rm GC2}'\cdot\frac{\p\log f_2'}{\p\bol{\zeta}_2}'}\cdot\frac{\p\log f_2'}{\p\bol{\zeta}_2'}\,d\bol{\zeta}_1d\bol{\zeta}_2'.
}
Noting that ${\Pi}_{11}={\Pi}_{11}'$, ${\Pi}_{12}'={\Pi}_{21}$, ${\Pi}_{22}={\Pi}_{22}'$, and 
${\Pi}_{21}'={\Pi}_{12}$,  
this expression can be rearranged as follows:
\eq{
\frac{dS_{12}}{dt}=&
\sum_{i,j=1}^2\frac{1}{2}\int f_if_jf_i'f_j'\mc{V}\lrs{\lr{\mathbb{P}^{\perp}_{ij}\cdot\int\p_{\bol{\zeta}_i}V_{ij}}\cdot\lr{
\mc{J}_{{\rm GC}j}'\cdot\frac{\p\log f_j'}{\p\bol{\zeta}_j'}-\mc{J}_{{\rm}i}\cdot\frac{\p\log f_i}{\p\bol{\zeta}_i}
}}^2\,d\bol{z}d\bol{z}'d\bol{z}''d\bol{z}'''
%\frac{1}{2}\tau_c^2\int f_1f_1'\bar{\Gamma}_{11}\lrs{\frac{\p V_{11}}{\p\bol{\zeta}_1}\cdot\lr{\mc{J}_{\rm GC1}'\cdot\frac{\p\log f_1'}{\p\bol{\zeta}_1'}-\mc{J}_{\rm GC1}\cdot\frac{\p\log f_1}{\p\bol{\zeta}_1}}}^2\,d\bol{\zeta}_1'd\bol{\zeta}_1\\&+
%\frac{1}{2}\tau_c^2\int f_1f_2'\bar{\Gamma}_{12}'\lrs{\frac{\p V_{12}}{\p\bol{\zeta}_1}\cdot\lr{\mc{J}_{\rm GC2}'\cdot\frac{\p\log f_2'}{\p\bol{\zeta}_2'}-\mc{J}_{\rm GC1}\cdot\frac{\p\log f_1}{\p\bol{\zeta}_1}}}^2\,d\bol{\zeta}_2'd\bol{\zeta}_1\\&+
%\frac{1}{2}\tau_c^2\int f_2f_2'\bar{\Gamma}_{22}\lrs{\frac{\p V_{22}}{\p\bol{\zeta}_2}\cdot\lr{\mc{J}_{\rm GC2}'\cdot\frac{\p\log f_2'}{\p\bol{\zeta}_2'}-\mc{J}_{\rm GC2}\cdot\frac{\p\log f_2}{\p\bol{\zeta}_2}}}^2\,d\bol{\zeta}_2'd\bol{\zeta}_2\\&+
%\frac{1}{2}\tau_c^2\int f_1'f_2\bar{\Gamma}_{21}'\lrs{\frac{\p V_{21}}{\p\bol{\zeta}_1}\cdot\lr{\mc{J}_{\rm GC1}'\cdot\frac{\p\log f_1'}{\p\bol{\zeta}_1'}-\mc{J}_{\rm GC2}\cdot\frac{\p\log f_2}{\p\bol{\zeta}_2}}}^2\,d\bol{\zeta}_2d\bol{\zeta}_1'
\geq 0,\label{dS12dt}
}
where in the last passage we assumed that $f_1,f_1',f_2,f_2'\geq0$. 

%recall metriplectic structure?

\section{Thermodynamic equilibria}
%The elastic scattering conditions \eqref{elsc}, \eqref{elsc1}, \eqref{elsc2}, \eqref{elsc3}, and \eqref{elsc4}, and 
The H-theorems \eqref{dSdt} and \eqref{dS12dt} can be used to derive thermodynamic equilibria (steady solutions of \eqref{ft} and \eqref{f1t2}). 
Indeed, if well-behaved thermodynamic equilibria exist for $f$, $f_1$, and $f_2$, they must satisfy $dS/dt=0$ and $dS_{12}/dt=0$. 
In particular, following the same steps as in the single-species case discussed in Sec. 3.5, one finds the equilibrium distribution functions 
\eq{
f_{\infty}=
%\lim_{t\rightarrow+\infty}f=
\frac{1}{Z}\exp\lrc{-\beta \lr{\mc{H}+\mc{Q}}-g\lr{\frac{\eta_1}{B_{01}},\frac{\eta_2}{B_{02}}}},
}
and 
\eq{
&f_{1\infty}=
%\lim_{t\rightarrow+\infty}f_1=
\frac{1}{Z_1}\exp\lrc{-\beta\lr{h_1+q_1\Phi_1+v_1}+g_1\lr{\frac{\eta_1}{B_{01}}}},\\&f_{2\infty}=
%\lim_{t\rightarrow+\infty}f_2=
\frac{1}{Z_2}\exp\lrc{-\beta\lr{h_2+q_2\Phi_2+v_2}+g_2\lr{\frac{\eta_2}{B_{02}}}},\label{f1f2eq}
}
where $g\lr{{\eta_1}/{B_0},{\eta_2}/{B_0}}$, $g_1\lr{\eta_1/B_0}$, and $g_2\lr{\eta_2/B_0}$ are functions of the magnetic moments which are determined by the initial conditions of the system (the shape of the distribution functions at $t=0$),  $\beta\in\mathbb{R}_{\geq 0}$ is the inverse temperature, and $Z$, $Z_1$, and $Z_2$ are normalization constants. 

We remark that the functions $g$, $g_1$, and $g_2$, combined with the nontrivial invariant measures \eqref{IMs} and \eqref{IM} are responsible for deviation from Maxwell-Boltzmann statistics. 
As a consequence, thermodynamic equilibrium may exhibit a self-organized inhomogeneous spatial density profile. For example, the spatial density $\rho_1\lr{\bol{x}_1}$ of the first species associated with $f_1$ is given by 
\eq{
\rho_1=\int f_1\,du_1d\eta_1=&\frac{1}{Z_1}\exp\lrc{-\beta\lr{q_1\Phi_1+v_1}}\int\exp\lrc{-\beta\lr{\frac{1}{2}m_1u_1^2-q_1u_1A_{\parallel 1}}}\,du_1\\&\int\exp\lrc{-\beta\eta_1+g_1\lr{\frac{\eta_1}{B_{01}}}}\,d\eta_1\\=&
\frac{1}{Z_1}\sqrt{\frac{2\pi}{\beta m_1}}\exp\lrc{-\beta\lr{q_1\Phi_1+v_1-\frac{q_1^2A_{\parallel 1}^2}{2m_1}}}\int\exp\lrc{-\beta\eta_1+g_1\lr{\frac{\eta_1}{B_{01}}}}\,d\eta_1.
}
In order to evaluate the integral on the right-hand side, we must specify the function $g_1$, which encapsulates the initial conditions (the initial distribution of magnetic moment). 
Let us consider the simplest case of a linear function of the magnetic moment $g_1=-\beta\gamma\eta_1/B_{01}$ where $\gamma>0$ is a physical constant (%if $g_1$ is small compared to $\beta\eta_1$, 
this can be thought as the first nontrivial term of a Taylor expansion of $g_1$).
Then, we obtain
\eq{
\rho_1=\frac{1}{Z_1}\sqrt{\frac{2\pi}{\beta m_1}}\exp\lrc{-\beta\lr{q_1\Phi_1+v_1-\frac{q_1^2A_{\parallel 1}^2}{2m_1}}}\frac{B_{01}}{\beta\lr{\gamma+B_{01}}}.\label{density1}
}
From this expression, it is clear that the conservation of the first adiabatic invariant results in deviation from Maxwell-Boltzmann statistics through the term involving $\gamma> 0$. In particular, the density distribution \eqref{density1} tends to be higher in regions of higher magnetic field strength $B_0$. 
A similar computation for a Gaussian-type profile $g_1=-\beta\gamma^2\frac{\eta_1^2}{B_{01}^2}$ leads to
\eq{
\rho_1=\frac{\pi B_{01}}{Z_1\beta\gamma\sqrt{2 m_1}}\exp\lrc{-\beta\lr{q_1\Phi_1+v_1-\frac{q_1^2A_{\parallel 1}^2}{2m_1}}}\lrs{1-{\rm erf}\lr{\frac{\sqrt{\beta}B_{01}}{2\gamma}}}\exp\lrc{\frac{\beta B_{01}^2}{4\gamma^2}}.
}

Finally, let us spend a few words on how to compute $g$ from the initial condition $f_{10}=f_1\lr{\bol{\zeta}_1,0}$, 
which we express in the form
\eq{
f_{0}=\frac{1}{Z}\exp\lrc{-\beta\lr{\mu B_{0}+\frac{1}{2}mu^2+q\Phi}+\chi\lr{\mu,\bol{x},u}}
=f^{\rm MB}e^{\chi},
}
where $f^{{\rm MB}}$ denotes the Maxwell-Boltzmann distribution, and $\chi=\chi\lr{\mu,\bol{x},u}$ is some given function of the phase space variables. 
We dropped the particle index for ease of notation, and assumed $A_{\parallel}=v=0$ to simplify the algebra.  
Now define the following function of 
the magnetic moment $\mu$,
\eq{
G=\int f_{0}B_0\,d\bol{x}du=\int f_{\infty}B_0\,d\bol{x}du, \label{Ggmu}
}
where 
\eq{
f_{\infty}=\frac{1}{Z}\exp\lrc{-\beta\lr{\mu B_0+\frac{1}{2}mu^2+q\Phi}+g\lr{\mu}}=f^{\rm MB}e^g,
}
denotes the distribution function at thermodynamic equilibrium, and the last equality in eq. \eqref{Ggmu} follows from the constancy of $\mu$ throughout time evolution.
This can be seen explicitly from the fact that the effective phase space velocity $\bol{Z}$ defined in \eqref{Zeff} has the form $Z=\mc{J}_{{\rm GC}}\cdot\bol{\Xi}$, where $\bol{\Xi}$ is a vector field. Hence, $Z^{\mu} = \bol{Z}\cdot\p_{\bol{z}}\mu = 0$, and the kinetic equation in the coordinate system $\lr{\bol{x},u,\mu}$
can be written as an exact divergence in the coordinates $\lr{\bol{x},u}$, leading to conservation of integrals of the type
$\int f B_0\,d\bol{x}du$.

We now examine the following cases:
\begin{enumerate}
\item $\chi=0$. In this case the initial condition $f_0=f^{\rm MB}$ is the Maxwell-Boltzmann distribution, while $f_{\infty}=f^{\rm MB}e^g$. 
From eq. \eqref{Ggmu}, it follows that  
\eq{
e^g=1\iff g=0,
}
implying that the system remains a Maxwell-Boltzmann distribution.

\item $\chi=\chi\lr{\mu}$. In this case the initial condition $f_0$ is separable in $\mu$, leading to
\eq{
e^{\chi}=e^g\iff \chi=g.
}

\item $\chi=-\beta\gamma\mu u^2$. In this setting, larger values of $\mu$ are initially penalized at large parallel kinetic energies through the constant $\gamma \geq 0$. From eq. \eqref{Ggmu}, it follows that
\eq{
\int_{\mathbb{R}}\exp\lrc{-\beta\lr{\frac{1}{2}m+\gamma\mu}u^2}\,du
= e^g\int_{\mathbb{R}}\exp\lrc{-\beta\frac{1}{2}mu^2}\,du
\iff  
g=\frac{1}{2}\log\lr{\frac{m}{m+2\gamma\mu}}.
}
\end{enumerate}
Above we treated some special cases, but for a general initial condition $f_0$ we may write
\eq{
g=\log \lr{\frac{\int f_0B_0\,d\bol{x}du}{\int f^{\rm MB}B_0\,d\bol{x}du}}.
}

\section{Metriplectic structure of the two-species guiding center collision operator}
Both equations \eqref{ft} and \eqref{f1t2} are endowed with a metriplectic structure such that the evolution of an observable $F$ takes the form
\eq{
\frac{dF}{dt}=\lrc{F,\mf{H}}+\lrs{F,S},
}
where $\lrc{\cdot,\cdot}$ is a Poisson bracket, $\lrs{\cdot,\cdot}$ a dissipative bracket, $\mf{H}$ the Hamiltonian (energy) of the system, and $S$ its entropy. 
For the algebraic  definition of the metriplectic bracket, we refer the reader to  
\cite{Morrison86, PJM84}.

To simplify the algebra, we limit our discussion of the metriplectic structure to the ``Vlasov--Poisson'' setting, where the magnetic perturbation is absent, i.e., $A_{\parallel} = 0$. However, we note that the more general case with $A_{\parallel} \neq 0$ can also be treated (see, e.g., \cite{PJMVM}).

The metriplectic structure associated with equation \eqref{ft} 
is determined by the brackets,
\eq{
\lrc{F,G}=\int f\frac{\p}{\p \bol{z}}\lr{\frac{\delta F}{\delta f}}\cdot\mf{J}_{\rm GC}\cdot\frac{\p }{\p\bol{z}}\lr{\frac{\delta G}{\delta f}}\,d\bol{z}
+\int f'\frac{\p}{\p \bol{z}'}\lr{\frac{\delta F}{\delta f'}}\cdot\mf{J}'_{\rm GC}\cdot\frac{\p }{\p\bol{z}'}\lr{\frac{\delta G}{\delta f'}}\,d\bol{z}',}
and 
\eq{
\lrs{F,G}=&\frac{1}{2}\int ff'\mc{V}\lrs{\delta\bol{z}'\cdot\frac{\p}{\p\bol{z}'}\lr{\frac{\delta F}{\delta f'}}+
\delta\bol{z}\cdot\frac{\p}{\p\bol{z}}\lr{\frac{\delta F}{\delta f}}}\lrs{
\delta\bol{z}'\cdot\frac{\p}{\p\bol{z}'}\lr{\frac{\delta G}{\delta f'}}
+\delta\bol{z}\cdot\frac{\p}{\p\bol{z}}\lr{\frac{\delta G}{\delta f}}
}\,d\bol{z}d\bol{z}'d\bol{z}''d\bol{z}''',
}
where $F,G$ are arbitrary and $\delta\bol{z}$ and $\delta\bol{z}'$ are given, mutatis mutandis, by equation \eqref{dzdz}. 
The generating functions of the system are given by the Hamiltonian $\mf{H}$ and the entropy $S$ defined in equations \eqref{Hft} and \eqref{S}. In particular, we have
\eq{
\frac{\p f}{\p t}=\lrc{f,\mf{H}}+\lrs{f,S}.
}

Similarly, the metriplectic structure associated with equation \eqref{f1t2} is determined by the brackets,
\eq{
\lrc{F,G}=&
\int f_1\frac{\p}{\p \bol{\zeta}_1}\lr{\frac{\delta F}{\delta f_1}}\cdot\mc{J}_{\rm GC1}\cdot\frac{\p }{\p\bol{\zeta}_1}\lr{\frac{\delta G}{\delta f_1}}\,d\bol{\zeta}_1
+\int f_2\frac{\p}{\p \bol{\zeta}_2}\lr{\frac{\delta F}{\delta f_2}}\cdot\mc{J}_{\rm GC2}\cdot\frac{\p }{\p\bol{\zeta}_2}\lr{\frac{\delta G}{\delta f_2}}\,d\bol{\zeta}_2\\
&+\int f_1'\frac{\p}{\p \bol{\zeta}_1'}\lr{\frac{\delta F}{\delta f_1'}}\cdot\mc{J}_{\rm GC1}'\cdot\frac{\p }{\p\bol{\zeta}_1'}\lr{\frac{\delta G}{\delta f_1}'}\,d\bol{\zeta}_1'
+\int f_2'\frac{\p}{\p \bol{\zeta}_2'}\lr{\frac{\delta F}{\delta f_2}'}\cdot\mc{J}_{\rm GC2}'\cdot\frac{\p }{\p\bol{\zeta}_2'}\lr{\frac{\delta G}{\delta f_2}'}\,d\bol{\zeta}_2'
,
}
and
\eq{
&\lrs{F,G}=\\&-\int f_1f_1'\lrs{\mc{J}_{\rm GC1}'\cdot\frac{\p}{\p\bol{\zeta}_1'}\lr{\frac{\delta F}{\delta f_1'}}-\mc{J}_{\rm GC1}\cdot\frac{\p}{\p\bol{\zeta}_1}\lr{\frac{\delta F}{\delta f_1}}}\cdot{\Pi}_{11}\cdot\lrs{\mc{J}_{\rm GC1}'\cdot\frac{\p}{\p\bol{\zeta}_1'}\lr{\frac{\delta G}{\delta f_1'}}-\mc{J}_{\rm GC1}\cdot\frac{\p}{\p\bol{\zeta}_1}\lr{\frac{\delta G}{\delta f_1}}}\,d\bol{\zeta}_1d\bol{\zeta}_1'\\
&-\int f_1f_2'\lrs{
\mc{J}_{\rm GC2}'\cdot\frac{\p}{\p\bol{\zeta}_2'}\lr{\frac{\delta F}{\delta f_2'}}-\mc{J}_{\rm GC1}\cdot\frac{\p}{\p\bol{\zeta}_1}\lr{\frac{\delta F}{\delta f_1}}
}\cdot{\Pi}_{12}'\cdot\lrs{
\mc{J}_{\rm GC2}'\cdot\frac{\p}{\p\bol{\zeta}_2'}\lr{\frac{\delta G}{\delta f_2'}}-\mc{J}_{\rm GC1}\cdot\frac{\p}{\p\bol{\zeta}_1}\lr{\frac{\delta G}{\delta f_1}}
}\,d\bol{\zeta}_1d\bol{\zeta}_2'\\
&-\int f_2f_2'\lrs{\mc{J}_{\rm GC2}'\cdot\frac{\p}{\p\bol{\zeta}_2'}\lr{\frac{\delta F}{\delta f_2'}}-\mc{J}_{\rm GC2}\cdot\frac{\p}{\p\bol{\zeta}_2}\lr{\frac{\delta F}{\delta f_2}}}\cdot{\Pi}_{22}\cdot\lrs{\mc{J}_{\rm GC2}'\cdot\frac{\p}{\p\bol{\zeta}_2'}\lr{\frac{\delta G}{\delta f_2'}}-\mc{J}_{\rm GC2}\cdot\frac{\p}{\p\bol{\zeta}_2}\lr{\frac{\delta G}{\delta f_2}}}\,d\bol{\zeta}_2d\bol{\zeta}_2'\\
&-\int f_2f_1'\lrs{
\mc{J}_{\rm GC1}'\cdot\frac{\p}{\p\bol{\zeta}_1'}\lr{\frac{\delta F}{\delta f_1'}}-\mc{J}_{\rm GC2}\cdot\frac{\p}{\p\bol{\zeta}_2}\lr{\frac{\delta F}{\delta f_2}}
}\cdot{\Pi}_{21}'\cdot\lrs{
\mc{J}_{\rm GC1}'\cdot\frac{\p}{\p\bol{\zeta}_1'}\lr{\frac{\delta G}{\delta f_1'}}-\mc{J}_{\rm GC2}\cdot\frac{\p}{\p\bol{\zeta}_2}\lr{\frac{\delta G}{\delta f_2}}
}\,d\bol{\zeta}_2d\bol{\zeta}_1'
.
}
The generating functions of the system are given by the Hamiltonian $\mf{H}_{12}$ and the entropy $S_{12}$ defined in equations \eqref{H12} and \eqref{S12}. It can be shown that
\eq{
\frac{\p f_1}{\p t}=\lrc{f_1,\mf{H}_{12}}+\lrs{f_1,S_{12}}.
}
Additional details on the verification of the Poisson bracket and dissipative bracket axioms, including boundary conditions,  can be found in \cite{SatoMorrison24}. 

%\section{Concluding remarks}

{\section{Grazing Coulomb collisions in guiding center phase space and linear theory}}
In this section, we aim to draw comparisons between the Landau collision operator \cite{Landau1936,Kampen,Lenard}, which describes grazing Coulomb collisions in canonical phase space, its gyrokinetic formulation \cite{Hir2,Hirv,Burby,Brizard04}, and the collision operator derived in our study (equation \eqref{f1t2}) for Coulomb collisions in guiding center phase space. Additionally, we will derive the linearized version of equation \eqref{f1t2}, highlighting its connection to linearized model collision operators commonly used in gyrokinetic theory \cite{Sugama09,Sugama19}. In its simplest form (see eq.~\eqref{dfdtlX} below), the linearized operator encodes the noncanonical Hamiltonian structure of guiding center phase space, while 
exhibiting a mathematical structure analogous to that of the linearized Landau operator and the gyrokinetic linearized Landau operator \cite{Pan20,Pan19}.

First, we note that the collision operator in equation \eqref{f1t2} reduces to the Landau collision operator when the 5-dimensional guiding center Poisson tensor \( \mathcal{J}_{{\rm GC}} \) is replaced by the canonical Poisson tensor (symplectic matrix) \( \mathcal{J}_c \) in the 6-dimensional canonical phase space \( \boldsymbol{\zeta} = (\boldsymbol{p}, \boldsymbol{q}) \), where \( \boldsymbol{p} \) and \( \boldsymbol{q} \) denote the momentum and position vectors, respectively. On this point, recall the discussion of Sec. 4. 
However, as demonstrated in Sec. 11 for the guiding center Poisson tensor, and in Sec. 3.5 for general Poisson tensors, the collision operator in a noncanonical phase space deviates significantly from the Landau operator. It exhibits a spatially nonlocal nature, accounts for generalized particle interactions, and leads to deviations from Maxwell-Boltzmann statistics due to a modified invariant phase space measure and conservation laws associated with interior Casimir invariants.

Secondly, while gyrokinetic collision operators represent the Landau collision operator expressed in gyrocenter variables under the gyrokinetic ordering—thus enabling the modeling of microturbulence at the gyroradius scale—the collision operator in equation \eqref{f1t2} differs fundamentally from the Landau operator expressed in guiding center variables. This difference arises from the intrinsic disparity between the guiding center Poisson tensor \( \mathcal{J}_{{\rm GC}} \) and the canonical Poisson tensor \( \mathcal{J}_c \). As a result, equation \eqref{f1t2} is not suitable for describing microturbulent regimes dominated by Coulomb collisions. Instead, it captures a different turbulent regime, characterized by the approximate conservation of the particle magnetic moment (see equation \eqref{cum}), which leads to the type of self-organized equilibrium states discussed in Section 11. These steady states are relevant to both laboratory and astrophysical plasmas, where Coulomb collisions are not dominant, and the characteristic spacetime scales are consistent with the approximate conservation of the first adiabatic invariant. Examples include non-neutral and pair plasmas confined by dipole magnetic fields \cite{Yoshida10,SatoPair}.

\vspace{0.3cm}

\subsection{Properties of the collision operator in a constant magnetic field}

In order to visualize the difference with the Landau operator, it is useful to
reduce the kinetic equation \eqref{f1t2} to the limit in which the magnetic field $B_0$ is a spatial constant and the potential energies $V_{11}$ and $V_{12}$ are approximately constant along the field lines, i.e., $\bol{B}^{\ast}_1\cdot\p_{\bol{x}_1}V_{11}\approx \bol{B}^{\ast}_1\cdot\p_{\bol{x}_1}V_{12}\approx 0$. Furthermore, we can simplify the collision operator by invoking the expansion  of the interaction tensor given in equation \eqref{IT2}, and by discarding the projectors (note, however, that with this second approximation energy conservation is no longer exact, but holds only up to second order in the collision time \( \tau_c \)). 
Substituting the explicit expression for the guiding center Poisson tensor \( \mathcal{J}_{{\rm GC}} \) (see equation \eqref{Jgc}) in this regime, we obtain:
\eq{
\frac{\p f_1}{\p t}=&
-\frac{\p}{\p\bol{\zeta}_1}\cdot\lrs{f_1\mc{J}_{{\rm GC1}}\cdot\frac{\p\lr{{h}_{1}+q_1\Phi_1+{v}_1}}
{\p\bol{\zeta}_1}}\\
&-\nabla_1\cdot\lrs{f_1\int\frac{1}{2}\tau_c^2\bar{\Gamma}_{11}f_1'\frac{\bol{b}_1\cp\nabla_{1} V_{11}}{q_1B_{01}}\lr{\frac{\bol{b}_1'\cp\nabla_1 V_{11}}{q_1B_{01}'}\cdot\nabla_1'\log f_1'-\frac{\bol{b}_1\cp\nabla_1 V_{11}}{q_1B_{01}}\cdot\nabla_1\log  f_1}\,d\bol{\zeta}_1'}\\
&-\nabla_1\cdot\lrs{f_1\int\frac{1}{2}\tau_c^2\bar{\Gamma}_{12}'f_2'\frac{\bol{b}_1\cp \nabla_1 V_{12}}{q_1B_{01}}\lr{
\frac{\bol{b}_2'\cp\nabla_1 V_{12}}{q_2B_{02}'}\cdot\nabla_2'\log f_2'-\frac{\bol{b}_1\cp\nabla_1 V_{12}}{q_1B_{01}}\cdot\nabla_1\log f_1
}\,d\bol{\zeta}_2'},\label{f1tgc}
}
where we introduced the gradient operators $\nabla_1=\p/\p\bol{x}_1$, $\nabla_1'=\p/\p\bol{x}_1'$, and $\nabla_2'=\p/\p\bol{x}_2'$,  
and the averaged scattering frequencies 
\eq{
&\bar{\Gamma}_{11}=\frac{1}{2}\int f_2f_2'\mc{V}\,d\bol{z}''d\bol{z}'''d\bol{\zeta}_2d\bol{\zeta}_2',\\
&\bar{\Gamma}_{12}'=\frac{1}{2}\int f_2f_1'\mc{V}\,d\bol{z}''d\bol{z}'''d\bol{\zeta}_2d\bol{\zeta}_1'.
}
Note that due to the localized nature of Coulomb collisions, we have $\bol{b}_1/B_{01}\approx\bol{b}_1'/B_{01}'$ in the first collision integral, as well as $\bol{b}_1/B_{01}\approx\bol{b}_2'/B_{02}'$ in the second one. Hence, introducing the notation
\eq{
\kappa_{11}=\frac{1}{2}\tau_c^2\bar{\Gamma}_{11},~~~~\kappa_{12}'=\frac{1}{2}\tau_c^2\bar{\Gamma}_{12}',
} 
and defining the $\bol{E}\cp\bol{B}$ drift velocities 
\eq{
\bol{v}_{E_{11}}=\frac{\bol{b}_1\cp\nabla_1V_{11}}{q_1B_{01}},~~~~\bol{v}_{E_{12}}=\frac{\bol{b}_1\cp\nabla_1V_{12}}{q_1B_{01}},\label{vE}
}
the collision operator 
$\mc{C}_1\lr{f_1,f_2}$
on the right-hand side of equation \eqref{f1tgc} can be approximated as
\eq{
\mc{C}_1\lr{f_1,f_2}=&-\nabla_1\cdot\lrs{ f_1\left\langle\kappa_{11}\bol{v}_{E_{11}}\bol{v}_{E_{11}}\cdot\lr{\nabla_1'\log f_1'-\nabla_1\log f_1}\right\rangle_1'}\\
&-\nabla_1\cdot\lrs{ f_1\left\langle\kappa_{12}'\bol{v}_{E_{12}}\bol{v}_{E_{12}}\cdot\lr{\frac{q_1}{q_2}\nabla_2'\log f_2'-\nabla_1\log f_1}\right\rangle_2'},
}
where $\langle\cdot\rangle_1'$ and $\langle\cdot\rangle_2'$  denote the ensemble averages with respect to the distribution functions $f_1'$ and $f_2'$. 
This expression demonstrates that the collisional particle flux is driven by gradients in the particle distributions along the \(\boldsymbol{E} \times \boldsymbol{B}\) velocities \eqref{vE}. This behavior aligns with the fact that, at equilibrium, the gradient of \(\log f\) tends to be parallel to the gradient of the electrostatic potential, and therefore perpendicular to the \(\boldsymbol{E} \times \boldsymbol{B}\) flow (see Section 11, equation \eqref{f1f2eq}).

%Let us now make the further assumption that $\kappa_{11}=\tau_c^2\bar{\Gamma}_{11}/2$ and $\kappa_{12}'=\tau_c^2\bar{\Gamma}_{12}'/2$ do not depend on the momentum variables $v_{\parallel s},\eta_s$, $s=1,2$. 

\subsection{The linearized collision operator}

We conclude this section by obtaining the linearized form of the 
kinetic equation \eqref{f1t2}. 
{To simplify the analysis let us consider the case with no magnetic perturbations, $A_{\parallel}=0$.} 
We divide the distribution functions   into a steady leading order component $f_{s0}\neq 0$ and a perturbative term $\delta f_s$ as
\eq{
f_s=f_{s0}+\delta f_s,~~~~s=1,2.
}
The steady components $f_{s0}$ are given by \eqref{f1f2eq}. 

At first-order, the  evolution equation for the perturbation $\delta f_1$ can be written as 
\eq{
\frac{\p\delta f_1}{\p t}=&-\frac{\p}{\p\bol{\zeta}_1}\cdot\lrs{\delta f_1\mc{J}_{{\rm GC}1}\cdot\frac{\p\lr{h_{10}+q_1\Phi_{10}+v_{10}}}{\p\bol{\zeta}_1}+f_{10}\mc{J}_{{\rm GC}1}\cdot\frac{\p\lr{\delta h_1+q_1\delta\Phi_1+\delta v_1}}{\p\bol{\zeta}_1}}\\
&+\frac{\p}{\p\bol{\zeta}_1}\cdot\lrs{\mc{J}_{{\rm GC}1}\cdot\int\lr{f_{10}\delta f_1'+f_{10}'\delta f_1}{
\Pi}_{11}\cdot\lr{\mc{J}_{{\rm GC}1}'\cdot\frac{\p\log f_{10}'}{\p\bol{\zeta}_1'}-\mc{J}_{{\rm GC}1}\cdot\frac{\p\log f_{10}}{\p\bol{\zeta}_1}}\,d\bol{\zeta}_1'}\\
&+\frac{\p}{\p\bol{\zeta}_1}\cdot\lrc{f_{10}\mc{J}_{{\rm GC}1}\cdot\int f_{10}'{\Pi}_{11}\cdot\lrs{\mc{J}'_{{\rm GC}1}\cdot\frac{\p}{\p\bol{\zeta}_1'}\lr{\frac{\delta f_1'}{f_{10}'}}
-\mc{J}_{{\rm GC}1}\cdot\frac{\p}{\p\bol{\zeta}_1}\lr{\frac{\delta f_{1}}{f_{10}}
}}\,d\bol{\zeta}_1'}\\
&+\frac{\p}{\p\bol{\zeta}_1}\cdot\lrs{\mc{J}_{{\rm GC}1}\cdot\int\lr{f_{10}\delta f_2'+f_{20}'\delta f_1}{
\Pi}_{12}'\cdot\lr{\mc{J}_{{\rm GC}2}'\cdot\frac{\p\log f_{20}'}{\p\bol{\zeta}_2'}-\mc{J}_{{\rm GC}1}\cdot\frac{\p\log f_{10}}{\p\bol{\zeta}_1}}\,d\bol{\zeta}_2'}\\
&+\frac{\p}{\p\bol{\zeta}_1}\cdot\lrc{f_{10}\mc{J}_{{\rm GC}1}\cdot\int f_{20}'{\Pi}_{12}'\cdot\lrs{\mc{J}'_{{\rm GC}2}\cdot\frac{\p}{\p\bol{\zeta}_2'}\lr{\frac{\delta f_2'}{f_{20}'}}
-\mc{J}_{{\rm GC}1}\cdot\frac{\p}{\p\bol{\zeta}_1}\lr{\frac{\delta f_{1}}{f_{10}}
}}\,d\bol{\zeta}_2'},
}
where $h_{10}+\delta h_1$,  $\Phi_{10}+\delta\Phi_1$,  and $v_{10}+\delta v_1$ represent the decompositions of $h_1$, $\Phi_1$, and $v_{1}$ into steady and perturbative parts. 
For ease of notation, we have also retained the same symbols for the interaction tensors, though they should be understood as representing their leading-order expansions: just replace the vector fields $\bol{\xi}_{ij}$ defining the projectors $\mathbb{P}_{ij}^{\perp}$ with their leading order expressions, e.g., replace $\bol{\xi}_{11}$ with \eq{\bol{\xi}_{11}^0=\mc{J}_{{\rm GC}}\cdot\p_{\bol{\zeta}}{\lr{h_0+q\Phi_0+v_0}}-\mc{J}'_{{\rm GC}}\cdot\p_{\bol{\zeta}'}{\lr{h_0'+q\Phi_0'+v_0'}}.}  
We have also used the fact that, due to the grazing scattering condition, perturbations $\delta\Pi$ in the interaction tensors constitute negligible higher-order corrections.  
Now recall that $h_s+v_s=mv_{\parallel s}^2/2+\eta_s+\lambda_s$ so that
$\delta \lr{h_s+q_s\Phi_s+v_s}=q_s\delta\Phi_s$. Furthermore, the steady states satisfy
\eq{
&\mc{J}_{{\rm GC}1}'\cdot\frac{\p\log f'_{10}}{\p\bol{\zeta}_1'}-\mc{J}_{{\rm GC}1}\cdot\frac{\p\log f_{10}}{\p\bol{\zeta}_1}\in{\rm ker}\lr{\Pi_{11}},\\
&\mc{J}_{{\rm GC}2}'\cdot\frac{\p\log f'_{20}}{\p\bol{\zeta}_2'}-\mc{J}_{{\rm GC}1}\cdot\frac{\p\log f_{10}}{\p\bol{\zeta}_1}\in{\rm ker}\lr{\Pi_{12}'}.
}
It follows that
\eq{
\frac{\p\delta f_1}{\p t}=&
-\frac{\p}{\p\bol{\zeta}_1}\cdot
\lrs{\delta f_1\mc{J}_{{\rm GC}1}\cdot\frac{\p\lr{h_{10}+q_1\Phi_{10}+v_{10}}}{\p\bol{\zeta}_1}+q_1f_{10}\mc{J}_{{\rm GC}1}\cdot\frac{\p \delta\Phi_1}{\p\bol{\zeta}_1}
}\\
&+\frac{\p}{\p\bol{\zeta}_1}\cdot \lrc{f_{10}\mc{J}_{{\rm GC}1}\cdot\int f_{10}'\Pi_{11}\cdot
\lrs{
\mc{J}'_{{\rm GC}1}\cdot \frac{\p}{\p\bol{\zeta}_1'}\lr{\frac{\delta f_1'}{f_{10}'}}-\mc{J}_{{\rm GC}1}\cdot\frac{\p}{\p\bol{\zeta}_1}\lr{\frac{\delta f_1}{f_{10}}}
}\,d\bol{\zeta}_1'}\\
&+
\frac{\p}{\p\bol{\zeta}_1}\cdot\lrc{f_{10}\mc{J}_{{\rm GC}1}\cdot\int f_{20}'\Pi_{12}'\cdot
\lrs{
\mc{J}'_{{\rm GC}2}\cdot \frac{\p}{\p\bol{\zeta}_2'}\lr{\frac{\delta f_2'}{f_{20}'}}-\mc{J}_{{\rm GC}1}\cdot\frac{\p}{\p\bol{\zeta}_1}
\lr{\frac{\delta f_1}{f_{10}}}
}\,d\bol{\zeta}_2'}.
}
In the case of collisions between particles of the same species, we have 
\eq{
\frac{\p\delta f}{\p t}=&-\frac{\p}{\p\bol{\zeta}}\cdot
\lrs{\delta f\mc{J}_{{\rm GC}}
\cdot\frac{\p\lr{h_{0}+q\Phi_{0}+v_{0}}}{\p\bol{\zeta}}+qf_{0}\mc{J}_{{\rm GC}}\cdot\frac{\p\delta\Phi}{\p\bol{\zeta}}
}\\
&+\frac{\p}{\p\bol{\zeta}}\cdot\lrc{
f_{0}\mc{J}_{{\rm GC}}\cdot\int f_{0}'\Pi_{11}\cdot\lrs{
\mc{J}_{{\rm GC}}'\cdot\frac{\p}{\p\bol{\zeta}'}\lr{\frac{\delta f'}{f_0'}}-\mc{J}_{{\rm GC}}\cdot\frac{\p}{\p\bol{\zeta}}\lr{\frac{\delta f}{f_0}}
}\,d\bol{\zeta}'
}.\label{df110}
%\frac{\p\delta f_1}{\p t}=&-\frac{\p}{\p\bol{\zeta}_1}\cdot\lrs{\delta f_1\mc{J}_{{\rm GC}1}\cdot\frac{\p\lr{h_{10}+q_1\Phi_{10}+v_{10}}}{\p\bol{\zeta}_1}+f_{10}\mc{J}_{{\rm GC}1}\cdot\frac{\p\lr{\delta h_1+q_1\delta\Phi_1+\delta v_1}}{\p\bol{\zeta}_1}}\\
%&+\frac{\p}{\p\bol{\zeta}_1}\cdot\lrs{\mc{J}_{{\rm GC}1}\cdot\int\lr{f_{10}\delta f_1'+f_{10}'\delta f_1}{
%\Pi}_{11}\cdot\lr{\mc{J}_{{\rm GC}1}'\cdot\frac{\p\log f_{10}'}{\p\bol{\zeta}_1'}-\mc{J}_{{\rm GC}1}\cdot\frac{\p\log f_{10}}{\p\bol{\zeta}_1}}\,d\bol{\zeta}_1'}\\
%&+\frac{\p}{\p\bol{\zeta}_1}\cdot\lrc{f_{10}\mc{J}_{{\rm GC}1}\cdot\int f_{10}'{\Pi}_{11}\cdot\lrs{\mc{J}'_{{\rm GC}1}\cdot\frac{\p}{\p\bol{\zeta}_1'}\lr{\frac{\delta f_1'}{f_{10}'}}
%-\mc{J}_{{\rm GC}1}\cdot\frac{\p}{\p\bol{\zeta}_1}\lr{\frac{\delta f_{1}}{f_{10}}
%}}\,d\bol{\zeta}_1'}.\label{df11}
}
However, the term involving $\delta\Phi$ is, on average, small. Indeed,
\eq{
\int \frac{\delta f}{f_0}\frac{\p}{\p\bol{\zeta}}\cdot\lr{qf_0\mc{J}_{{\rm GC}}\cdot\frac{\p\delta\Phi}{\p\bol{\zeta}}}\,d\bol{\zeta}=
\int \delta f\delta f'\frac{\p\log f_0}{\p\bol{\zeta}}\cdot\mc{J}_{{\rm GC}}\cdot\frac{\p V_{11}}{\p\bol{\zeta}}\,d\bol{\zeta}=\frac{\beta}{2}\int \delta f\delta f'\bol{\xi}_{11}^0\cdot\frac{\p V_{11}}{\p\bol{\zeta}}=O\lr{\delta f^2\epsilon}. 
}
%where $\bol{\xi}_{11}^0=\mc{J}_{\rm GC}\cdot\p_{\bol{\zeta}}\lr{h_0+q\Phi_0+v_0}-\mc{J}_{\rm GC}'\cdot\p_{\bol{\zeta}'}\lr{h_0'+q\Phi_0'+v_0'}$. 
At leading order, the governing equation can thus be reduced to
\eq{
\frac{\p\delta f}{\p t}=\frac{\p}{\p\bol{\zeta}}\cdot\lrc{\mc{J}_{{\rm GC}}
\cdot
\lrs{-\delta f\frac{\p\lr{h_{0}+q\Phi_{0}+v_{0}}}{\p\bol{\zeta}}
+
f_0\int f_0'\Pi_{11}\cdot\lr{
\mc{J}_{{\rm GC}}'\cdot\frac{\p}{\p\bol{\zeta}'}\lr{\frac{\delta f'}{f_0'}}-\mc{J}_{{\rm GC}}\cdot\frac{\p}{\p\bol{\zeta}}\lr{\frac{\delta f}{f_0}}
}\,d\bol{\zeta}'
}}.\label{df11}
}
Let us briefly discuss the properties of equation \eqref{df11}. 
First, we observe that for practical purposes (e.g., numerical simulation), 
it could be convenient to express the interaction tensor $\Pi_{11}$ as follows
\eq{
\Pi_{11}=\mathbb{P}^{\perp}_{11}\cdot\ms{I}\cdot\mathbb{P}^{\perp}_{11},~~~~\mathbb{P}_{11}^{\perp}=I-\frac{\bol{\xi}_{11}^0\bol{\xi}_{11}^0}{{\bol{\xi}_{11}^{02}}},~~~~\ms{I}=\frac{1}{2}\int\mc{V}_{11}\lr{\int_{\tau_c}\p_{\bol{\zeta}}V_{11}\,dt}\lr{\int_{\tau_c}\p_{\bol{\zeta}}V_{11}\,dt}d\bol{\zeta}''d\bol{\zeta}''', 
}
where the tensor $\ms{I}$ is a given tensor expressing the characteristic phase space displacement caused by a collision. Furthermore, noting that $\p_{\bol{\zeta}}V_{11}=\lr{\p_{\bol{x}}V_{11},0,0}$, 
in many cases one would expect the spatial diagonal terms to be dominant in $\ms{I}$, 
leading to a further simplification, analogous to that encountered with the Landau operator,  
$\Pi_{11} = D\, \mathbb{P}_{11}^{\perp}\cdot I_x\cdot\mathbb{P}^{\perp}_{11}$,   
with $D\lr{\bol{\zeta},\bol{\zeta}'}$ a given function,  
and $I_x$ the $5\times 5$ matrix whose spatial block is the $3\times 3$ identity matrix, 
and all other entries are zero.

If, as in the Landau operator, we assume that collisions are spatially localized, 
the function $D$ will include a delta function, 
$D= \mc{D}\lr{u,\eta,u',\eta'}\delta\lr{\bol{x}-\bol{x}'}$, 
and the simplified expression for equation \eqref{df11} becomes
\eq{
\frac{\p\delta f}{\p t}=&-\frac{\p}{\p\bol{\zeta}}\cdot\lrs{\delta f\mc{J}_{{\rm GC}}
\cdot\frac{\p\lr{h_0+q\Phi_0+v_0}}{\p\bol{\zeta}}}\\
&+\frac{\p}{\p\bol{\zeta}}\cdot\lrc{f_0\mc{J}_{{\rm GC}}\cdot\int f_0'\mc{D}
\mathbb{P}^{\perp}_{11}\cdot I_x\cdot\mathbb{P}^{\perp}_{11}
\cdot\lrs{\mc{J}_{{\rm GC}}'\cdot\frac{\p}{\p\bol{\zeta}'}\lr{\frac{\delta f'}{f_0'}}-
\mc{J}_{{\rm GC}}\cdot\frac{\p}{\p\bol{\zeta}}\lr{\frac{\delta f}{f_0}}}\,du'\,d\eta'},\label{df11simple}
}
where 
all quantities are evaluated at the same $\bol{x}$, and 
now the prime symbol $'$ indicates evaluation at the same spatial position, 
but at different $u$ and $\eta$, e.g., $\delta f'=\delta f\lr{\bol{x},u',\eta'}$.

Next, let us examine conservation laws and entropy growth for equation \eqref{df11} (these properties apply to \eqref{df11simple} as well). 
The perturbation of the particle number, 
\eq{
\delta N = \int \delta f \, d\boldsymbol{\zeta} 
%+ \int \delta f_1' \, d\boldsymbol{\zeta}_1',
}
is conserved under suitable boundary conditions because equation \eqref{df11} is in divergence form.  
Next, the perturbed interior Casimir invariant
\eq{
\delta\mathscr{M}=\int \delta f\, g\lr{\frac{\eta}{B_0}}\,d\bol{\zeta},
}
where $g$ is an arbitrary function of the magnetic moment $\mu=\eta/B_0$, 
is a constant of motion. Indeed, the rate of change can be written as 
\eq{
\frac{d\delta\mathscr{M}}{dt}=
\int \lrs{...}\cdot\mc{J}_{{\rm GC}}\cdot\frac{\p g}{\p\bol{\zeta}}\,d\bol{\zeta}=0.
} 
The perturbed energy is given by
\eq{
\delta\mf{H}=\int \lrs{\lr{h_0+\frac{1}{2}q\Phi_0+v_0}\delta f+\frac{1}{2}qf_0\delta\Phi}\,d\bol{\zeta}.
}
As usual, performing integration by parts and eliminating boundary terms, we have
\eq{
\frac{d\delta\mf{H}}{dt}=&
\int\frac{\p\delta f}{\p t}\lr{h_0+q\Phi_0+v_0}
\,d\bol{\zeta}\\
=&\int \lrc{
%qf_0\mc{J}_{{\rm GC}}\cdot\frac{\p\delta\Phi}{\p\bol{\zeta}}
-f_0\mc{J}_{{\rm GC}}\cdot\int f_0' \Pi_{11}\cdot\lrs{\mc{J}_{{\rm GC}}'\cdot\frac{\p}{\p\bol{\zeta}'}\lr{\frac{\delta f'}{f_0'}}
-\mc{J}_{{\rm GC}}\cdot\frac{\p}{\p\bol{\zeta}}\lr{\frac{\delta f}{f_0}}
}\,d\bol{\zeta}'
}\cdot\frac{\p\lr{h_0+q\Phi_0+v_0}}{\p\bol{\zeta}}\,d\bol{\zeta}.
}
%Now not that, since $f_0=f_0\lr{h_0+q\Phi_0+v_0,\eta/B_0}$ and the phase space flow is measure preserving, the first term in the integrand can be written in divergence form:
%\eq{
%f_0\lrs{\mc{J}_{{\rm GC}}\cdot\frac{\p\delta\Phi}{\p\bol{\zeta}}}\cdot\frac{\p\lr{h_0+q\Phi_0+v_0}}{\p\bol{\zeta}}=
%-\frac{\p}{\p\bol{\zeta}}\cdot\lr{f_0\delta\Phi\mc{J}_{{\rm GC}}\cdot\frac{\p\lr{h_0+q\Phi_0+v_0}}{\p\bol{\zeta}}}.
%}
We thus find
\eq{
\frac{d\delta\mf{H}}{dt}=\frac{1}{2}\int f_0f_0'\bol{\xi}_{11}^0\cdot\Pi_{11}\cdot\lrs{\mc{J}_{{\rm GC}}'\cdot\frac{\p}{\p\bol{\zeta}'}\lr{\frac{\delta f'}{f_0'}}-\mc{J}_{{\rm GC}}\cdot\frac{\p}{\p\bol{\zeta}}\lr{\frac{\delta f}{f_0}}}\,d\bol{\zeta}d\bol{\zeta}'=0,
}
where we used the fact that  $\bol{\xi}_{11}^0\cdot\Pi_{11}=0$. Hence, the perturbed energy $\delta\mf{H}$ is a constant of motion. 

Let us now derive the H-theorem for the linearized equation \eqref{df11}. 
%the conservation laws and the \( H \)-theorem satisfied by the parent equation \eqref{f1t2} are inherited by equation \eqref{df11} to the order of the expansion. To illustrate this, let us consider entropy growth and expand the system's entropy \( S_1 \) as follows:
%\eq{
%S_1 = S_{10} + \delta S_1 + O\big(\delta f_1^2\big),
%}
%where
%\eq{
%\begin{aligned}
%S_1 &= -\int f_1 \log f_1 \, d\boldsymbol{\zeta}_1 - \int f_1' \log f_1' \, d\boldsymbol{\zeta}_1', \\
%S_{10} &= -\int f_{10} \log f_{10} \, d\boldsymbol{\zeta}_1 - \int f_{10}' \log f_{10}' \, d\boldsymbol{\zeta}_1', \\
The perturbed entropy is a second order functional of  $\delta f$, arising from the expansion of $S$, and it can be 
conveniently expressed as 
\eq{
\delta S = -\frac{1}{2}\int \frac{\delta f^2}{f_0}\,d\bol{\zeta}-\frac{1}{2}\int\frac{\delta f'{}^2}{f_0'}\,d\bol{\zeta}'.
%-\int \delta f \lr{ 1 + \log f_{0} } \, d\boldsymbol{\zeta} - \int \delta f' \lr{ 1 + \log f_{0}' } \, d\boldsymbol{\zeta}'.
%\end{aligned}
}
Using again the functional form  of $f_0$ and eliminating boundary integrals, it can be shown that 
\eq{
\frac{d\delta S}{dt}=&
-\int\frac{\delta f}{f_0}\frac{\p\delta f}{\p t}\,d\bol{\zeta}-\int\frac{\delta f'}{f_0'}\frac{\p\delta f'}{\p t}\,d\bol{\zeta}'\\
=&\int f_0f_0'\lrs{
\mc{J}_{\rm GC}'\cdot\frac{\p}{\p\bol{\zeta}'}\lr{\frac{\delta f'}{f_{0}'}}-\mc{J}_{\rm GC}\cdot\frac{\p}{\p\bol{\zeta}}\lr{\frac{\delta f}{f_0}}}\cdot\Pi_{11}\cdot\lrs{
\mc{J}_{\rm GC}'\cdot\frac{\p}{\p\bol{\zeta}'}\lr{\frac{\delta f'}{f_{0}'}}-\mc{J}_{\rm GC}\cdot\frac{\p}{\p\bol{\zeta}}\lr{\frac{\delta f}{f_0}}}\,d\bol{\zeta}d\bol{\zeta}'\geq 0.
%\\
%&-q\int \delta f\frac{\p\delta\Phi}{\p\bol{\zeta}}\cdot\mc{J}_{\rm GC}\cdot\frac{\p\log f_0}{\p\bol{\zeta}}\,d\bol{\zeta}
%-q\int \delta f'\frac{\p\delta\Phi'}{\p\bol{\zeta}'}\cdot\mc{J}_{\rm GC}'\cdot\frac{\p\log f_0'}{\p\bol{\zeta}'}\,d\bol{\zeta}'
%\\
%=&\int f_0f_0'\lr{\frac{\mc{J}_{\rm GC}'}{f_0'}\cdot\frac{\p\delta f'}{\p\bol{\zeta}'}-\frac{\mc{J}_{\rm GC}}{f_0}\cdot\frac{\p\delta f}{\p\bol{\zeta}}}\cdot\Pi_{11}\cdot\lr{\frac{\mc{J}_{\rm GC}'}{f_0'}\cdot\frac{\p\delta f'}{\p\bol{\zeta}'}-\frac{\mc{J}_{\rm GC}}{f_0}\cdot\frac{\p\delta f}{\p\bol{\zeta}}}\,d\bol{\zeta}d\bol{\zeta}'\\
%&\beta\int\delta f\delta f'\frac{\p V_{11}}{\p\bol{\zeta}}\cdot\bol{\xi}_{11}\,d\bol{\zeta}d\bol{\zeta}'.
}

%Since \( \frac{dS_1}{dt} \geq 0 \) and \( \frac{dS_{10}}{dt} = 0 \), it follows that:
%\eq{
%\frac{d\delta S_1}{dt} + O\big(\delta f_1^2\big) \geq 0,
%}
%where the time derivative is evaluated with respect to the full dynamics (equation \eqref{f1t2}). However, substituting \eqref{f1t2} with the first-order linear equation \eqref{df11} introduces only second-order corrections, leading to:
%\eq{
%\left( \frac{d\delta S_1}{dt} \right)_{{\rm L}} + O\big(\delta f_1^2\big) \geq 0,
%}
%where the subscript \({\rm L}\) indicates that the evolution of \( \delta S_1 \) is now evaluated with respect to the linearized system \eqref{df11}.

%Equation \eqref{df11} has the familiar structure
%\eq{
%\frac{\p\delta f_1}{\p t}=\mc{C}_1\lrs{\delta f_1,f_{10}}+\mc{C}_1\lrs{}
%}

\subsection{Summary of governing equations for the linearized guiding center collision operator}
We conclude this paper with a short self-contained summary of the simplest form for the linearized equations governing Coulomb collisions in guiding center phase space. This summary should be useful when implementing these equations numerically. 

We consider the simplest setting where magnetic perturbations are absent, $A_{\parallel}=0$. 
The phase space coordinates are 
$\bol{\zeta}=\lr{x,y,z,u,\eta}$, where $\bol{x}=\lr{x,y,z}$ denotes the spatial position of the guiding center, $u$ the velocity along the magnetic field, and $\eta=\mu B_0$ the energy of cyclotron motion, with magnetic moment $\mu$ and background vacuum magnetic field $\bol{B}_0$ of strength $B_0$. 

Denoting the distribution function of the system as $f\lr{\bol{\zeta},t}=f_0\lr{\bol{\zeta}}+\delta f\lr{\bol{\zeta},t}$, where $f_0$ is the equilibrium part and $\delta f$ is the perturbation, the governing equation for the fluctuation $\delta f$ in its simplest form is given by
\eq{
\frac{\p\delta f}{\p t}=&-\frac{\p}{\p\bol{\zeta}}\cdot\lrs{\delta f\mc{J}_{{\rm GC}}
\cdot\frac{\p\lr{h_0+q\Phi_0}}{\p\bol{\zeta}}}\\
&+\mc{D}\frac{\p}{\p\bol{\zeta}}\cdot\lrc{f_0\mc{J}_{{\rm GC}}\cdot\int f_0'
\mathbb{P}^{\perp}_{11}\cdot I_x\cdot\mathbb{P}^{\perp}_{11}
\cdot\lrs{\mc{J}_{{\rm GC}}'\cdot\frac{\p}{\p\bol{\zeta}'}\lr{\frac{\delta f'}{f_0'}}-
\mc{J}_{{\rm GC}}\cdot\frac{\p}{\p\bol{\zeta}}\lr{\frac{\delta f}{f_0}}}\,du'\,d\eta'},
\\
\mathbb{P}^{\perp}_{11}=&I-\frac{\bol{\xi}_{11}^0\bol{\xi}_{11}^0}{\bol{\xi}_{11}^{02}},\\
\bol{\xi}_{11}^0=&\mc{J}_{{\rm GC}}\cdot\p_{\bol{\zeta}}\lr{h_0+q\Phi_0}-\mc{J}_{{\rm GC}}'\cdot\p_{\bol{\zeta}'}\lr{h_0'+q\Phi_0'},\\
f_0=&\frac{1}{Z}\exp\lrc{-\beta\lr{h_0+q\Phi_0}+g\lr{\frac{\eta}{B_{0}}}},\label{dfdtlX}
}
where $h_0+q\Phi_0=mu^2/2+\mu B_0+q\Phi_0$ is the guiding center energy,
$q\Phi_0=\int f_0'V_{11}\,d\bol{x}'$ is the equilibrium electrostatic potential energy,
$V_{11}=q^2/4\pi\epsilon_0\abs{\bol{x}-\bol{x}'}$ is the binary interaction potential energy, 
$m$ and $q$ are the particle mass and charge, respectively, 
$\mc{D}\approx \lr{q^2/4\pi\epsilon_0\ell^2}^2\ell^3\tau_c$ is a physical constant representing the strength of collisions bearing dimensions of ${\rm N^2 \cdot s\cdot m^3 }$, with $\tau_c$ the collision time and $\ell$ the characteristic spatial scale of the interaction, 
$I$ is the $5\times 5$ identity matrix, $I_x$ is the $5\times 5$ matrix whose spatial block is the $3\times 3$ identity matrix and all other entries are zero, 
$Z$ is a normalization constant, $\beta$ is the inverse temperature, 
and $g$ is any function of $\mu=\eta/B_0$. 
The prime symbol $'$ indicates evaluation at $\bol{\zeta}'=\lr{\bol{x},u',\eta'}$, e.g., $\mc{J}_{{\rm GC}}'=\mc{J}_{{\rm GC}}\lr{\bol{\zeta}'}$. 
The guiding center Poisson tensor $\mc{J}_{{\rm GC}}$ is given by
\begin{equation}
\begin{split}
\mc{J}_{\rm GC}=\begin{bmatrix}
0&-\frac{b_{z}}{qB_{0}}&\frac{b_{y}}{qB_{0}}&\frac{B_{x}^{*}}{mB_{0}}&\frac{\eta}{qB_0^2}\bol{b}\cp\nabla B_0\cdot\nabla x\\
\frac{b_{z}}{qB_{0}}&0&-\frac{b_{x}}{qB_{0}}&\frac{B_{y}^{\ast}}{mB_{0}}&\frac{\eta}{qB_0^2}\bol{b}\cp\nabla B_0\cdot\nabla y\\
-\frac{b_{y}}{qB_{0}}&\frac{b_{x}}{qB_{0}}&0&\frac{B_{z}^{\ast}}{mB_{0}}&\frac{\eta}{qB_0^2}\bol{b}\cp\nabla B_0\cdot\nabla z\\
-\frac{B_{x}^{\ast}}{mB_{0}}&-\frac{B_{y}^{\ast}}{mB_{0}}&-\frac{B_{z}^{\ast}}{mB_{0}}&0&-\eta\frac{\bol{B}^{\ast}\cdot\nabla B_0}{mB_0^2}\\
\frac{\eta}{qB_0^2}\bol{b}\cp\nabla x\cdot\nabla B_0&\frac{\eta}{qB_0^2}\bol{b}\cp\nabla y\cdot\nabla B_0&\frac{\eta}{qB_0^2}\bol{b}\cp\nabla z\cdot\nabla B_0&\eta\frac{\bol{B}^{\ast}\cdot\nabla B_0}{mB_0^2}&0
\end{bmatrix},\label{Jgc2}
\end{split}
\end{equation}
with $\bol{B}^{*}=\bol{B}_0+\frac{mu}{q}\nabla\cp\bol{b}$, 
$B_{\parallel}^{*}=\bol{B}^{*}\cdot\bol{b}$, and $\bol{b}=\bol{B}_0/B_0$. 

Equation \eqref{dfdtlX} preserves the perturbation of the particle number $\delta N$, the perturbed total magnetic moment $\delta\ms{M}$, and the perturbed energy $\delta\mf{H}$, and maximizes the entropy $\delta S$:
\eq{
&\delta N=\int \delta f\,d\bol{\zeta},~~~~\delta\ms{M}=\int \delta f\,w\lr{\frac{\eta}{B_0}}\,d\bol{\zeta},~~~~\delta\mf{H}=\int \delta f\lr{h_0+q\Phi_0},\\
&\delta S = -\int \frac{\delta f^2}{f_0}\,d\bol{\zeta},
}
where $w$ is any function of $\mu=\eta/B_0$.

\section{Concluding remarks}

In this work, we have developed a collision operator for weakly collisional guiding center plasmas within the drift-kinetic framework. Our focus has been on long-wavelength, low-frequency turbulence, which governs large-scale, self-organizing phenomena in both laboratory and astrophysical plasmas. By restricting our analysis to weakly collisional regimes, we derived a five-dimensional kinetic equation that incorporates Coulomb scattering without involving the cyclotron phase, offering a significant reduction in complexity while preserving the essential dynamics of the system.

The guiding center collision operator formulated here is fully determined by the noncanonical Hamiltonian structure of guiding center dynamics and the Coulomb interaction potential. Importantly, the evolution equation for the guiding center distribution function exhibits a metriplectic structure, ensuring the conservation of particle number, momentum, energy, and interior Casimir invariants. This structure provides a robust thermodynamic foundation, allowing for the derivation of an H-theorem that governs the system's approach to equilibrium.

One of the key physical insights derived from this work is that the conservation of interior Casimir invariants, particularly the total magnetic moment, introduces phase space constraints that lead to deviations from Maxwell-Boltzmann statistics in the thermodynamic equilibrium. This can result in self-organized, inhomogeneous density distributions, providing a deeper understanding of the role of collisions in turbulence and transport in plasmas. The derived collision operator thus offers a valuable tool for exploring large-scale plasma dynamics and provides a computationally efficient model for numerical simulations of self-organizing plasma phenomena.

Finally, it is worth noting that the theory can, in principle, be extended to the gyrokinetic framework, allowing for the derivation of a collision operator capable of describing microturbulence. However, such an extension would require the inclusion of finite Larmor radius (FLR) effects in all relevant expressions, which would considerably increase the complexity of the formulas. This would likely reduce the benefits of the gyrocenter reduction, making a return to the full six-dimensional kinetic theory more appealing. Given these challenges, we do not pursue this extension in the present work.

\section*{Statements and declarations}

\subsection*{Data availability}
Data sharing not applicable to this article as no datasets were generated or analysed during the current study.

\subsection*{Funding}
The research of NS was partially supported by JSPS KAKENHI Grant No. 25K07267, No. 22H04936,  and No. 24K00615. 
PJM acknowledges support from the DOE Office of Fusion Energy
Sciences under DE-FG02-04ER-54742. 
%This work was partly supported by MEXT Promotion of Distinctive Joint Research Center Program JPMXP0723833165. 

\subsection*{Competing interests} 
The authors have no competing interests to declare that are relevant to the content of this article.

\end{document}